\documentclass[letterpaper,12pt]{article}
\usepackage[utf8]{inputenc}
\DeclareUnicodeCharacter{0229}{\c{e}}
\usepackage[T1]{fontenc}
\usepackage{amsmath,amssymb,physics}
\usepackage{graphicx,wrapfig}
\usepackage{hyperref}
\usepackage{authblk}

\hypersetup{
  colorlinks=true,
  hypertexnames=false,
  linktocpage,
  colorlinks=true, 
  urlcolor=magenta!90!black,    
  linkcolor=blue!60!black, 
  citecolor=black!60 
}
\usepackage[backend=biber, style=phys, biblabel=brackets, sorting=none,eprint=true]{biblatex}
\DeclareSourcemap{
  \maps[datatype=bibtex]{
    \map{
      \step[fieldset=abstract, null]
      \step[fieldset=file, null]
      \step[fieldset=annote, null]
      \step[fieldset=keywords, null]
    }
  }
}
\usepackage[capitalize, compress]{cleveref}
\usepackage{xcolor}
\usepackage{enumitem}
\usepackage{geometry}
\usepackage{float}
\usepackage{subcaption}

\usepackage{xcolor}
\title{Strategic Plan for \\
Neutral Atom Quantum Computation \\}

\hypersetup{
  colorlinks=true,
  hypertexnames=false,
  linktocpage,
  colorlinks=true,
  urlcolor=magenta!90!black,
  linkcolor=blue!60!black,
  citecolor=black!60
}
\usepackage{geometry}
\usepackage{ulem}
\title{Strategic Plan for \\
Neutral Atom Quantum Computation \\
}

\makeatletter
\renewcommand{\maketitle}{%
  \begingroup
  \renewcommand{\thefootnote}{\fnsymbol{footnote}}%
  \def\@makefnmark{\hbox{\@textsuperscript{\normalfont\@thefnmark}}}%
  \long\def\@makefntext##1{\parindent 1em\noindent
        \hb@xt@1.8em{\hss\@textsuperscript{\normalfont\@thefnmark}}##1}%
  \begin{flushleft}%
    {\LARGE\@title\par}%
    \vskip 1em%
    {\large\@author\par}%
    \vskip 0.3em%
    \vskip 0em%
    {\footnotesize\noindent\tiny $^*$Corresponding author (\href{mailto:amenssen@mit.edu}{amenssen@mit.edu}). Section contact authors are \uline{underlined}:\\[3pt]
    \quad Defining Practical Quantum Advantage: M.~Gullans (\href{mailto:mgullans@umd.edu}{mgullans@umd.edu})\\
    \quad Neutral Atom Quantum Processor Hardware: T.~Wang (\href{mailto:toutwang@fas.harvard.edu}{toutwang@fas.harvard.edu}), T.~Manovitz (\href{mailto:tom.manovitz@weizmann.ac.il}{tom.manovitz@weizmann.ac.il})\\
    \quad Scalable Photonic Control Technologies: A.\,J.~Menssen (\href{mailto:amenssen@mit.edu}{amenssen@mit.edu})\\
    \quad Quantum Error Correction:J.~Taylor (\href{mailto:jmtaylor@umd.edu}{jmtaylor@umd.edu}) \\
    \quad Compilation of Quantum Circuits: J.~Cong (\href{mailto:cong@cs.ucla.edu}{cong@cs.ucla.edu})\\
    \quad Networking Quantum Processors: J.~Sinclair (\href{mailto:josiah.sinclair@wisc.edu}{josiah.sinclair@wisc.edu})%
    \par}%
  \end{flushleft}%
  \endgroup
}
\makeatother

\author[1]{\uline{Adrian J. Menssen}$^*$}
\author[2,3]{\uline{Tout Wang}}
\author[4]{\uline{Michael Gullans}}
\author[5]{\uline{Tom Manovitz}}
\author[4]{\uline{Jacob M. Taylor}}
\author[6]{\uline{Jason Cong}}
\author[7]{\uline{Josiah Sinclair}}
\author[1]{Ziv Aqua}
\author[8]{Daniel J. Blumenthal}
\author[2]{J. Pablo Bonilla Ataides}
\author[2]{Johannes Borregaard}
\author[9]{Antoine Browaeys}
\author[1]{Paola Cappellaro}
\author[1]{Soonwon Choi}
\author[10]{Alexandre Cooper}
\author[11]{Robin C\^ot\'e}
\author[12,13]{Jacob P. Covey}
\author[14]{Alexandre Dauphin}
\author[15]{Ivana Dimitrova}
\author[16]{Matt Eichenfield}
\author[1]{Dirk Englund}
\author[16]{Jacob Freedman}
\author[17]{Akihisa Goban}
\author[2]{Brandon Grinkemeyer}
\author[2]{Andi Gu}
\author[1]{Ruonan Han}
\author[18,19]{Dominik Hangleiter}
\author[1]{Aram W. Harrow}
\author[13]{Liang Jiang}
\author[20]{Eun-ah Kim}
\author[1]{Felix W. Knollmann}
\author[21]{Aleksander Kubica}
\author[9]{Thierry Lahaye}
\author[14]{Lucas Lassabliere}
\author[2]{Joonho Lee}
\author[22]{Bingzhao Li}
\author[22]{Mo Li}
\author[6]{Wan-Hsuan Lin}
\author[2]{Mikhail D. Lukin}
\author[2]{Varun Menon}
\author[1]{Thomas Propson}
\author[7]{Akbar Safari}
\author[7,23]{Mark Saffman}
\author[14]{Pascal Scholl}
\author[24,25]{Alexander Schuckert}
\author[26]{Giulia Semeghini}
\author[27]{Jonathan Simon}
\author[1,28]{David Spierings}
\author[2]{Daniel Bochen Tan}
\author[1]{Shai Tsesses}
\author[1]{Vladan Vuletic}
\author[6]{Hanrui Wang}
\author[6]{Hanyu Wang}
\author[2]{Susanne Yelin}
\author[29,30,31]{Johannes Zeiher}
\author[1]{Hengyun Zhou}
\affil[1]{Massachusetts Institute of Technology, 77 Massachusetts Avenue, Cambridge, MA 02139, USA}
\affil[2]{Harvard University, Cambridge, MA 02138, USA}
\affil[3]{Present address: QuEra Computing Inc., 1284 Soldiers Field Road, Boston, MA 02135, USA}
\affil[4]{Joint Center for Quantum Information and Computer Science, NIST/University of Maryland, College Park, College Park, MD 20742, USA}
\affil[5]{Weizmann Institute of Science, Rehovot 7610001, Israel}
\affil[6]{University of California, Los Angeles, 405 Hilgard Avenue, Los Angeles, CA 90095, USA}
\affil[7]{University of Wisconsin–Madison, Department of Physics, University of Wisconsin-Madison, 1150 University Avenue, Madison, WI 53706, USA}
\affil[8]{University of California Santa Barbara, Santa Barbara, CA 93106, USA}
\affil[9]{Universit\'e Paris-Saclay, Institut d'Optique Graduate School, CNRS, Laboratoire Charles Fabry, 91127 Palaiseau Cedex, France}
\affil[10]{Institute for Quantum Computing, University of Waterloo, Waterloo, ON N2L 3G1, Canada}
\affil[11]{University of Massachusetts Boston, 100 Morrissey Blvd., Boston, MA 02125, USA}
\affil[12]{The University of Illinois at Urbana-Champaign, Urbana, IL 61801, USA}
\affil[13]{The University of Chicago, 5801 S. Ellis Ave., Chicago, IL 60637, USA}
\affil[14]{PASQAL SAS, 24 Rue Emile Baudot, 91120 Palaiseau, France}
\affil[15]{Northeastern University, 147 S. Bedford St, Burlington, MA 01803}
\affil[16]{University of Colorado Boulder, Boulder, CO 80309, USA}
\affil[17]{Nanofiber Quantum Technologies, Inc. (NanoQT), 1-22-3 Nishiwaseda, Shinjuku-ku, Tokyo, 169-0051, Japan}
\affil[18]{Simons Institute for the Theory of Computing, University of California, Berkeley, CA 94720, USA}
\affil[19]{ETH Z\"urich, R\"amistrasse 101, 8092 Z\"urich, Switzerland}
\affil[20]{Cornell University, 616 Thurston Ave., Ithaca, NY 14853, USA}
\affil[21]{Yale University, New Haven, CT 06520, USA}
\affil[22]{University of Washington, Seattle, WA 98195, USA}
\affil[23]{Infleqtion, Madison, WI, 53703}
\affil[24]{Quera Computing UK Ltd., Unit 2, Tech Foundry 2, Didcot OX11 0QQ, UK}
\affil[25]{DIENS, École Normale Supérieure, PSL University, CNRS, INRIA, 45 rue d’Ulm, Paris 75005, France}
\affil[26]{Harvard John A. Paulson School of Engineering and Applied Sciences, Harvard University, Cambridge, MA 02138, USA}
\affil[27]{Stanford University, 382 Via Pueblo, Stanford, CA 94305, USA}
\affil[28]{Elmore Family School of Electrical and Computer Engineering, Purdue University, West Lafayette, IN 47907, USA}
\affil[29]{Ludwig-Maximilians-Universit\"at M\"unchen (LMU Munich), Geschwister-Scholl-Platz 1, 80539 Munich, Germany}
\affil[30]{Max Planck Institute of Quantum Optics (MPQ), Hans-Kopfermann-Str. 1, 85748 Garching, Germany}
\affil[31]{planqc GmbH, Lichtenbergstrasse 8, 85748 Garching bei M\"unchen, Germany}

\date{\today}
\begin{document}

\maketitle

\begin{abstract}
\noindent 
We present a strategic plan for neutral atom quantum computation, bringing together hardware development and theory advancements to achieve the goal of practical quantum advantage. The concept of practical quantum advantage is defined, along with how to verify claims of advantage, and approaches to designing quantum algorithms that deliver practical advantage. Future directions for neutral quantum processor hardware are described: scaling-up system size,  Qubit encodings and atomic platforms,  going further below threshold with neutral-atom logical-qubit performance, continuous reloading of qubits, and fast readout. We also explore opportunities for scalable integrated photonic control technologies. Alongside hardware advancements, new developments in quantum error correction and compilation of quantum circuits are proposed. Finally, we examine the opportunity of networking multiple neutral atom quantum processors together to perform distributed quantum computing and overcome possible limitations of a single system.
\end{abstract}

\newpage
\tableofcontents

\section*{Introduction} 
\label{sec:intro}
Quantum computing architectures have matured significantly over the past decade, with several qubit platforms in a close race toward a common goal: quantum computation beyond the noisy intermediate-scale quantum (NISQ) regime, reaching practical utility.
Recent experiments have demonstrated below-threshold error correction in neutral atoms for the first time~\cite{Bluvstein_2025FT}. This progress was crucially enabled by improvements to neutral atom gate fidelity~\cite{Evered_2023,Tsai2025Benchmarking,Evered:2026jrf} and by dynamically reconfigurable arrays, in which coherent atom transport provides long-range entanglement and effectively all-to-all connectivity \cite{bluvstein2022quantum}.

Similarly, significant advances have been made in  quantum error correction theory that drastically reduce physical resources to achieve desired protection against errors \cite{bravyi2024high,xu2024,Webster:2026lag}. 
Together, experimental and theoretical progress makes the goal of reaching practical quantum utility in neutral atoms seem attainable. However, significant challenges remain around scaling and controlling systems of the required size. 

In this strategic plan, we present a roadmap for scaling up the neutral atom quantum computing platform, based on the NSF Town Hall on Advancing Quantum Computing with Neutral Atoms that was convened at MIT Endicott House in January 2025. The concept of practical quantum advantage is defined in Section~\ref{sec:practicalAdvantage}, in particular what it means, how to verify claims of advantage, and what algorithms give practical advantage. In Section~\ref{sec:atomControl}, we describe the path forward for scaling up neutral atom quantum processor hardware. Section~\ref{sec:lightControl} details approaches to scalable photonic control. Sections~\ref{sec:QEC} and~\ref{sec:compilation} address the topics of quantum error correction and quantum circuit compilation, respectively. Going beyond the limits of a single neutral atom quantum processor, we explore in Section~\ref{sec:network} the prospects for networking multiple quantum processors for distributed quantum computation.
\begin{figure}
\centering
\includegraphics[width=1\linewidth]{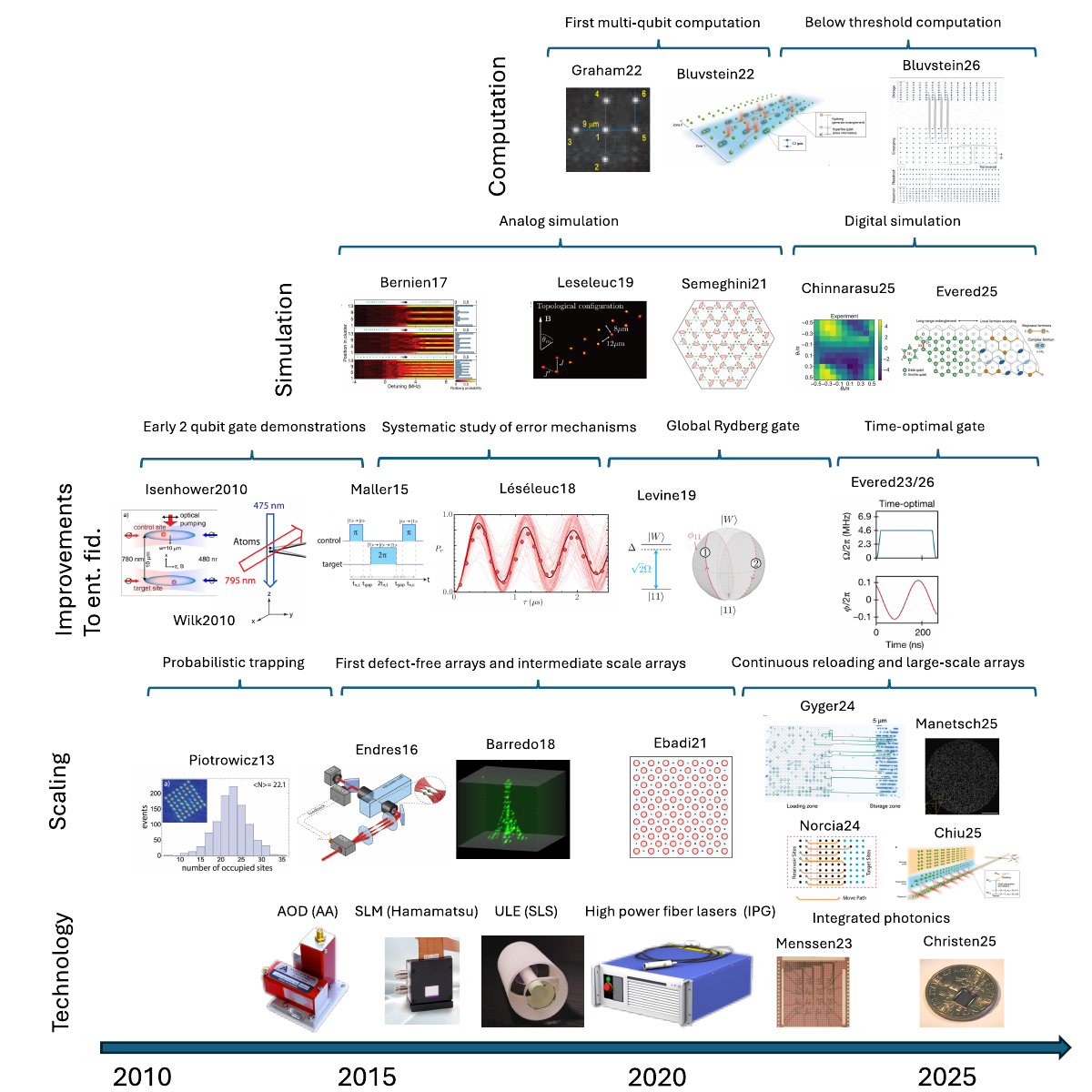}
\caption{Overview of the evolution of neutral atom quantum computation.
Neutral-atom efforts are shown along five categories-Computation, Simulation, Improvements to entanglement fidelity, Scaling, and Technology stack-with representative papers.
Computation: first multi-qubit computation~\cite{Graham_2022,bluvstein2022quantum} $\rightarrow$ below-threshold computation~\cite{Bluvstein_2025FT}.
Simulation: analog Rydberg simulators~\cite{bernien2017probing,de_Leseleuc_2019,Semeghini_2021} $\rightarrow$ digital simulation \cite{Chinnarasu2025,Evered_2025_kitaev}.
Improvements to entanglement fidelity: early two-qubit gate demonstrations~\cite{Isenhower_2010,Wilk_2010} $\rightarrow$ systematic study of error mechanisms~\cite{Maller_2015,leseluec2018} $\rightarrow$ global Rydberg gate~\cite{Levine_2019} time-optimal gate~\cite{Evered_2023,Evered:2026jrf}.
Scaling: probabilistic trapping ~\cite{Piotrowicz_2013} $\rightarrow$ first defect-free and intermediate-scale arrays ~\cite{endres2016atom,Barredo_2018,Ebadi_2021} $\rightarrow$ continuous reloading and large-scale arrays~\cite{norcia2024iterative,Gyger_2024,Manetsch_2025,Chiu_2025}.
Technology stack: commercial AOD (AA-optoelectronics), SLM (Hamamatsu), Ultra Low Expansion Cavity - ULE (Stable Laser Systems), and high-power fiber lasers (IPG photonics); integrated photonic modulators~\cite{menssen2023scalable,christen2025integrated}.}
\label{fig:overview}
\end{figure}

Figure~\ref{fig:overview} presents an overview of the field and its core scientific advancements together with the technologies that enabled them. 

Digital quantum computation begins with the first demonstrations of entangling and single qubit gates on multiple atoms. Local optical addressing and atom-shuttling-mediated long-range entanglement enabled the first simple algorithms.  Increases in system scale and improved entangling operations culminated in the first below-threshold error-corrected computations. 

Quantum simulation traces a parallel development: from small-scale analog simulators probing the dynamics of strongly interacting Rydberg systems, to analog simulation of topological phases of matter. Finally, the two fields converge in the first demonstrations of digital quantum simulations. 

The evolution of Rydberg entangling gates has seen several notable improvements, from the early pioneering work using individually addressed gate pulses, to the globally addressed CZ gates and time-optimal gates. 

The ease with which qubits can be scaled in neutral atoms is a key advantage of this platform. The number of trapped atoms using optical tweezers or optical lattice trapping in free space is limited primarily by available laser power. Early experiments loaded small arrays probabilistically; feed-forward-assisted sorting then assembled defect-free arrays of hundreds of atoms. Finally, continuously reloaded systems solved the problem of resupplying atoms lost during gate operations: A remote source (e.g., an active MOT) replenishes lost atoms and delivers them to the computation region.

The technology stack supporting these developments encompasses several key components: Acousto-optic deflectors (AODs) enabled for the first time assembling defect-free arrays of atoms. Liquid Crystal on Silicon (LCOS) spatial light modulators have made large trap arrays possible by enabling holographic projection of spot arrays generating thousands of tweezer traps. Utilizing methods from clock-research - laser stabilization to ultra-low expansion cavities (ULE-cavities) - has critically enabled improvements in two qubit gate fidelity. The adoption of low-noise, high-power fiber-amplifier lasers has enabled scaling tweezer trapping powers and lasers for quantum gates.
Future scaling of this architecture to the regime of quantum utility will likely require novel approaches to light control and generation, harnessing the inherent scalability of integrated optics for light control.

\begin{figure}[h!]
\centering
\includegraphics[width=0.75\linewidth]{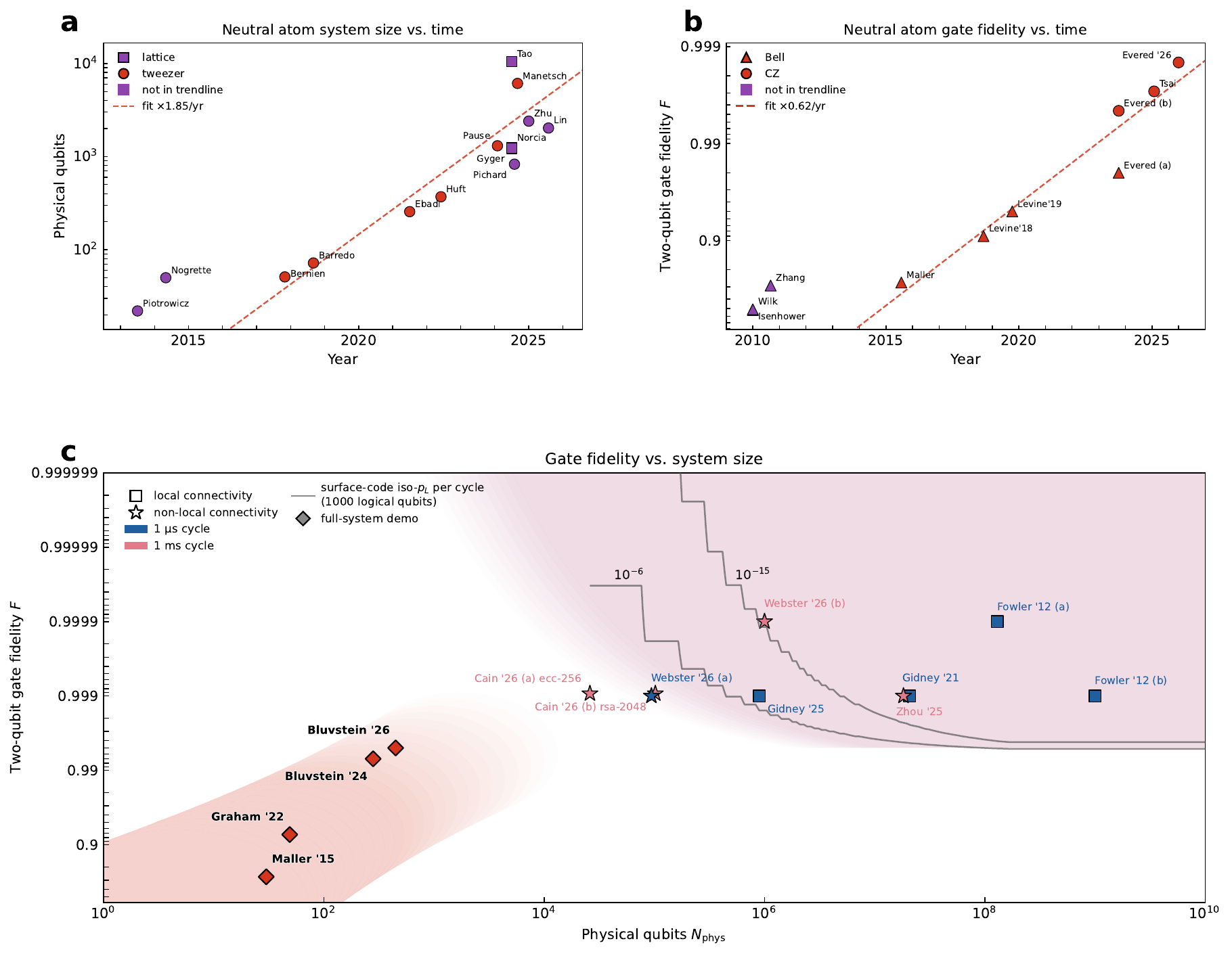}
\caption{Plot of scaling trends, best in class results. (a) Trends for qubit number over time. Fit through current tweezer architecture data-points, early pioneering results were excluded from the trend-line. Optical lattice architectures are shown, but not included in the trend-line. Several tweezer architectures close to best in class are shown but not fitted.
(b) Two-qubit fidelity improvement over time. Early pioneering work not included in trend-line.
(c) Advancing fronts of experimental and theoretical progress in fidelity vs. physical qubit space. Shaded regions indicate experimental (lower left) and theoretical advances (upper right). Full-system experimental demonstrations and several theoretical results for different error correction architectures are given: surface codes, assuming nearest-neighbor connectivity (Fowler, Gidney).
Grey curves represent physical qubit cost-curve for 1000 logical qubits with minimum indicated per-cycle  logical qubit error of $10^{-6} $ and $10^{-15} $ using the surface code scaling model from \cite{fowler2012surface}.
Theory data points are estimates for 2048 bit RSA (256 ECC) resource requirements.
Cain reference assumes a neutral atom system and qLDPC architecture. Webster assumes a qLDPC architecture and full qubit connectivity.
\textit{Neutral atom qubit number:}~\cite{Piotrowicz_2013,Nogrette_2014,Bernien_2017,Barredo_2018,Ebadi_2021,Huft_2022,Pause_2024,Manetsch_2025,Gyger_2024,tao2024high,Pichard_2024,Norcia_2024,lin2024aienabled,Zhu_2025}.
\textit{Neutral atom fidelity:}~\cite{Isenhower_2010,Wilk_2010,Zhang_2010,Maller_2015,Levine_2019,Evered_2023,Tsai_2025,Evered:2026jrf,Levine_2018}.
\textit{Theory anchors (panel c):}~\cite{Fowler_2012,gidney_how_2025,Gidney_2021,Cain:2026rmb,Webster:2026lag,Zhou:2025jmi}.
\textit{Full-system demonstrations (panel c):}~Neutral atoms --- Maller '15~\cite{Maller_2015}, Graham '22~\cite{Graham_2022}, Bluvstein '24~\cite{Bluvstein_2023}, Bluvstein '26~\cite{Bluvstein_2025}.}
\label{fig:utility_race}
\end{figure}
In Figure~\ref{fig:utility_race} we illustrate some general trends in system size
and gate fidelity improvements over time. In the spirit of Moore's law, we fit the
increase in qubit count (a) and two-qubit fidelity (b) to exponential scaling laws
through a set of hallmark experimental demonstrations.
We only fit to the best results for each category (size and fidelity) at each point in time. 
The number of physical qubits has roughly increased by a factor of $\simeq$ 1.8 per year, while gate errors have decreased by a factor of $\simeq$ 0.6 per year. This can of course only provide a coarse estimate. For entangling gates conventions changed over time. While earlier results have predominantly cited Bell fidelities, more recent results have used CZ gates as benchmark, making comparisons more challenging.
Research into improving system size and gate fidelity often represents independent
efforts, with state-of-the-art system sizes not usually implementing
state-of-the-art gate fidelity in the same system, or only realizing a restricted
set of the full control degrees of freedom. The requirement of any new quantum gate
implementation is however that it is scalable in principle to the full system size.
This gap must be closed by engineering suitable scalable quantum gate controllers.

In Panel (c),
gate fidelity is plotted vs.\ physical qubit number.
We show selected ``full system demonstrations'', which represent systems with
the largest demonstrated physical qubit count as well as single- and two-qubit gate
control.
Additionally we show hallmark theoretical results realizing RSA 2048-bit integer
factorization / (ECC-256) and their respective resource requirements, progressing
from surface codes and their improvements to newer qLDPC architectures. See
Section~\ref{sec:QEC} for a detailed discussion of these developments. The grey
curves represent a surface code benchmark of 1000 error-corrected logical qubits at per cycle error rates of $10^{-6}$ and $10^{-15}$. 

The utility threshold and physical hardware capability are converging, as more resource-efficient codes and improved architectures advance from both directions.
Under the scaling rates of the past decade, neutral atoms could reach quantum utility within the next decade. Whether this timeline can be maintained will depend crucially on the ability to continually scale qubit counts to reach the $\sim10^5-\sim10^6$ regime. The rate of improvement in gate fidelity may slow down due to either physical limits or reduced pressure once error thresholds for fault tolerant operations are sufficiently surpassed. Another crucial challenge will be the ability to scalably control quantum processors of significantly increased qubit count compared with current demonstrations.

\section{Defining Practical Quantum Advantage} \label{sec:practicalAdvantage}

The ultimate goal of building a neutral-atom quantum computer is to demonstrate quantum advantage—solving a computational task beyond the reach of classical hardware.
In analog quantum simulators with restricted programmability, tasks at the edge or beyond the current capabilities of current classical numerical techniques have been achieved as early as ten years ago~\cite{choi2016exploring,shaw_benchmarking_2024,king_beyond-classical_2025,andersen_thermalization_2025,chen_spectroscopy_2025}.
In more recent years, we have seen first demonstrations of computations with robust evidence for quantum advantage on universal, programmable devices ~\cite{arute2019quantum,wu_strong_2021,decross_computational_2024,morvan_phase_2024,gao_establishing_2025}. 
However, so far they have been achieved for highly tailored tasks designed for the demonstration of  quantum advantage. 
Ultimately, for a quantum computer to be useful, quantum advantage must be achieved for a practically relevant task. 
We propose a somewhat minimal notion of \emph{practical quantum advantage}: 
\begin{quote}
    A \emph{practical quantum advantage} is achieved when a quantum computation has been performed such that the following are true.
\begin{itemize}
    \item  (\emph{Correctness}) The computation correctly solves a computational task.\footnote{To be meaningful, the task specification must include a targeted precision.} 
    \item (\emph{Advantage}) That task cannot be solved on the available classical hardware at the time. 
    \item (\emph{Scaling}) The quantum algorithm has a scaling advantage compared to the best classical algorithms.
    \item (\emph{Practicality}) The task is relevant to a community external to the builders of the hardware. 
\end{itemize}

\end{quote}

Our definition of quantum advantage is deliberately strong in requiring the task to be not solvable \emph{at all} on current classical hardware.
Weaker forms of advantage, for example, an advantage in terms of runtime or energy consumption of a quantum computer may be a next step. 
However, such notions of quantum advantage are fragile in that they may only require small improvements in classical algorithms to be removed. Moreover, our definition of practical quantum advantage naturally extends the existing notion of quantum advantage or ``quantum supremacy'' and thus ensures that its achievement is a step forward from random circuit sampling experiments rather than a completely different form of advantage claim.

Naturally, (practical) quantum advantage is not an absolute notion. For instance, the runtime of comparable classical algorithms remains a moving target, and so does the scaling of the quantum and classical hardware. 
Likewise, what it means to be ``practical'' necessarily remains vague. 
Quantum advantage must therefore be assessed on a case-by-case basis according to different criteria.
\begin{enumerate}

    \item For different problems, \emph{correctness} is more clear-cut and verifiable than for others. On the one hand, whether or not an integer has been factored is checkable in a split-second. On the other hand, knowing whether a computed spectrum of a molecule is correct may not be verifiable at all except by comparison to the measured spectrum, which may be a difficult task. Therefore, tasks for which verification is either possible and/or where the performance of a quantum computer can be guaranteed are to be preferred over tasks where this is not true.

    Determining correctness is also a tricky point in the context of analog simulators, which may be allowed to approximately solve a computational task in the presence of experimental noise. By contrast to fault-tolerant quantum computations using algorithms with performance guarantees, determining the precision of analog simulation must be done on a case-by-case basis using characterization techniques for analog dynamics.
    
    \item How confident are we in the validity of the \emph{advantage claim}? For example, are there complexity-theoretic arguments for the classical hardness of the task achieved on the quantum computer? How extensively has the task been studied in the literature? Has it actually been tried to solve the task on classical hardware?  A custom-tailored task may bear the risk of being hard classically only because it has not been studied.  
    
    \item To what extent can the correctness of the computation or the \emph{scaling} advantage claim be verified? 
    Different levels of rigor in a notion of verification can convince a certain level of skepticism.
    Is extrapolation involved? What model of a classical adversary is used? How statistically significant is the advantage?

    \item Finally, one may be tempted to define the \emph{practicality} of a task in terms of dollar value of solving the task. However, we think this notion is unfit for measuring scientific progress (e.g. what is the dollar value of predicting properties of a heavy-ion collision?) and therefore propose to simply consider the importance or degree of investigation a task has experienced in a community external to the builders of the hardware.  
\end{enumerate}
To define different regimes of quantum advantage, we need a quantitative measure of the ``size'' of the quantum computer. We define the size by the number of logical qubits and the number of quantum operations ``quops''. In fault-tolerant quantum computers, a quop is typically measured as the number of magic gates such as Toffoli or $T$ gates. However, we propose a slightly more nuanced definition: we define a quop as an operation which can be done on a single or two qubits within one error-correction cycle. This definition means that the maximum number of quops a quantum computer can perform is given by the inverse of the logical error rate per syndrome extraction cycle and qubit. This definition takes into account the operations required to create magic gates and the non-trivial number of non-magic gates involved in many algorithms.  In reconfigurable architectures, a transversal gate requires roughly one cycle and therefore, a transversal gate is one quop. The Toffoli count is a lower bound on the Quop count and the exact overhead is architecture and algorithm dependent, but is typically a factor of 10-100. Therefore, a circuit with $10^4-10^5$ Toffoli/T gates can be a megaquop circuit in our definition.

Based on the criteria outlined above, we expect there to be roughly three meaningful regimes of (practical) quantum advantage corresponding to the size of the available quantum computers.

\begin{itemize}
    \item \emph{Weak unverifiable quantum advantage} can be achieved using $10^3 - 10^6$ (k/M) Quops (quantum operations). 
    For example, the recent random circuit sampling experiments have used a low number of kQuops \cite{arute2019quantum,wu_strong_2021,decross_computational_2024,morvan_phase_2024,gao_establishing_2025}, and are not verifiable in the advantage regime. 

    \item \emph{Early practical advantage} can plausibly be achieved using $10^6 - 10^9$ (M/G) Quops. This includes the generation of classically certifiable random numbers \cite{brakerski_cryptographic_2018,aaronson_certified_2023,liu_certified_2025}, and specific quantum simulation problems \cite{King2025,haghshenas_digitalquantummagnetismfrontier_2025,alam2025fermionic,alam2025programmable,granet2025superconducting}. 
    In this regime, quantum advantage claims can also be classically verified using protocols such as tailored factoring or proofs of quantumness \cite{brakerski_simpler_2020,kahanamoku-meyer_classically_2022,liu_depth-efficient_2022,zhu2021interactive,kahanamoku-meyer_jacobi_2024}. 

    \item \emph{Broad practical quantum advantage} is what we hope to achieve eventually in the regime of $10^9 - 10^{12}$ (G/T) Quops for tasks such as quantum chemistry \cite{Abrams1997,aspuru-guzik2005,mcardle2020quantum}, material science \cite{Babbush2018,Bauer2020} nuclear physics \cite{watson_quantum_2023}, cryptography \cite{gidney2019how,litinski_how_2023} or solving optimization problems \cite{farhi2001quantum,jordan_optimization_2024}. 
\end{itemize}

We outline the challenges, opportunities, and research directions in the field of practical quantum advantage according to the criteria above, that is, related to (1) \textbf{Definition} of quantum advantage, (2) \textbf{Verification} of quantum advantage, and (3) \textbf{Algorithms} for practical quantum advantage. 

\subsection{Challenges and Opportunities}

\paragraph{Definition of quantum advantage}
The key challenge in the context of delineating a quantum advantage is that we need to operate in a context of moving parts---quantum and classical hardware are developing rapidly, and so do classical algorithms in simulating quantum computation in general, or specific quantum advantage tasks in particular. 
This challenge is particularly pressing in the near term, when we expect quantum advantage to be achieved for tasks that are to some extent tailored towards quantum computers in general, or even specific quantum hardware. 
More generally, we are facing the problem that the field of simulating general quantum computations is only recently developing \cite{markov_simulating_2008,van_den_nest_simulating_2011,bravyi_improved_2016,zhou_what_2020,pan_simulation_2022,pan_solving_2022,kalachev_classical_2021,kalachev_recursive_2021,gray_hyperoptimized_2024}, and simulations of tailored problems such as transversal logical circuits \cite{hangleiter_fault-tolerant_2024,maslov_fast_2024,codsi_classically_2023}, decoded quantum interferometry \cite{jordan_optimization_2024} or certain quantum simulations have not received much attention in the algorithms literature at all. 
This results in the danger of advantage claims on classically unstudied problems, for which the target can change drastically from one day to the next. 
A recent prime example of this is the experiment on the kicked Ising model, performed by the IBM team \cite{kim_evidence_2023} on a previously unstudied lattice geometry, which enabled its subsequent classical simulation \cite{tindall_efficient_2024,begusic_fast_2024,shao_simulating_2024,kechedzhi_effective_2024,rudolph_classical_2023}. By contrast, recent studies on Quantinuum and Google hardware have focused on the digital simulation of the lattice Ising  model in 2D~\cite{haghshenas_digitalquantummagnetismfrontier_2025} and the simulation of the Fermi-Hubbard  model~\cite{Hemery2024,nigmatullin2024,alam2025fermionic,alam2025programmable,granet2025superconducting}. 
These have been paradigmatic benchmarks for new classical numerical methods, and have been studied extensively using numerical methods and  are therefore more suitable as a benchmark for quantum advantage.
Another problem arises when comparing advantage claims achieved on different problems and different hardware, as in the context of random circuit sampling with different gate sets and connectivities, IQP sampling, and different variants of boson sampling \cite{arute2019quantum,decross_computational_2024,bluvstein_logical_2024,wang_boson_2019,zhong_quantum_2020,madsen_quantum_2022}. 
Again, this problem arises in particular in the near term when problems remain tailored in order to reduce overhead as much as possible, but we expect such optimizations to remain crucial in the foreseeable future. 

As a result of this, the development of classical simulation techniques and algorithms as well as a better complexity-theoretic understanding of quantum computation are required for establishing practical quantum advantages. 
These efforts can help us obtain a better understanding of ``exploits'' that classical techniques can use and therefore delineate further the regime in which we can hope for quantum speedups. 
It can help us understand better which scenarios quantum computers are useful in. 
A community-agreed standard of what constitutes a quantum advantage and metrics to measure their quality on different scales can contribute to the more accurate comparison of computations performed on different experimental platforms, and thereby also the computational properties of those platforms themselves. 

\paragraph{Verification}
A key aspect of performing computations with quantum advantage is that they must (initially) be verified in order to gain trust in the correct functioning of the quantum computer. 
This is a challenge since, after all, we are aiming to perform computations which cannot be performed on alternative tested and trusted classical hardware. 
Unfortunately, the status of verification of quantum advantage remains in a limbo of sorts: on the one hand, the current demonstrations based on random circuit sampling are not efficiently verifiable and extrapolations are necessary to gain some trust in the correctness of the computations in the advantage regime.
What is more, the available benchmarks are vulnerable to classical attacks \cite{gao_limitations_2024}. 
On the other hand, the smallest verifiable tests of quantumness involve performing arithmetic operations in superpositions---be it for cryptographic proofs of quantumness \cite{brakerski_cryptographic_2018,kahanamoku-meyer_classically_2022}, factoring \cite{shor_algorithms_1994,gidney_how_2021,kahanamoku-meyer_jacobi_2024}, or code-based schemes \cite{yamakawa_verifiable_2022}---for which current algorithms require at least MQuops. 
What is more, there is a lack of efficient techniques for verifying structured computational problems in the regime of early practical advantage such as quantum simulation. 

The development of efficient verification techniques for quantum advantages can first and foremost generate an increased trust in the viability of speedups on quantum computers, and the possibility of building such devices. 
More concretely, such techniques will be applicable in generating certifiably secure random numbers generated at high rates. 
These find applications in cryptography as well as randomized algorithms. 

\paragraph{Quantum algorithms}
The arguably most important challenge in the context of practical quantum advantages is, first, that there are only few practical quantum algorithms with a clear exponential quantum advantage and that those algorithms have---in the cases that we know---immense physical resource requirements on the order of GQuops \cite{gidney_how_2021,litinski_how_2023}.
In the context of quantum simulation of materials, chemistry, and high-energy physics these resource requirements are in part due to the high overhead involved in mapping fermions and bosons to qubits (see e.g.\ Ref.~\cite{Derby2021,crane2024hybrid}) and the need to perform small-angle rotations which in the fault-tolerant setting require $10-100$ Quops per rotation.

Ultimately, the search for new quantum algorithms can broaden the range of useful applications for large-scale fault-tolerant quantum computers. 
In the specific setting of practical quantum advantages, there is scope for interesting algorithms in the regime of early practical advantage that may be implemented within the next five years. 
Those may use new algorithmic ideas as well as reductions of physical resources using the parallelizability and long-range connectivity which are specific resources of neutral-atom quantum computers. \\

Altogether, a concerted effort to clearly identify target problems with practical advantages bears the promise to focus and speed-up research on the three dimensions of practical advantages, bringing practical advantages within closer reach. 

\subsection{Research Directions}

\begin{figure}
\centering
    \includegraphics[width=\linewidth]{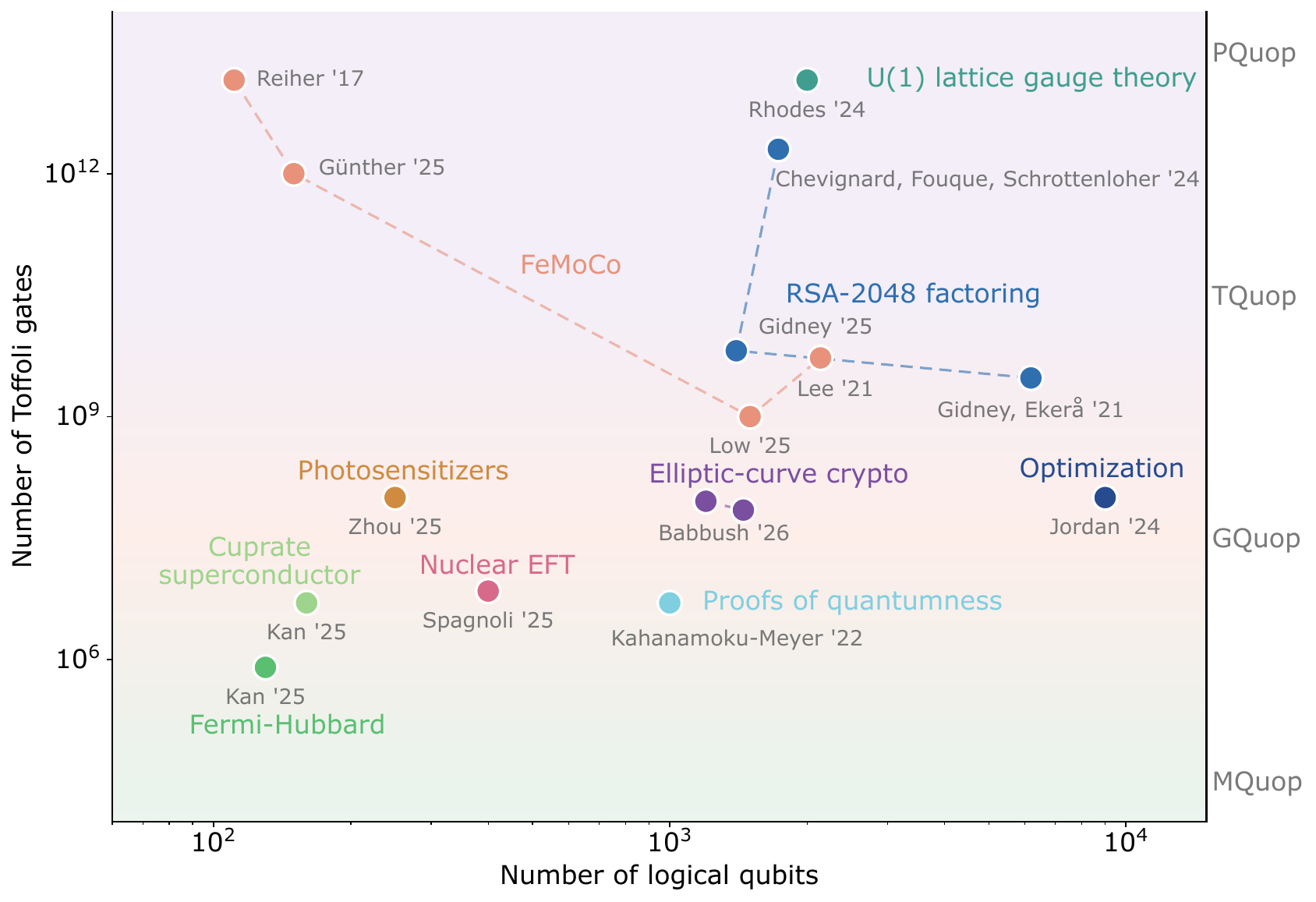}
    \caption{Examples of explicit resource estimates for key applications. Each marker is an explicit resource estimate in terms of number of logical qubits and Toffoli or T gates (using 1 Toffoli = 7 T gates). The right vertical axis restates that count
in Quops: one Quop is an operation that can be performed within one syndrome extraction cycle. The maximum Quop count of a QEC code is therefore roughly its inverse logical error rate per logical qubit. The
Toffoli count is a lower bound on the Quop count and the exact overhead (roughly
10 to 100, depending on the Clifford count per Toffoli and on the magic-state
factory) is architecture-dependent. To give a rough impression for the quop regimes  we estimated quop=10$\times$ Toffoli and indicated rough MQuop/GQuop regimes by shaded colored bands.
Counts are as reported by each reference: proofs of quantumness \cite{kahanamoku-meyer_classically_2022}
($1000$; $\sim\!5\times10^{6}$); Fermi-Hubbard and cuprate superconductor models
\cite{kan_resource-optimized_2025} ($123$ ($7\times 7$ lattice), $8\times10^{5}$; and $160$($8\times 8$ lattice),
$5\times10^{6}$); optimization by decoded quantum interferometry
\cite{jordan_optimization_2024} ($9\times10^{3}$; $10^{8}$); first-quantization
nuclear dynamics in pionless EFT \cite{spagnoli_quantum_2025} (few hundred; tens
of millions T gates); a 2D U(1) lattice gauge theory \cite{rhodes_exponential_2024}
($2000$; $\sim\!10^{14}$~T); BODIPY photosensitizers \cite{zhou_quantum_2025}
($180$ to $350$; $10^{7}$ to $10^{9}$). Dashed lines trace one problem across
methods. FeMoCo: Reiher \cite{reiher_elucidating_2017} ($111$; $10^{14}$~T),
single-ancilla \cite{guenther_phase_2025} ($150$; $\sim\!10^{12}$), tensor
hypercontraction \cite{lee_even_2021} ($2142$; $5.3\times10^{9}$), spectrum
amplification \cite{low_fast_2025} ($\sim\!1500$; $10^{9}$). RSA-2048
\cite{chevignard_reducing_2024,gidney_how_2025,Gidney_2021}: ($1730$;
$2\times10^{12}$), ($\sim\!1400$; $6.5\times10^{9}$), ($6189$; $3\times10^{9}$).
Elliptic-curve discrete log (256-bit) \cite{babbush_securing_2026}: two operating
points near $1200$ to $1450$ qubits, $7$ to $9\times10^{7}$ Toffoli.
Four qubit counts are not stated by their references; the value plotted is an estimate and based on: proofs of quantumness, $1000$, the register size for modular squaring
on a $\sim\!1000$-bit modulus; U(1) lattice
gauge theory, $2000$, the low end of the $10^{3}$ to $10^{4}$ logical-qubit range
reported; and single-ancilla FeMoCo, $150$,
the $108$-orbital active space plus one ancilla and compilation overhead, read
from Fig.~1 of Ref.~\cite{guenther_phase_2025}.}
    \label{fig:advantage_overview}
\end{figure}

We propose to focus research efforts on the problems described in the following sections.  A qualitative illustration of the trajectory we envision for neutral atom hardware is shown in Fig.~\ref{fig:advantage_overview}.  We expect specialized examples of practical quantum advantage are achievable at the MQuop scale that is achievable based on current hardware demonstrations, but going beyond this to much larger scale quantum processors with broader practical advantage remains an open problem.

\subsubsection{Definition of Quantum Advantage}

\paragraph{Standard and metric development} 
Currently, the most solid evidence for quantum advantage has been achieved using variants of quantum random circuit sampling \cite{hangleiter_computational_2023}. 
Due to the differences in gate sets, architecture connectivity, and quality of the operations of the different experiments (compare, e.g., Refs.~\cite{arute2019quantum,decross_computational_2024,bluvstein_logical_2024,wang_boson_2019,zhong_quantum_2020,madsen_quantum_2022,ransford2025helios98qubittrappedionquantum}), even claims for this comparably small and well-defined set of problems are hard to compare in terms of the advantage they provide over classical computers.
As quantum advantages will be achieved for more and more diverse problems, the problem of comparing different claims becomes even more pressing. 
This is why we suggest an effort to develop a set of standards and metrics for quantum advantage claims. 
Those should involve and indicate the degree of the advantage and our confidence in it, both in terms of the complexity-theoretic evidence and experimental validation or verification of the successful achievement of the task. 

One metric towards this end is the \emph{quantum volume} \cite{cross_validating_2019}, but other measures that may be better suited to sampling have been introduced recently such as a mixed state entanglement proxy \cite{shaw_benchmarking_2024}, or the complexity density \cite{decross_computational_2024}. 
It may also be advisable to assess an advantage on several different scales. 

\paragraph{Paradigmatic simulation problems} 
In the context of quantum simulation, a significant issue for quantum advantage claims is the sheer diversity of the problems proposed, and the lack of a systematic search for classical algorithms.
We therefore suggest to identify and establish a community consensus on a small number of ``key challenge problems''. 
Those problems should have a long history in classical numerical studies and be amenable to a near-term practical quantum advantage. Examples could be new molecular systems, transport dynamics, reaction dynamics, or the simulation of time evolution in the 2D Fermi-Hubbard model. 
The latter has served as a recent frontier benchmark of Noisy Intermediate-Scale Quantum (NISQ) devices: a $2\times8$ lattice of a spinful Hubbard model has been simulated in 2023 for two Trotter steps~\cite{Hemery2024}, growing to a $6\times6$ spinless model for four Trotter steps~\cite{nigmatullin2024} in 2024 and a $4\times4$ spinful model for three steps~\cite{Evered_2025_kitaev} in the beginning of 2025 (with approximate particle-number conservation). Most recently, $6\times6$ and $7\times4$ spinful Fermi-Hubbard models have been evolved for up to four Trotter steps, which reaches the edge of what classical methods can achieve~\cite{alam2025fermionic,alam2025programmable,granet2025superconducting}. 
When defining such a challenge problem it is essential to also set the required accuracy, since there are trade-offs between accuracy and runtime. 
One might seek faster quantum simulations for the same accuracy as classical simulations or more accurate quantum simulations for the same computational runtime as classical simulations.

To make such challenge problems open and systematic, we also encourage the organization of competitions analogous to the NIST competitions involved in the process of establishing post-quantum-secure cryptographic standards. 
In those competitions, both classical and quantum computing communities tackle these problems and could thus provide compelling evidence of superior quantum performance.

\paragraph{Classical simulation techniques} 
The key of any quantitative quantum advantage claim is the comparison to classical simulation algorithms. There are broadly-speaking two ways to perform a classical simulation of a problem targeted with quantum computation. On the one hand, the circuit run on the quantum computer can be simulated classically. On the other hand, the problem can be solved directly.

For the simulation of universal random circuits,  we have seen a flurry of optimized techniques  \cite{markov_simulating_2008,van_den_nest_simulating_2011,bravyi_improved_2016,zhou_what_2020,pan_simulation_2022,pan_solving_2022,kalachev_classical_2021,kalachev_recursive_2021,gray_hyperoptimized_2024}. However, more recently proposed settings in which a (practical) quantum advantage can be achieved have been less explored. 
In particular, we have no detailed understanding of the simulation complexity of transversal logical circuits (IQP circuits) \cite{dalzell_how_2020,codsi_classically_2023,hangleiter_fault-tolerant_2024,maslov_fast_2024}, the recently proposed problem of decoded quantum interferometry (DQI) \cite{jordan_optimization_2024}, the original proposal of Yamakawa and Zhandry \cite{yamakawa_verifiable_2022}, or cryptographic proofs of quantumness \cite{brakerski_cryptographic_2018,brakerski_simpler_2020,kahanamoku-meyer_classically_2022}.

For direct classical simulation, many problems have been attempted to be solved for decades, even before the advent of quantum computers. For instance, the most competitive method for ground-state energy estimation is the density-matrix renormalization group~\cite{schollwock2011densitymatrix}.

It is important that quantum computations are not only compared to circuit-simulation methods, but also the best-available classical method for targeting the problem directly. In particular, it is important to compare resources to achieve the same precision or error. 

\subsubsection{Verification}

There are several possible methods for verifying the output of a quantum algorithm.  In each case, let $x$ denote the input to a quantum computer and let $y$ denote its (possibly random) output.  

\begin{itemize}
\item {\bf Efficient classical verification using public information.}  In this case, a classical verifier can efficiently check whether a given input-output pair $(x,y)$ is valid.  For example, $x$ might be an integer, and $y$ a nontrivial factor.  If coming up with valid $y$ given $x$ is believed to be hard for classical computers (say based on the assumption that factoring is hard classically), then this is an ideal form of verifiable quantum advantage.  Shor's algorithm works here, as do some simpler schemes~\cite{zhu2021interactive,kahanamoku-meyer_classically_2022}.  However, all known protocols of this form rely on performing arithmetic in coherent superposition.

\item {\bf Efficient classical verification using private information.} If the goal is to demonstrate quantum advantage, then we can choose to generate $x$ in a helpful way. In particular, suppose the classical verifier generates a private random string $r$ and then uses it to generate the input $x$ to the quantum computer, which outputs $y$.  The verifier then checks the triple $(r,x,y)$ for validity. Again we need to assume that classical computers cannot efficiently produce valid $y$ from $x$.  Here the key difference is that the quantum computer (or any classical spoofer) only gets to see $x$ and not $r$.  An idea using verifier-private information was proposed in the original IQP paper~\cite{shepherd_temporally_2009}, but this particular family of IQP instances was mostly dequantized in \cite{kahanamoku-meyer_forging_2023,gross_secret_2025}.  Private information is also a key ingredient of certified random number generation~\cite{aaronson_certified_2023,brakerski_cryptographic_2018,mahadev_efficient_2022}, with a recent first demonstration achieved on a trapped-ion processor \cite{liu_certified_2025}. Indeed private information can be used to efficient certify any quantum computation~\cite{mahadev_verify_2018}, assuming the hardness of LWE.  The only caveat is that the quantum computer need to be able to  perform arithmetic in superposition.

\item {\bf Quantum verification.}
If the output $y$ is nearly deterministic then it can be verified by running the quantum circuit multiple times.  Here the verification could be carried out on a different quantum computer, or a quantum device of a different form, such as an analog quantum computer or even an experiment.  These choices are inevitably limited by systematic errors, which in turn can be controlled by comparing answers across multiple inputs and/or computing platforms.  Most quantum simulation tasks naturally fall into this category.

A recent example of this form of verification is in Google's OTOC (out-of-time-order correlator) experiment~\cite{abanin2025observation}, which carried out a detailed analysis of what is necessary for a quantum simulation to be both useful and verifiable.  In that case, the quantum simulation has application in NMR ~\cite{zhang2025quantum}, which could in principle serve as an independent experimental validation of the results.

\item {\bf Inefficient classical verification.}
Since quantum computers can always be simulated classically with exponential overhead, inefficient verification is always possible. For tasks such as boson sampling and random circuit sampling, this is the only known verification method.  Indeed for tasks with anticoncentration and no private verifier information, verifying a distribution is at least as hard as spoofing it~\cite{SFG_verify_22}, so inefficient verification appears inevitable in these cases.
\end{itemize}

An implicit requirement for verifiability is that the quantum task to be performed is well specified and success or failure is clearly identifiable.  This is straightforward for most toy problems, but realistic useful quantum computations are likely to be part of hybrid classical-quantum workflows, possibly including AI models.  As complexity of the overall algorithm increases, it can become more difficult to measure relation between the quantum subroutine  and overall algorithm performance.

\subsubsection{Algorithms}

\paragraph{Algorithms with exponential advantage} 
The gold standard for quantum computers is algorithms with an exponential advantage. 
Finding practical algorithms with such exponential advantages beyond the quantum simulation and hidden subgroup problem (HSP) paradigm has proven to be extremely challenging, and yet it remains one of the most pressing and important problems in the field today. 
Some promising recent suggestions involve the quantum exploitation of structure in classical codes \cite{yamakawa_verifiable_2022,jordan_optimization_2024}, but the classical hardness of the problems has not been heavily studied.
The pace of experimental progress in quantum computers is rapid enough that it may outpace our theoretical understanding of their practical benefits.  This situation represents both an exciting scientific opportunity and a risk for the field, as highlighted by the recent breakthroughs on compilation strategies in qLDPC codes \cite{cain_shors_2026} and factoring circuits \cite{babbush_securing_2026}.

We therefore strongly encourage the community to focus significant efforts on understanding the relations between existing algorithms with a (claimed) exponential advantage, and identifying new problems for which there may be such an advantage. 
In the context of post-quantum cryptography, this also involves in particular the study of quantum algorithms for lattice problems \cite{liu_uncertainty_2023,liu_post-quantum_2024,chen_quantum_2022,chen_quantum_2024}. 

\paragraph{Lowering the cost of quantum algorithms via co-design}
A further important direction of research is to lower the practical cost of implementing quantum algorithms in logical quantum processors. 
This involves developing more efficient techniques to implement classical computations quantumly, including classical arithmetic~\cite{gidney_windowed_2019,gidney_asymptotically_2019,kahanamoku-meyer_fast_2024}, more specific subroutines necessary for factoring and discrete logarithms \cite{cleve_fast_2000,regev_efficient_2023,ekera_quantum_2017,li_efficient_2012,kahanamoku-meyer_jacobi_2024}, as well as hardware-specific routines such as routing \cite{constantinides_optimal_2024}. 
Importantly, one may also  consider co-designing the  error-correcting code with algorithmic subroutines via the available naturally fault-tolerant gate set, an example being the fault-tolerant compilation of quantum-advantage experiments~\cite{hangleiter_fault-tolerant_2024}. 
Going beyond, one may hope for exotic error-correcting codes \cite{kubischta_family_2023} in which arithmetic or subroutines important for quantum simulation and material science can be  implemented naturally. These efforts can be further accelerated through the development of accurate resource estimates co-designed with the hardware.  Such estimates can help guide where to make the most significant gains through further optimization.

\paragraph{Hybrid classical-quantum algorithms}
Quantum computers are exponentially more effective than classical computers on some tasks, but have little-to-no speedup on others. Combined with their much greater cost, this means that it is important to use quantum computers selectively.  This is true even within a larger algorithmic goal, such as simulating a quantum system.  Here the right way to use a quantum computer will be to find subtasks on which we can exploit the exponential speedups of quantum computing while using classical computers for as much pre-processing, post-processing, and iterative feedback as possible.

The most famous hybrid algorithms are the variational quantum algorithms, in which a classical computer is used as an outer loop of a gradient descent algorithm where a quantum computer is used for the inner loop to estimate the cost function and/or gradients.  However, more sophisticated classical-quantum integrations are also possible, e.g. for quantum-assisted quantum Monte Carlo-\cite{Huggins2022,wan2022matchgate} or by using quantum computers to train machine learning models~\cite{q4bio-Copenhagen}.  There is enormous scope for designing new hybrid algorithms that take full advantage of the complementary capabilities of quantum and classical hardware.

\paragraph{Algorithm development for thermal states and dynamics}  
Many digital quantum simulations so far have focused on preparing ground states. However, it is of yet unclear whether there is a practically interesting range of systems for which the ground state problem is classically hard and quantumly easy. By contrast, recent results suggest that finite-temperature state preparation for the practically relevant case of constant (as a function of system size) temperatures and local Hamiltonians is classically hard but quantumly easy \cite{Bergamaschi2024,rajakumar2024}. In addition, simulating the dynamics of quantum-many-body systems is expected to be challenging for classical numerics. We therefore suggest building on previous works~\cite{Poulin2009,temme_quantum_2011,motta_determining_2020,Zhu2020,Lu2021,Ghanem2023,chen2023,schuckert2025} to developing algorithms for the simulation of static and dynamical properties of thermal states, which include practically relevant applications such as vibrational spectroscopy of molecules, prediction of transport coefficients of materials, and the pump-probe spectroscopy of solids. 

For calculating local observables in thermal states, usage of heuristic assumptions about the structure of the Hamiltonian, first and foremost the eigenstate thermalization hypothesis (ETH), may provide practical performance improvements at the expense of provable performance guarantees, see e.g.~\cite{Lu2021,Schuckert2023}. We also suggest preparing the microcanonical state, i.e., a state 
with $\langle{H} \rangle=E$ and $\sqrt{\langle(H-E)^2\rangle}/E\stackrel{L\rightarrow\infty}{\rightarrow}0$. Microcanonical ensembles are easier to prepare than Gibbs states~\cite{Lu2021,garratt2024,schuckert2025}, but yield the same observables under the assumption of ETH.
To prove quantum advantage, we must also understand the complexity of computing thermal observables as well as preparing microcanonical states of local models.

In terms of dynamical simulations, we suggest focusing on spin transport or electron transport simulations since those are quantumly simple tasks beyond the scope of classical simulators~\cite{yi-thomas_comparing_2024}. A simple target task is the determination of infinite-temperature diffusion coefficients, for example important for NMR experiments~\cite{Seetharam2023}, and quantumly simple by either z-product state sampling~\cite{Joshi2022}, using a 1-design (random product states; this would also enable usage of the fact that Trotter error is reduced for 1-design states~\cite{Qi2022}) or non-equilibrium initial states \cite{haghshenas_digitalquantummagnetismfrontier_2025}.

Finally, we suggest to analyse the possibility of integrating native noise into a quantum simulation along the lines of \cite{trivedi_quantum_2024}. 
Perhaps, native noise can drive a dissipative algorithm towards a solution rather than obstructing it. Some thermal observable algorithms have such noise resilience~\cite{Ghanem2023}. Another example is the simulation of NMR spectra~\cite{Seetharam2023,zhang2025quantum}.

\textbf{Efficient fermion-to-qubit mappings} For developing efficient encodings of fermions into neutral atoms, we suggest two avenues.
For local models such as the Fermi-Hubbard model, we suggest using low-weight encodings~\cite{Verstraete_2005,Derby2021}. They enable near-term Trotter simulation with little gate overhead and some error-detecting capabilities as part of the encoding. This comes at the cost of  some ancillas (for the currently most feasible encoding in ~\cite{Derby2021}, about $N/2$ ancillas for $N$ fermionic modes). However, neutral-atom platforms are usually not qubit limited, but gate-fidelity limited. In addition, low-weight encodings require little atom movement, which reduces atom-loss errors. Finally, neutral-atom dual-species arrays in conjunction with multi-qubit gates can be used for further efficiency gains~\cite{Hu2025}. 
For non-local models such as quantum chemistry and materials, we suggest using recent circuit-compression methods~\cite{maskara2025fast}, which reduce the depth overhead to $O(\log(N))$ with $O(N)$ ancillas. The need for ancillas can be removed at the cost of now using  $O(\log^2(N))$ depth~\cite{constantinides2025lowdepth,maskaraprivate}. While asymptotically efficient, these methods still require a large prefactor overhead in terms of entangling gates. This overhead can be removed completely by using fermionic atoms to encode fermions~\cite{Hartke2022,Yan2022,walter2023,gonzalez-cuadra2023fermionic,Chalopin2025}. Such a fermion-to-fermion encoding can also be made fault-tolerant~\cite{schuckert2024} or error corrected non-fault-tolerantly~\cite{ott2024}. 

\textbf{Analog simulators}
Rydberg-atom arrays can be used both for digital quantum computation and analog simulation. Analog quantum simulators allow the study of complex quantum many-body dynamics by mimicking their microscopic Hamiltonians~\cite{Browaeys2020}. While having restricted programmability, they allow for the simulation of continuous time dynamics with no need for Trotterization. One may thus think of analog simulation as a specific, hardware-tailored algorithm.
In the last decade, analog quantum simulators have been used to simulate many-body systems including the quantum Ising model to probe coarsening dynamics~\cite{manovitz2025quantum,andersen_thermalization_2025} or to reproduce the properties of the low-dimensional frustrated quantum magnet TmMgGaO4~\cite{leclerc2026one}, or the XY Hamiltonian to probe elementary excitations through spectroscopic measurements~\cite{chen_spectroscopy_2025} or transport coefficients~\cite{Joshi2022}. Furthermore, the recent implementation of a digital analog protocol, combining digital state preparation and measurements with an analog evolution, has allowed one to reveal signatures of the Rokhsar-Kivelson state~\cite{geim2026engineering}.
These problems are already at the edge of or beyond the current capabilities of classical numerical methods and become increasingly challenging as system size and evolution time increase.

A better understanding of simulation errors and certifiable techniques for characterizing errors in analog systems are the key challenges that need to be solved in the context of analog simulations.
In this direction, recent theory works have shown that certain observables, in particular local observables, can remain stable under deterministic and stochastic errors~\cite{daley_22,trivedi_quantum_2024,PhysRevX.15.021017,cai_stochastic_2023}. 
Furthermore, recent experiments have achieved more systematic characterization and control of the noise in analog simulations and thus demonstrate a quantum advantage similar to random circuit sampling~\cite{shaw_benchmarking_2024}.
Combined with large-scale classical benchmarks, such techniques will be required to demonstrate practical quantum advantage in problems related to many-body physics or material science~\cite{vovrosh2025simulating,vovrosh2025resource}.

\section{Neutral Atom Quantum Processor Hardware} \label{sec:atomControl}

Arrays of neutral atoms trapped in optical tweezers and coherently excited to Rydberg states have emerged as a leading platform for quantum computation, with system sizes reaching thousands of qubits~\cite{Pause2024,Gyger2024,norcia2024iterative,Pichard_2024,lin2024aienabled,manetsch2024tweezer,holman2026trapping,Chiu2025Coherent}, two-qubit entangling gate fidelities exceeding 99.5\% \cite{Evered_2023,radnaev2025universal,Tsai2025Benchmarking,Evered:2026jrf}, and a dynamically reconfigurable quantum processor architecture that allows arbitrary two-qubit gate connectivity~\cite{bluvstein2022quantum}. Recent demonstrations of logical qubit algorithms \cite{bluvstein_logical_2024,rodriguez_experimental_2024,reichardt2024logical,mathiot2026} have involved up to 48 logical qubits (up to 280 physical qubits) executing complex circuits \cite{bluvstein_logical_2024}. This has brought us into the era of error-corrected logical quantum computing, with the exciting prospect of further scaling up the system in the near-term to operate in the deep-circuit regime with hundreds of logical qubits or more.

In parallel, these advances consolidate the existing foundation of analog quantum computing, which continues to evolve through recent demonstrations exploring new interaction models and large-scale many-body phenomena~\cite{bernien2017probing,chen2022continuous,darcangelo2024,manovitz2025quantum,gonzalez-cuadra2025,Qiao2025,Emperauger2025}. These developments are accompanied by a clearer identification of the analog algorithms and operating regimes most likely to outperform classical approaches~\cite{lanes2025}. Together, these combined progresses strengthen the analog approach and naturally open the way to hybrid digital-analog computation that combines high-fidelity gate operations with well-controlled global Hamiltonian evolution~\cite{bluvstein2022quantum,Guseynov2022,Lu2024,andersen_thermalization_2025,katz2025}.

\subsection{Challenges and Opportunities}

Current estimates suggest that algorithms with practical quantum advantage such as Shor's algorithm for factoring will take at least thousands of logical qubits (e.g. distance $d=27$ surface codes operating at the 0.1\% level for two-qubit gate error). Recent work \cite{cain_shors_2026} utilizing high-rate qLDPC codes estimates 10.000 - 100.000 physical neutral-atom qubits as a lower threshold to run Shor's algorithm. For the neutral atom platform to advance towards this regime requires substantial further scaling-up of current system sizes of a few thousand qubits, primarily through overcoming laser power limitations and developing scalable light control methods for qubit control (see Section \ref{sec:lightControl}). 

At the same time, improving two-qubit gate errors even further below threshold for quantum error correcting codes will enable higher encoding rates (logical per physical qubits), improving system scaling and enabling the execution of deeper quantum circuits.

As logical error rates go down and quantum algorithms start to reach the deep circuit regime, quantum operations will start being limited by the small but finite amount of atom loss due to imperfections in the two-qubit entangling Rydberg gate, finite vacuum lifetime, and losses during mid-circuit readout \cite{Evered_2023}, requiring the continuous replacement of lost qubits to allow quantum circuits running on long-lived logical qubits to reach the megaquop regime and beyond. Experiments running such deep quantum circuits will also greatly benefit from being able to further reduce the cycle time for quantum error correction, which is dominated by the time required for qubit state readout via imaging. Faster cycle times will in the near-term allow easier calibration of long experiment sequences, and in the long-term make running truly practical quantum algorithms feasible on human timescales.

We note that these challenges are shared with analog quantum computing and hybrid analog-digital computing. In particular for analog computing, recent works suggest that improving hardware capabilities with faster cycle times, larger qubits number and better control may lead to quantum advantage~\cite{cazals2025identifying}.

A standard approach to optical trapping of neutral atoms
uses holographic imaging of optical tweezers, employing phase-only holograms. These holograms can be generated using Liquid Crystal on Silicon (LCOS) displays, that can manage $\sim$ kW of optical powers for near infrared wavelengths. 
Light efficiencies of these devices, range around 50-70 $\%$. Liquid crystal response times typically in the 100s of milliseconds limit these devices to programmable but static trap arrangements. Dielectric micro-mirror devices (DMDs)
provide faster switching <100 $\mu s$ but are restricted to binary on-off holographic patterns, resulting in poor light efficiency. 

Acousto-optic deflectors (AODs) in two axes can deliver atomic motion on a time scale commensurate with requirements for quantum operations. This capability complements static trapping (e.g. for qubit storage or as atomic reservoir), with dynamic operations (e.g. for assembly of defect free arrays and long range entanglement operations). Together this static and dynamic trapping has been the backbone of the most recent demonstrations of error corrected codes in tweezer array systems \cite{bluvstein_logical_2024,Bluvstein_2025FT}.  A limitation of 2D acousto-optic devices is that the optical field in the image plane has to be a separable function of x and y coordinates i.e. $I(x,y)\propto f(x)g(y)$. 

Qubit readout is a critical capability for enabling fault-tolerant quantum processors, particularly in the face of finite qubit coherence times and the existence of decay channels. Fault tolerance in quantum computation is dependent on a reliable measurement to correct errors \cite{hastrup2021analysis}, and even more so on repeated reliable measurements performed during the quantum circuit operation, to verify its end result \cite{baranes2025leveraging}. Therefore, it is imperative to possess a fast, high-fidelity, non-destructive and preferably state-selective readout of qubits in order to provide the necessary infrastructure for error-correction. 

Currently, readout in atomic arrays (which form the basis for the neutral-atom quantum processor) is most commonly implemented by taking their fluorescence image \cite{endres2016atom}. Each atom in one qubit state appears bright, whereas atoms in the other qubit state remain dark. Photons are collected by a high-numerical aperture lens and routed in free-space to single-photon sensitive cameras, converting fluorescence photons into a suitable electronic signal. While this method reaches fidelities above 99\%, and can sustain over 99\% survival of atoms after readout, this is achieved at the cost of long integration times, routinely spanning 5-20ms \cite{radnaev2025universal,norcia2024iterative,zhang2024high,tao2024high}, making it the most time-consuming of all processor operations.

\subsection{Research Directions}

\subsubsection{Scaling Qubit Arrays}\label{subsec:scaleup}
\begin{figure}
\centering
    \includegraphics[width=1\linewidth]{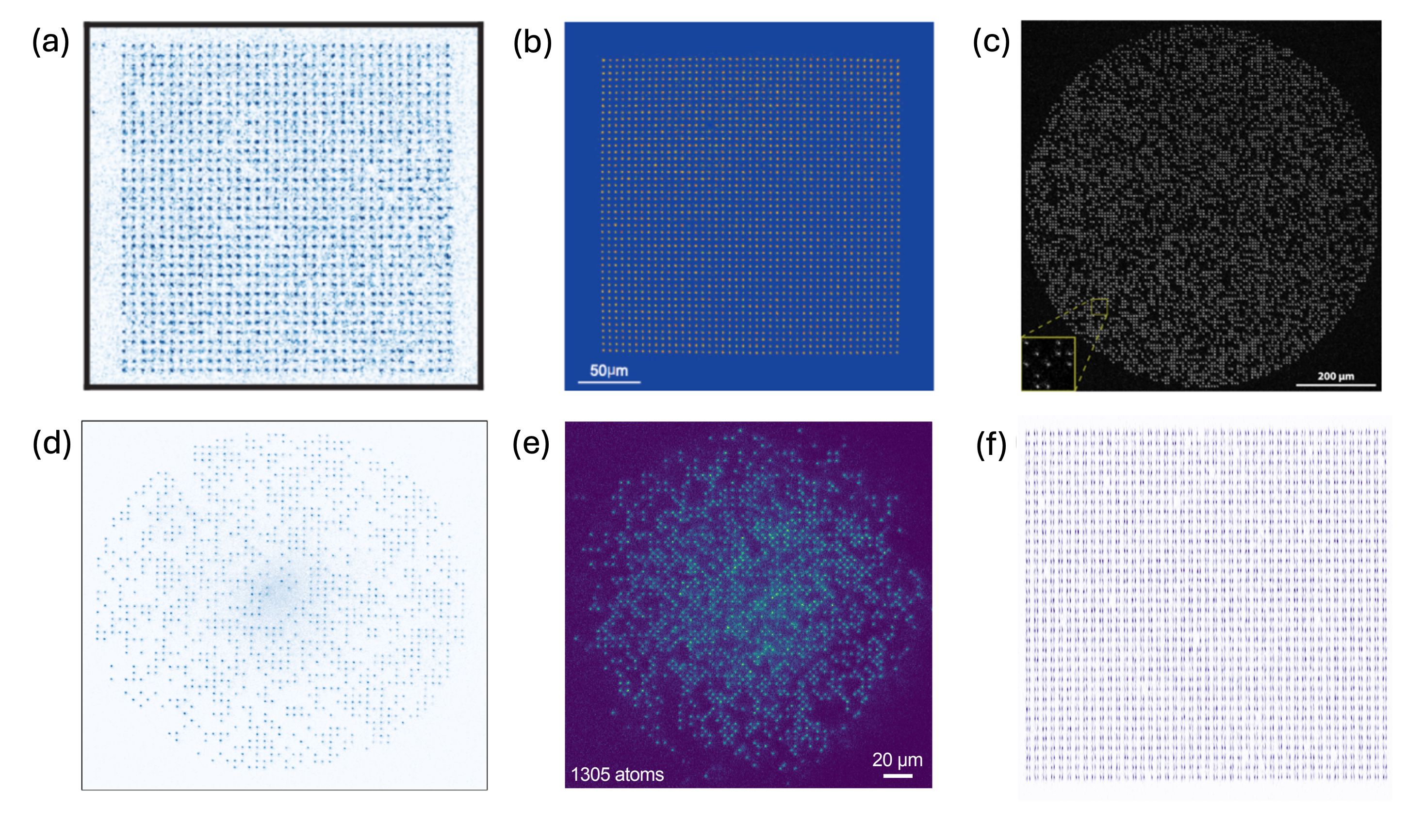}
    \caption{Arrays of thousands of neutral atom qubits: (a) Atom Computing >1200 $^{171}$Yb atoms \cite{norcia2024iterative}, (b) Jian-Wei Pan group (USTC) >2000 $^{87}$Rb atoms \cite{lin2024aienabled}, (c) Endres group (Caltech) >6100 $^{133}$Cs atoms \cite{manetsch2024tweezer}, (d) Pasqal >1100 $^{87}$Rb atoms~\cite{Pichard_2024}, (e) Birkl group (TU Darmstadt) >1300 $^{85}$Rb atoms~\cite{Pause2024},
    and (f) Lukin group (Harvard) >3200 $^{87}$Rb atoms \cite{Chiu2025Coherent}.}
    \label{fig:atomarray3000}
\end{figure}

Several neutral atom array experiments have already demonstrated system sizes of thousands of atomic qubits (Fig.~\ref{fig:atomarray3000}) ~\cite{Pause2024,Gyger2024,norcia2024iterative,Pichard_2024,lin2024aienabled,manetsch2024tweezer,holman2026trapping,Chiu2025Coherent}, with the main limitation being available laser power. Further scaling-up of system sizes to the 10,000 - 100,000 qubit level can be accomplished by a combination of developing higher power laser systems, incorporating methods for beam combining, and exploring optical lattice trapping to take advantage of optical power recycling. 

\paragraph{Scaling two-dimensional arrays}\

The development of high power, low noise, single-frequency laser sources, in particular fiber amplifiers, has already been critical to trapping large numbers of atomic qubits. For example, the Lukin group (Harvard) has been able to build arrays of >3,000 Rb atoms with 15~W of laser power around 850~nm \cite{Chiu2025Coherent}, delivered by a commercial system that utilizes sum-frequency generation of two longer wavelength fiber amplifier laser systems. Even higher power levels of 30-40~W are already becoming available at these wavelengths, with further increases possible with dedicated engineering efforts from laser vendors. 

Thus, we expect that system sizes for neutral atom quantum processors will soon exceed 10,000 qubits.

Looking ahead, single-frequency output powers of up to kW levels have been demonstrated at certain wavelengths for fiber amplifiers \cite{Ahmadi2020Generating}, with the main challenge being thermal handling of fiber components and suppressing stimulated Brillouin scattering induced high-frequency intensity noise as the output power increases. Another promising avenue are scalable integrated 2D arrays of VCSEL lasers which are available for NIR wavelengths and can deliver several mW of single mode light per device. Arrays of several 1000 VCSELs are already routinely manufactured. 
A combined laser technology development effort from both industry and academia to deliver laser systems at the kW level will allow trapping of up to 100,000 atoms. 
At such high levels of laser powers, issues such as microscope objective thermal lensing and optical coating damage must also be carefully managed.

Beyond the limits of available power from a single laser system, the number of trapped qubits can be further scaled-up by beam combining, either spectral beam combining or coherent beam combining. 
Spectral beam combining uses a wavelength filter optic to merge two input laser beams at different wavelengths into a single output beam with multiple spectral components. The density of spectral components is limited by the spectral resolution of the filter, which for typical dielectric-coated filters is a few nm. Sharper filters such as Bragg gratings can offer spectral resolutions well under 1~nm, with the main challenges being output beam quality and small angular separations of the input beams. Integrated photonic filters that are controlled using precision piezoelectric stress optic actuators will offer spectral resolution in the sub-GHz regime down to MHz with extremely high extinction ratios exceeding 100 dB. Such technologies will enable a more powerful, flexible degree of spectral engineering and control. The Endres group (Caltech) has already used two different wavelengths at 1055 and 1061~nm to demonstrate an array of more than 6000 atomic qubits with $^{133}$Cs atoms~\cite{manetsch2024tweezer}. Similarly, Pasqal has combined two close wavelengths at 813 and 818~nm to obtain arrays of more than 1100 $^{87}$Rb atoms in a cryogenic environment~\cite{Pichard_2024}. With careful planning of optical beam paths, spectral beam combining can allow scaling up by at least a factor of 5-10 from the array size limits of a single laser system.

Coherent beam combining is a method of scaling up laser power at a single output wavelength, achieving this by merging two input laser beams at a beam splitter to deliver a single output with twice the power \cite{Chiow2012Harmonic}. This requires maintaining phase coherence and mode-matching between the input laser beams, and can be extended to combining more than two input laser sources, with the main challenge being to maintain relative phase coherence between an increasing number of beam paths. In high-power optical systems where fiber coupling and spatial filtering are not viable solutions, astigmatism and ellipticity can be corrected using a compact three-cylindrical-lens scheme, in which two cylindrical lenses mounted one behind the other at a relative angle form a tunable biaxial lens~\cite{Khoubyarian2025}. Coherent beam combining can be viewed as complementary to spectral beam combining. The former increases power available at a single wavelength, while the latter allows multiplexing of trapping beams at different wavelengths, with maximum possible scaling-up of system sizes achieved by incorporating both approaches.

Recently advances have been made in using meta surfaces \cite{holman2026trapping} for holography in tweezer-trap array systems. In this approach a static holographic mask is used to generate the tweezer array. While this does not offer the flexibility of a programmable hologram, a higher resolution hologram at sub-wavelength features promise to deliver higher power handling, better light efficiency and larger number of resolvable spots.

\paragraph{Three-dimensional structures}
Extending optical tweezer platforms from two to three dimensions is a longstanding goal for neutral-atom systems. Beyond simply increasing the number of available trapping sites, three-dimensional architectures can improve packing density, enable richer interaction graphs, and offer native geometries that are attractive for quantum simulation and fault-tolerant quantum computing. Early work by the Browaeys group demonstrated the trapping and assembly of atoms in arbitrary three-dimensional geometries~\cite{Barredo2018}.

A particularly important potential advantage of 3D arrays is improved use of optical power. In conventional 2D tweezer arrays, the number of trapping sites scales linearly with optical power, $N\propto P$. In contrast, certain 3D architectures can surpass this scaling~\cite{Schlosser2023}. 

One prominent route to 3D trapping is based on the Talbot effect, in which a periodic intensity pattern self-images at integer multiples of the Talbot distance,
$z_T = \frac{2d^2}{\lambda}$, where $d$ is the array period and $\lambda$ is the wavelength. Using a microlens array, Schlosser et al.~\cite{Schlosser2023} generated a three-dimensional lattice of optical tweezers and trapped single atoms in multiple Talbot planes simultaneously. In this approach, the total number of available trapping sites can scale as $N\propto P^{3/2}$, offering a more favorable power scaling than conventional single-plane arrays.

Three-dimensional geometries are also attractive from the perspective of quantum error correction, since they can provide enhanced connectivity and may allow certain 3D codes to be implemented more naturally, with less overhead from atom transport than would be required when compiling such codes onto a purely 2D architecture~\cite{bluvstein_logical_2024}.

Despite this promise, most neutral-atom experiments still operate predominantly in two spatial dimensions. While 3D structures have been generated and rearranged~\cite{Barredo2018,Lin2025}, their use in quantum computation and quantum simulation has so far remained comparatively limited~\cite{Kim2020}. Nevertheless, access to the third spatial dimension has become a major driver of recent technical progress, and several approaches based on acousto-optic deflectors  have recently enabled dynamical reconfiguration of atoms in all three spatial dimensions~\cite{Picard2025,Guo2025,Lu2026}.

\textbf{Optical Lattices} 

\begin{figure}[h!]
    \centering
    \includegraphics[width=0.75\linewidth]{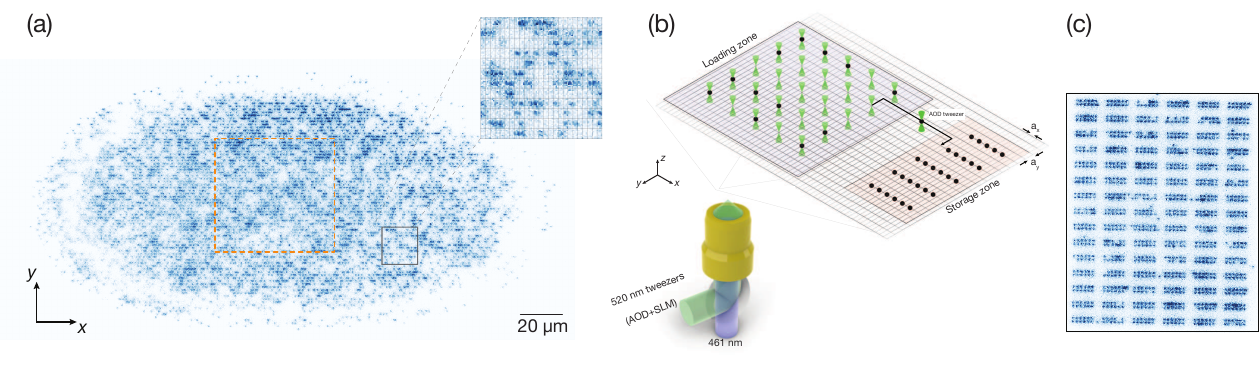}
    \caption{Lattice arrays for scaling quantum registers. \textbf{(a)} Using a single, strongly elliptical folded beam allows to create regular trap arrays with more than $25.000$ lattice sites that can be loaded from a magneto-optical trap and imaged with high fidelity and low loss~\cite{tao2024high}. \textbf{(b)} Zone-based architecture in a lattice array, where a loading-zone is iteratively loaded from a magneto-optical trap, whereas atoms in the storage zone are protected by shelving in a metastable state. \textbf{(c)} Ordered array of close to $1000$ atoms in an optical lattice.}
    \label{fig:placeholder}
\end{figure}

A distinct approach to scaling-up system size complementary to optical tweezers are optical lattices \cite{Gyger2024,norcia2024iterative}. In this approach, multiple interfering laser beams produce a periodic trapping potential for atoms. The interfering beams can come from a single beam folded multiple times to efficiently recycle laser power, leading to simple designs.
Recently, the Zeiher/Bloch group (MPQ/LMU) demonstrated that initial loading from a magneto-optical trap and high-fidelity, low-loss imaging of more than 10.000 atoms in about 25.000 sites in optical lattices is achievable with a single elliptical trapping beam in a bow-tie configuration~\cite{tao2024high}. Crucially, and in contrast to the scaling for optical tweezers, the number of trappable sites in such optical lattice scales with the square of the available laser power. With off-the shelf, high-power single-frequency laser sources in the ytterbium amplifier band, the approach demonstrated in~\cite{tao2024high} straightforwardly scales to hundreds of thousand trappable sites without major changes in the architecture.
Even further power recycling is possible via the integration of enhancement cavities, as has been demonstrated by the Blatt/Bloch group~\cite{park2021cavity} and Atom Computing \cite{norcia2024iterative}. Challenges in the optical lattice approach include having confinement along all three spatial axes, and the difficulties that the lattice potential modulation presents for coherently moving atoms in between lattice sites. Both of these can be addressed in special purpose asymmetric bow-tie configurations and by transporting atoms between lattice sites~\cite{tao2024high,Gyger2024}.
The combination of optical lattice trapping and high-fidelity qubit manipulations and the execution of quantum circuits has yet to be demonstrated. However, once that milestone is achieved, optical lattice arrays in combination with reconfigurable optical tweezers can be a powerful approach to scaling-up neutral atom quantum processors.

\subsubsection{Qubit encoding and atomic platforms}

\paragraph{Alkaline earth(-like) atoms (AEAs)}\
 
\begin{figure}
    \centering
\includegraphics[width=0.6\textwidth]{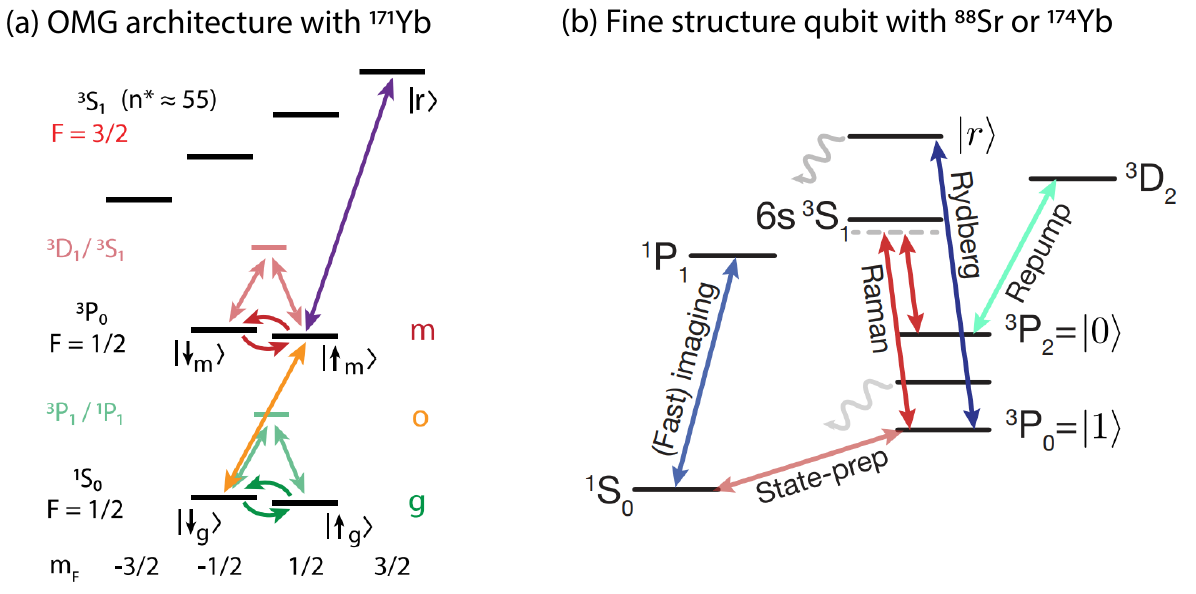}
    \caption{Alkaline earth(-like) atoms (AEAs). The unique level structure due to the divalance of AEAs opens many new opportunities for neutral atom quantum computing. The most salient feature is the presence of long-lived optically metastable manifolds. (a) Ytterbium-171 offers access to the ``Optical, Metastable, Ground" (``OMG") architecture in which qubits can be encoded in the ground or metastable nuclear spin-1/2 degree of freedom or the optical atomic clock transition. Nuclear spins can be manipulated with stimulated Raman transitions, and Rydberg excitations can be performed from the metastable clock state. (b) The metastable fine structure qubit encoding with bosonic AEAs such as $^{88}$Sr and $^{174}$Yb. Qubits are encoded in $^3P_0$ and $^3P_2$ ($m_J=0$) and Rydberg excitation is performed from $^3P_0$. Like the OMG architecture, readout is performed from the ground state. Figure adapted from~\cite{Tao2025}.}
     \label{fig:aea}
\end{figure}

Alkaline earth(-like) atoms (AEAs) offer a substantially richer level structure than alkali species while still providing convenient transitions for fast cooling and high-fidelity measurement. Strontium and ytterbium have been widely used for optical lattice clocks for two decades~\cite{Ludlow2015}, but optical tweezer arrays of these species were first realized in 2018 and 2019~\cite{Cooper2018,Norcia2018,Saskin2018,covey2019sr}. A unique feature of AEAs is their narrow $^1S_0\leftrightarrow^3P_1$ ``intercombination" line that offers new cooling mechanisms such as Sisyphus cooling~\cite{Cooper2018,covey2019sr}, resolved sideband cooling~\cite{Cooper2018,Norcia2018}, as well as gray molasses and electromagnetically-induced transparency cooling for isotopes with non-zero nuclear spin~\cite{lis2023midcircuit,li2025parallelized}. From there, ``atomic array optical clocks" in which sub-Hz frequency metrology combined with programmable single-site control and readout in an array were realized in 2019~\cite{Madjarov2019,Norcia2019}. 

Additionally, Rydberg excitation was realized in strontium and ytterbium arrays in 2019~\cite{madjarov2020high,wilson2022trapping}. AEAs offer two new opportunities for Rydberg excitation: 1) a relatively strong single-photon excitation pathway to an S-series Rydberg state via the metastable clock state~\cite{madjarov2020high}, and 2) access to a core ion with optically-accessible transitions and optical polarizability~\cite{madjarov2020high,wilson2022trapping,burgers2021controlling}. The former was crucial for early demonstrations of fast and high-fidelity clock-Rydberg Rabi oscillations and two-atom entanglement. The latter was crucial for autoionization-based high-fidelity readout and optical tweezer trapping of Rydberg states. More broadly, metastable qubit or clock-Rydberg encodings offer the possibility to perform erasure error conversion via readout of the ground state~\cite{wu2022erasure,scholl2023erasure,ma2023high}, and the broad blue $^1S_0\leftrightarrow^1P_1$ transition enables readout in only $\approx20$ $\mu$s~\cite{scholl2023erasure,ma2023high}.

Despite these successes, bosonic isotopes (with zero nuclear spin, $I=0$) have an enormous potential weakness: they do not immediately offer a qubit encoding involving the electronic ground state that is well suited with the needs of quantum computation. The most obvious choice is the optical clock transition. However, the clock transition has negligible linewidth for $I=0$ isotopes and can only be ``activated" by applying a large external magnetic field~\cite{Madjarov2019,Norcia2019,madjarov2020high}. Additionally, an optically-encoded qubit suffers from phase stability of the optical reference, temperature and Doppler effects of the trapped atoms, and the fundamental incompatibility with mid-circuit measurement when probing from the ground state. Although each of these challenges could potentially be overcome and largely have been with the advent of fast motional-state-preserving clock pulses~\cite{lis2023midcircuit,Zhang2024} as well as motional qubit encoding and ``shelving" architectures~\cite{Scholl2023hyper}, the attention of the AEA community has largely turned to either the $I=1/2$ isotope of ytterbium ($^{171}$Yb) or to metastable fine structure qubit encodings. These approaches are illustrated in Fig.~\ref{fig:aea} and we now briefly describe them both.   

\textbf{The OMG architecture with nuclear spin qubits in $^{171}$Yb}. The first arrays of fermionic ($I\neq0$) AEAs were realized in 2022 for $^{87}$Sr ($I=9/2$)~\cite{barnes2022assembly} and $^{171}$Yb ($I=1/2$)~\cite{jenkins2022ytterbium,ma2022universal}. Compared to the bosonic isotopes, the fermionic isotopes have hyperfine structure which affects the differential polarizability of the intercombination transition in the tweezer due to the interplay with tensor light shifts. This in turn adds some complexity, but also some new opportunities, for imaging and cooling fermionic AEAs in tweezer arrays. For instance, it was shown in 2023 that quantum non-demolition measurement of the ground nuclear spin qubit can be realized simply by operating at a modest magnetic field for which the Zeeman splitting on the intercombination probe transition is sufficiently large~\cite{huie2023repetitive,norcia2023midcircuit}.

The trapped ion community coined the term ``OMG", which is an abbreviation of ``optical, metastable, ground"~\cite{Allcock2021}. This is a blueprint based on the flexibility to encode qubits in either the hyperfine/Zeeman structure of the ground state or the long-lived metastable state, or to utilize the optical transition spanning the ground-metastable transition. Essentially, this blueprint focuses on the programmable inter-operability for disparate operations such as mid-circuit measurement and qubit reset/recooling and continuous reloading that are performed simultaneously with logical operations. $^{171}$Yb offers an excellent system to implement the OMG architecture~\cite{chen2022analyzing,wu2022erasure}. Several experimental groups have taken steps in this direction~\cite{lis2023midcircuit,ma2023high,li2025parallelized} and are building towards realizing full quantum computing architectures~\cite{muniz2024high,muniz2025repeated,Li2024}. 

\textbf{Metastable fine-structure qubits with bosonic AEAs}
Recently, an alternative qubit encoding between the excited fine-structure states $^3P_2$ ($m_J=0$) and $^3P_0$ has been explored in AEAs~\cite{Pucher2024,Unnikrishnan2024,Ammenwerth2025,Tao2025,Lib2026}. Compared to nuclear and hyperfine qubits, the decoupling of the qubit states from atomic motional degrees of freedom is more challenging. However, using magic-angle tuning of the magnetic field with respect to the trap light polarization, allows to realize magic conditions at various wavelengths~\cite{Pucher2024,Unnikrishnan2024}. Despite the added complexity, it was shown that coherence times of several hundred milliseconds can be reached~\cite{Ammenwerth2025}.
The fine-structure (FS) qubit brings several interesting features that are currently being explored in first experiments: Similar to the metastable nuclear qubits, Rydberg states can be coupled on single-photon transitions, avoiding scattering or AC-stark shifts from intermediate states and potentially higher two-qubit gate fidelities. Furthermore, errors leading to population leakage outside the metastable qubit subspace can be detected mid-circuit and converted to erasures~\cite{Tao2025,Lib2026}. Finally, driving the metastable using two phase-stable lasers and drawing on developments on the clock qubit~\cite{Shaw2024} allows for novel control by converting atomic placement to single-qubit manipulations~\cite{Lib2026}. Compared to optical transitions, the FS-qubit significantly relaxes the requirements on positional accuracy, and has reduced sensitivity to photon recoil for co-propagating coupling beams~\cite{Tao2025}.

\paragraph{Dual Species Architecture}\label{sec:dualspecies} A single system that utilizes two different neutral atom species enables crosstalk-free control and readout of two subsets of qubits due to the wavelength separation of relevant atomic transitions \cite{singh2022dual,beterov2015rydberg,CFang2025}. This can be leveraged in multiple ways for quantum error correction. First, by encoding data and ancilla qubits in distinct atomic species, we can perform mid-circuit ancilla readout in situ --- i.e., without moving atoms to a separate readout zone, or invoking hiding operations \cite{graham2023mid},—since photon scattering from one species does not affect the other. A first demonstration of this idea has already been performed utilizing a dual-species array of Rb and Cs, where phase errors on Rb atoms were detected with ancillary Cs atoms \cite{singh2023mid}.  This architecture also provides control over three distinct interaction terms: two intra-species and one inter-species. By selecting Rydberg states with suitable F\"orster resonances and applying external electric fields for additional tunability, one can access regimes with different relative strengths among these interactions~\cite{beterov2015rydberg,petrosyan2024fast}. Experimentally, interspecies interactions and entangling gates have been demonstrated in  Rb and Cs arrays \cite{anand2024dual,White2026,Miles2026}. The interspecies gates have been used for what are effectively syndrome measurements on groups of two and three atoms. 

The rubidium isotopes $^{85}$Rb and $^{87}$Rb  can also be used for the same purpose~\cite{YZeng2017}, although the 
isotopic shift of the transition between the  upper hyperfine levels of the $5s$ to $5p_{3/2}$ states is only about 1.2 GHz, which provides only a partial suppression of crosstalk for species-selective measurements. A system using two isotopes of ytterbium, one fermionic and one bosonic, has also been demonstrated \cite{Nakamura2024}. In this case the $^{171}$Yb fermionic isotope provides a nuclear spin qubit and the $^{174}$Yb bosonic  isotope acts as an ancilla qubit on an optical transition which provides the possibility of low crosstalk mid-circuit measurements. 

Of particular interest is the regime where inter-species interactions are strong while intraspecies couplings are negligible, which is ideal for multi-qubit controlled gates. This regime ensures strong control-target coupling while minimizing unwanted target-target and control-control interactions~\cite{Young2021}. A simple implementation of an N-qubit entangling gate builds upon the original two-qubit controlled-phase gate proposal by Jaksch et al.~\cite{Jaksch2000}. Here, a sequence of three Rydberg pulses, selectively applied to control and target atoms, induces the required phase accumulations. A key advantage of the dual-species architecture is that this gate can be implemented without requiring any local control to selectively address control and target atoms. Moreover, the gate time remains independent of N, being solely determined by the single-qubit Rabi frequencies. This represents a significant advantage over implementations relying on two-qubit gate decompositions. 
Multi-qubit gates are critical for implementing high-dimensional QEC codes, which offer increased robustness against local errors by spreading logical information over a large number of qubits. Efficient error detection in these codes relies on the ability to measure high-weight stabilizers, facilitated by
multi-qubit gates \cite{terhal2015quantum}.

\begin{figure}
    \centering
\includegraphics[width=0.65\textwidth]{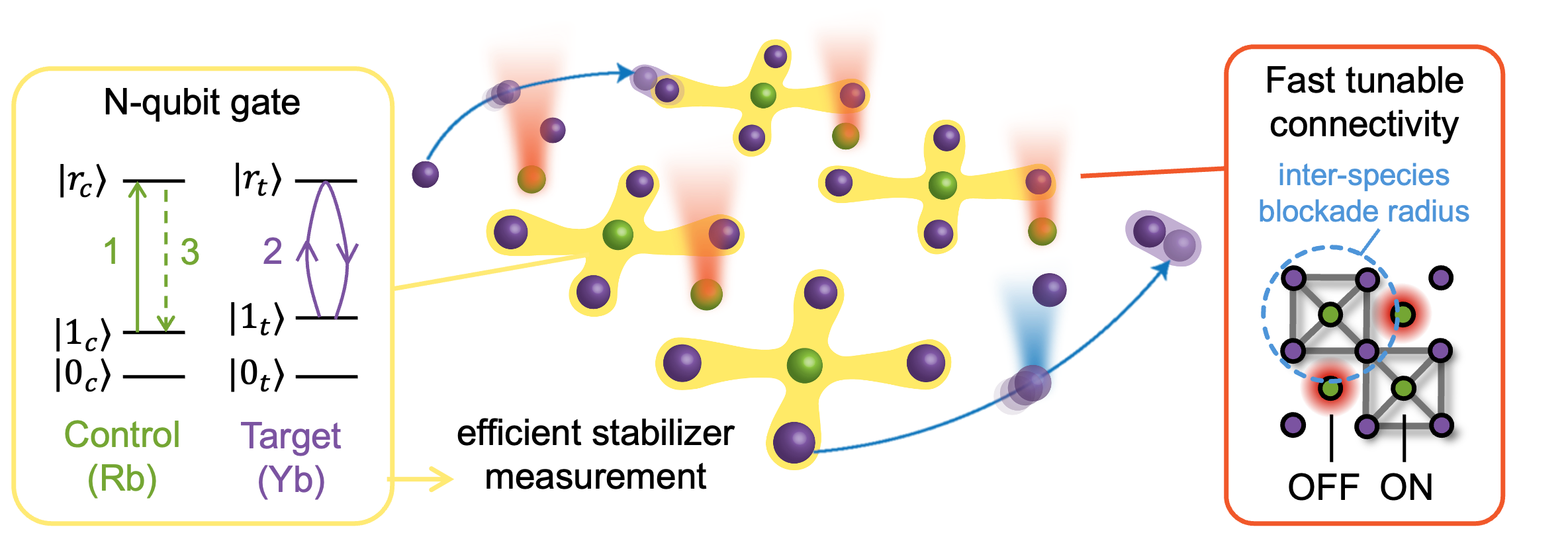}
    \caption{Dual-species arrays: new tools for efficient quantum error correction. N-qubit gates (left inset) can be efficiently applied in dual-species arrays thanks to the tunability of control-target versus target-target/control-control interactions. These can be combined with local light shifts (red) to enable fast switching of local interactions (right inset). These tools can improve quantum circuit efficiency by enabling stabilizer measurements with a single multi-qubit entangling gate (yellow) and fast local-connectivity changes.}
     \label{fig:tunnelinggate}
\end{figure}

Multi-qubit gates can also be used to realize fast switching of short-range atom connectivity.
In this scheme, atoms in species 1 do not interact directly with each other but through the mediation of an auxiliary atom in species 2. Using this auxiliary atom in combination with local light shifts one can achieve fast tunable connectivity~\cite{Jeong2022}. The effective interaction between data qubits can be turned on and off by using local addressing beams to shift the connecting auxiliary qubit in and out of resonance of its Rydberg transition. When a local light shift brings this atom off resonance, it cannot be excited to the Rydberg state and thus does not interact with its neighboring data qubits, which in turn cannot get entangled with each other. In this way, the local connectivity can be tuned without having to move atoms between different configurations, and atom shuffling is only necessary to connect atoms at distant locations in the array. Given that local gates typically make up a large fraction of the gates in a quantum circuit, especially during state preparation and syndrome extraction, this approach is expected to result in a significant increase in clock speed. 

Note that a natural alternative approach to tunable local connectivity can be achieved by using local Rydberg excitation beams~\cite{Graham_2022,manovitz2025quantum,radnaev2025universal,Chinnarasu2025}, addressing only subsets of atoms that need to be entangled. This strategy is  less robust, since it is challenging to ensure negligible crosstalk and good homogeneity across different beams when addressing pairs of atoms. On the contrary, when using local addressing beams to light shift single atoms, crosstalk is less significant because the beams can be focused down to sizes comparable with the trapping tweezers and homogeneity is not a strict requirement since the atoms only need to be shifted out of resonance. This scheme is also compatible with parallel implementation of entangling gates.

The dual-species approach also enables the engineering of both coherent and dissipative dynamics involving multi-qubit operators~\cite{weimer2010rydberg}. This capability enables the exploration of alternative methods to fault tolerance, such as passive error correction~\cite{terhal2015quantum}. In this approach, errors are not actively detected and corrected as in QEC but are mitigated by coupling an encoded logical qubit to an environment that corrects errors through controlled dissipation. In this approach, one species would encode the logical qubit, while the other would act as the dissipative bath. 

The distinct electronic structures of alkali and AEL atoms offer unique advantages for encoding and manipulating quantum information. Recent experiments have demonstrated Yb's potential as an ideal memory qubit due to its nuclear spin qubit's long coherence times~\cite{jenkins2022ytterbium,ma2022universal}. Yb's metastable states, with lifetimes of several seconds, provide additional nuclear spin qubits on top of the one in the ground-state manifold. This opens possibilities for advantageous quantum computing architectures~\cite{lis2023midcircuit}, including shelving of quantum information in protected states, ququart encoding~\cite{jia2024architecture}, and erasure conversion~\cite{ma2023high,scholl2023erasure}. Additionally, Yb atoms can be trapped in optical tweezers while in the Rydberg state~\cite{wilson2022trapping}, allowing for extended coherence times during entangling operations. On the other hand, alkali atoms, with their simpler electronic structure, offer their own advantages. Hyperfine qubit encoding in Rb has yielded relatively long coherence times, while coherent excitations to Rydberg states have enabled fast high-fidelity entangling gates~\cite{Evered_2023} and their integration into complex logical algorithms~\cite{bluvstein_logical_2024}. Alkali atoms tend to have larger polarizabilities at accessible wavelengths, which are also crucial for scaling up the number of qubits using reasonable optical power for the tweezer array.
By combining atomic species with different features in the same array, one can seek to create a programmable quantum platform that leverages each species' strengths for roles best suited to their unique features.

\paragraph{Circular Rydberg states}\
Circular Rydberg states are a promising alternative to conventional low-angular-momentum Rydberg qubits because of their exceptionally long radiative lifetimes and strong dipolar interactions, which make them attractive for both analog quantum simulation and quantum computation~\cite{Nguyen2018,Cohen2021}. Their radiative lifetimes can reach tens of milliseconds for moderate principal quantum numbers and may be further extended to the minute range through spontaneous-emission inhibition in suitable microwave structures~\cite{Hulet1985,Nguyen2018,Cohen2021,Wu2023}. Recent progress in generating and trapping circular Rydberg states in optical tweezers, including in alkaline-earth systems, makes this direction increasingly credible~\cite{Holzl2024}. The main challenges remain high-efficiency state preparation, very precise electric-field control, and the integration into scalable architectures with local control and non-destructive state-selective readout~\cite{Machu2025}.

\subsubsection{Gates and High-Fidelity Control}

\begin{figure}
\centering
    \includegraphics[width=1\linewidth]{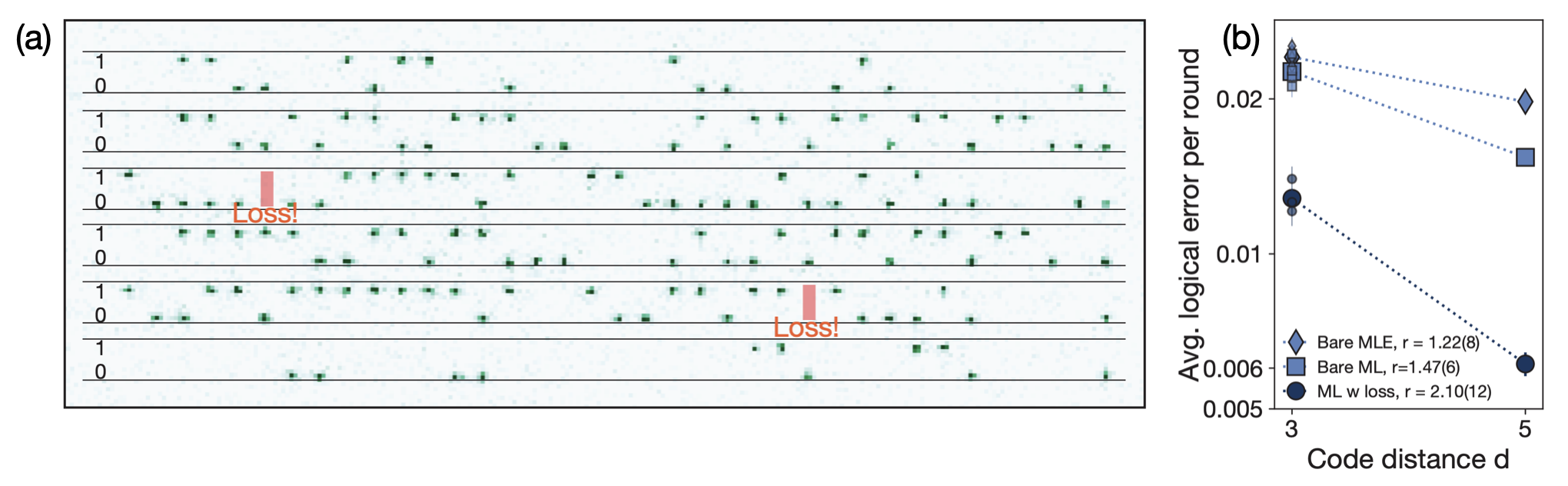}
    \caption{ (a) State-selective 1D lattice pins qubit state $|0\rangle$ in its optical tweezer while another moving tweezer transfers qubit state $|1\rangle$ to adjacent empty tweezer. Qubit state is read out by imaging which site atom is on, with loss detected as no atom on either site. After imaging, atoms in qubit state $|0\rangle$ are reset back to original optical tweezer. (b) With loss detection, distance-5 surface code operates with $>2\times$ less error than distance-3. Data from the Lukin group (Harvard)\cite{Bluvstein_2025}.
    }
    \label{fig:lossdetection}
\end{figure}

Recent logical qubit benchmarking from the Lukin group (Harvard) has already demonstrated the surface code operating at $>2\times$ below threshold when including detection of atom loss (Fig.~\ref{fig:lossdetection}) \cite{Bluvstein_2025FT}. Reaching further below threshold will result in strong exponential suppression of logical qubit error rates with increasing code distance $d$, allowing circuit depths beyond 1 megaquop before uncorrected error.

\paragraph{Strategies for improving two-qubit gate fidelities
}\ 

One ingredient to going further below threshold is to improve two-qubit gate fidelities, which are already $>99.5\%$ on a number of platforms \cite{Evered_2023,muniz2024high,radnaev2025universal,Tsai2025Benchmarking}. For both alkali and alkaline-earth atom qubits, this is primarily a function of increasing available laser power. For example, a factor of 10 increase in the available laser power at 1015~nm for the Rb Rydberg transition was crucial for the demonstration of $>99.5\%$ two-qubit entangling gate fidelity in the Lukin group (Harvard) \cite{Evered_2023}. For alkali atom qubits such as Rb, the main challenge of integrating higher power lasers to reach higher gate fidelities is managing the correspondingly larger off-resonant light shifts coming from the two-photon Rydberg excitation scheme \cite{Evered_2023}. This is primarily a challenge of measuring these light shifts directly using the atoms themselves, and then iteratively feeding back on the spatial light modulator (SLM) in the optical beam path to improve light shift homogeneity to the $<0.5\%$ level required to reach $99.9\%$ two qubit gate fidelity.

Alkaline-earth atoms such as Sr and Yb have single-photon excitation pathways available that do not experience the same light shift issues, but the corresponding wavelength is deep in the ultraviolet (UV), where the current lack of good beam modulation technologies like SLM's is a limitation for scaling-up Rydberg entangling gates to large system sizes \cite{muniz2024high,reichardt2024logical}. 
A possible pathway toward efficient UV modulation is laid out in Section \ref{sec:lightControl}. Here we give examples for architectures enabling scalable light modulation based on integrated photonic circuits that use materials compatible with DUV wavelengths. 

Another path to improve two-qubit gate fidelities is to increase the Rydberg lifetime by inhibiting black-body-radiation decay, utilizing for example a cryogenic environment to suppress blackbody radiations~\cite{Schymik2021,Pichard_2024,zhang2024high}, or by suppressing the relevant blackbody modes via the Purcell effect~\cite{pultinevicius2025long}.

Recent efforts~\cite{Evered:2026jrf} at improving the gate fidelity of two-qubit entangling gates in Rubidium have brought down the error to the $10^{-3}$ range at 99.85 \% raw and 99.94\% post-selected CZ gate fidelity.

\paragraph{Error bias and erasure}\
In parallel with improving two-qubit gate fidelities, it is also possible to increase the effective thresholds of quantum error correcting codes such as the surface code, to operate further below threshold. In alkaline-earth atoms like Yb, a significant fraction of the error budget comes from decay into other atomic states that can be optically detected and converted into erasures \cite{wu2022erasure,ma2023high,sahay2023high}. Similarly, in alkali atoms, a significant fraction of errors correspond to atom loss and can be detected (Fig.~\ref{fig:lossdetection}) as delayed erasure errors. Leveraging erasure errors can increase the effective threshold of the surface code from $1\%$ to $2\%$ or higher \cite{wu2022erasure,baranes2025leveraging}, especially if additional measures are taken to convert other errors into erasure errors. Once leakage/loss is accounted for as erasures/delayed erasures, the remaining Pauli errors are themselves biased -- dephasing ($Z$-type) being the dominant contribution -- and this bias can be exploited to boost the threshold further with codes tailored to biased noise~\cite{aliferis2008fault,tuckett2018ultrahigh,tuckett2019tailoring,tuckett2019fault,bonillaataides2021xzzx}.

\paragraph{Individually-Addressed Gates}\
While a key advantage of the neutral atom quantum computing platform is the possibility of globally addressing all atoms simultaneously with the same laser beams for gates, and thus reducing the complexity of classical controls required, there are several possible advantages with individually-addressed single and two-qubit gates. For instance where single qubit gates should only be applied to only a subset of physical qubits, while in other places the same gates are applied to all qubits simultaneously (e.g. XY dynamical decoupling pulses or global Hadamards).
It thus becomes unrealistic to move atoms large distances in and out of a globally addressed single-qubit gate beam during gates that must occur on only a subset of physical qubits, and here individual-addressing becomes critical. Currently, single-qubit gate fidelities up to 99.9\% have been demonstrated with individual-addressing beams generated by an acousto-optical modulator~\cite{radnaev2025universal}. 

Two-qubit gates feature native local control, because the gate action can be turned off by separating individual atom pairs sufficiently to turn off the Rydberg interaction, which falls off rapidly as a function of distance ($1/r^6$). Nevertheless, there are scenarios in which small laser spots individually focused on atom pairs can have advantages over global addressing. 

For one, at moderate system sizes this individual focusing is more power-efficient than global addressing, and can be a route to higher fidelity two-qubit gates approaching 99.9\%.   

For larger system sizes global addressing, in which a single laser beam passes through all atoms simultaneously, becomes more power efficient. However, the fidelity achievable over a large array of atoms using a global beam is limited by the beam inhomogeneity across the array. In practice very high gate fidelity may only be achieved over a small section of the array. Local addressing opens up the possibility of site-wise calibrating every gate pulse, overcoming this limitation. On the other hand, individual addressing for Rydberg excitations means that nearby atoms that do not experience the gate do not have to be moved far enough away to turn off the Rydberg interaction, thus possibly increasing the packing density of qubits within the active region defined by the microscope field of view. In this case, one must be careful about still placing nearby atoms sufficiently far apart to suppress cross-talk from the individual-addressing laser beam. 
Conventional bulk components have to great success been applied to control systems containing hundreds of atoms \cite{bluvstein_logical_2024}.
At larger system sizes, new platforms for scalable optical quantum control e.g. photonic integrated circuits  (see Section \ref{sec:lightControl}) show a possible avenue towards increasing the optical control degrees of freedom for individual addressing.

\paragraph{Analog and hybrid architectures}
Neutral-atom processors are not only a promising platform for high-fidelity digital quantum gates, but also a natural platform for programmable many-body dynamics. Analog operation has already proved useful across a broad range of applications, from quantum simulation~\cite{greiner2002quantum,bernien2017probing,scholl2021microwave,ebadi2021quantum,chen2022continuous,darcangelo2024,Xu2025_Hubbard,gonzalez-cuadra2025,Qiao2025,Emperauger2025,manovitz2025quantum,leclerc2026} to combinatorial optimization problems~\cite{ebadi2022quantum,cazals2025identifying}. As control over native dynamics improves, these capabilities are extending beyond simple Ising Hamiltonians~\cite{scholl2021microwave,chen2022continuous,Qiao2025} and making hybrid analog-digital schemes an increasingly promising direction~\cite{katz2025,geim2026engineering}.

Analog operation requires control capabilities similar to those needed for high-fidelity digital gates, including high available laser power, good spatial homogeneity, and low intensity and phase laser noise. These requirements become particularly stringent in quantum simulation experiments carried out in regimes where the Rabi frequencies, detunings, and interaction energies are all comparable, so that even small calibration errors directly distort the target Hamiltonian. This requires precise calibration of the Rabi frequencies and detunings at the $\sim$0.1\% level. Accurate control of the interaction energies is also needed in such regimes, placing strong requirements on atom positioning, trap quality and stability, and atomic temperature. Higher experimental repetition rates will also play an important role both for applications that require many shots and for maintaining this level of precision through more frequent calibration updates.

To access new regimes of many-body physics, recent efforts have focused on tailoring the native dynamics of neutral-atom platforms beyond the simplest globally driven Ising Hamiltonians. A first step in this direction is to enrich the native Ising model with local Hamiltonian terms~\cite{leseleuc2017,geim2026engineering}, for example through local detunings used either to encode weights in optimization problems or to compensate boundary effects in analog state-preparation protocols. More broadly, recent experiments have demonstrated access to other interaction models, including XY and XXZ Hamiltonians~\cite{scholl2021microwave,chen2022continuous,Qiao2025}, as well as Floquet-engineered effective models~\cite{goldman2014periodically,bluvstein2021controlling,geier2021floquet,geim2026engineering}. Periodic driving is particularly attractive in this context, as it provides a route to engineer a broader class of Hamiltonians and to tune parameters that are not directly available in the native dynamics. The next challenge is to make these richer interaction models precise, robust, and rapidly calibratable at scale.

Beyond extending the range of accessible Hamiltonians, an important next step is to integrate analog evolution with digital control and measurement. Hybrid analog-digital architectures combine the efficiency of analog many-body evolution with the flexibility of digital control. A particularly promising route to such architectures is based on coherent mapping between interacting Rydberg manifolds and long-lived qubit encodings~\cite{geim2026engineering}. This enables fully programmable state preparation and measurement, local operations, non-destructive loss-resolved readout, and fast experimental cycle rates, while analog evolution implements many-body interactions that would otherwise require deep gate sequences. More generally, hybrid operation opens the way to interleaved layers of analog evolution and digital control, greatly expanding the range of applications that can be implemented on a neutral-atom processor~\cite{Zhou_2020,geim2026engineering,Lu2024,katz2025_Hybrid}. However, in the absence of full error correction, hybrid protocols will remain sensitive to noise accumulated across both analog and digital layers. This makes low error rates per qubit and per control block essential. It also requires a robust analog-digital interface, with high-fidelity coherent mapping and a clear understanding of the regimes in which such mapping remains valid for arbitrary layouts and interaction patterns.

\subsubsection{Continuous Reloading of Neutral Atom Qubits}

In the deep circuit regime, it becomes important to continuously replace qubits lost to Rydberg gate imperfections and finite vacuum lifetime. Specifically, one must cool and load fresh atoms, verify successful loading, perform state initialization, and move the initialized qubit back into the correct position, all with a sufficiently high rate to sustain large-scale quantum computation while not causing decoherence on the remaining computational qubits. One difficulty with realizing the above is that many of the above operations (cooling, imaging, state initialization) require scattering photons from atoms, and these scattered photons can cause decoherence on the remaining qubits. Moreover, even if the laser beams for qubit reloading are spatially localized, stray light from these beams can still cause decoherence on the computational qubits.

\begin{figure}
\centering
    \includegraphics[width=1\linewidth]{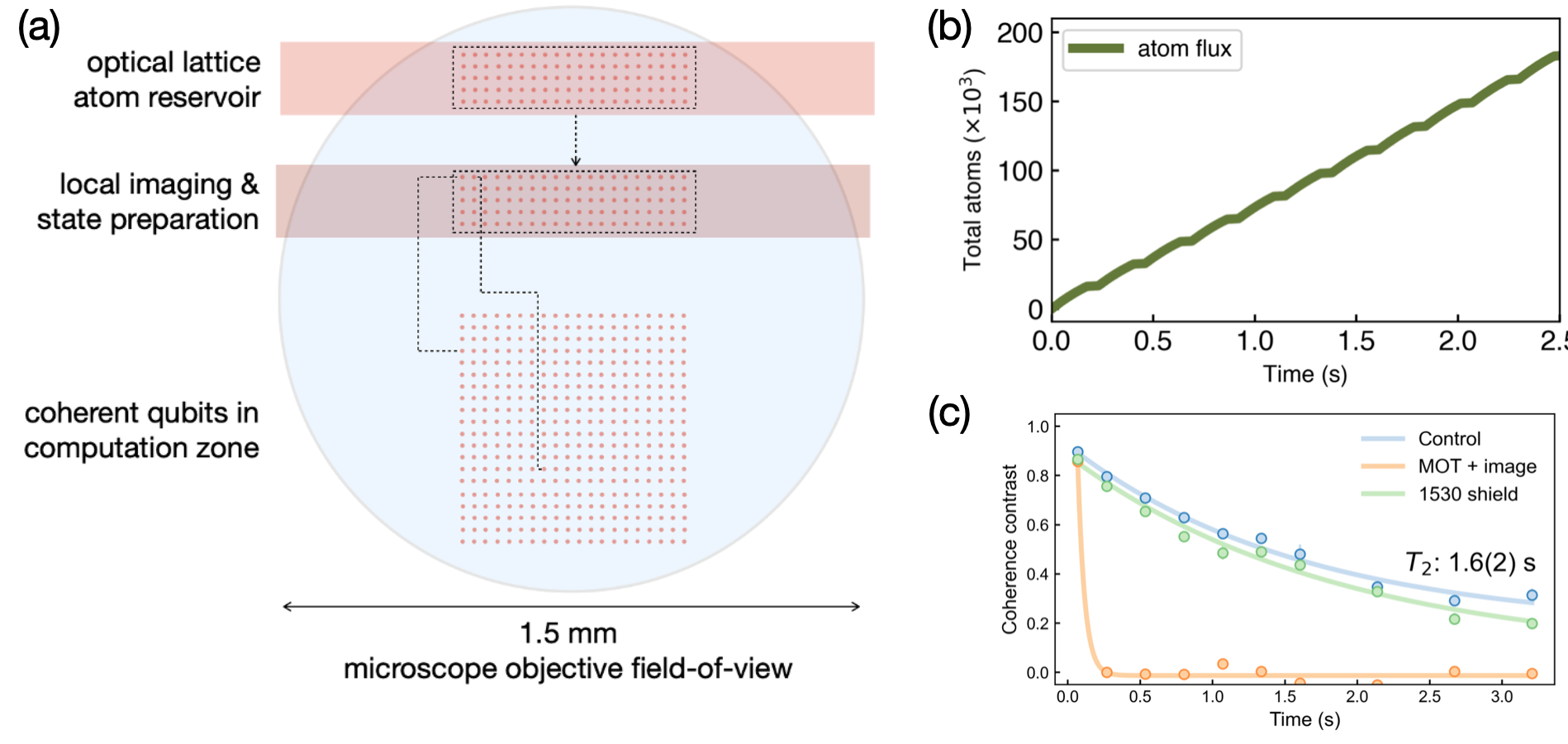}
    \caption{(a) Continuous reloading architecture consisting of optical lattice reservoir, reloading zone for imaging and state preparation, and computational zone, with moving AOD tweezers transporting atoms in between  zones. (b) Rate of continuous atom reloading from the optical lattice reservoir. (c) Effect of reloading operations on coherence of data qubits with and without 1530 nm light shift beam
    }
    \label{fig:reloading}
\end{figure}

One approach to realizing the above is with a dual species architecture \cite{singh2022dual,singh2023mid} (see Section \ref{sec:dualspecies} below), where the spectral separation of the two different atomic species ensures negligible decoherence from photon scattering and from stray laser light. Continuous reloading is also possible with only a single atomic species, but great care must be taken to spatially separate reloading and computational zones, in order to have laser beams for reloading localized to that zone, with vanishing overlap with the computational zone. The spatial separation also reduces the probability that a photon scattered during reloading will cause decoherence on a qubit in the computational zone. In addition, state shelving techniques available in alkaline-earth atoms can eliminate effect of scattered photons on computational qubits. 

Fresh atoms can be continuously delivered to the active region by an optical lattice or a 2D Magneto-Optical Trap (MOT) atomic beam.
The 2D MOT method was explored by the Bernien group (Chicago / Innsbruck) \cite{singh2022dual}. A transport-cavity based approach for Ytterbium was demonstrated by the Thompson group \cite{li25continuous}.
Several groups have demonstrated continuous optical lattice delivery of fresh atoms \cite{norcia2024iterative,Gyger2024}. Recent work in the Lukin group (Harvard) has shown an optical lattice method with very high rate of delivery of atoms ($\sim$100,000 / s) combined with maintaining the coherence of an array of up to $\sim$3000 atoms during reloading \cite{Chiu2025Coherent}. This work combines (Fig.~\ref{fig:reloading}) optical lattice transport of atoms, a large field-of-view objective to allow sufficient spatial separation between reloading and computational zones, a novel, laser-free method of dissipatively loading optical tweezers from the optical lattice reservoir, high resolution acousto-optical deflectors for long-distance atom transport, and a strong light-shift beam from a telecom C-band laser that locally changes the atomic transition frequency in Rb by $\sim$GHz without affecting the hyperfine ground state qubit, thereby suppressing response to near-resonant scattered photons. In combination with the non-destructive, atom loss detecting readout method demonstrated in the same group (Fig.~\ref{fig:lossdetection}) \cite{Bluvstein_2025}, all the ingredients are in place to sustain continuous operation of neutral atom quantum processors with >10,000 physical qubits.

\subsubsection{Fast, High Fidelity Readout of Neutral Atom Qubits}

\begin{figure}[h]
    \centering
    \centering
   \includegraphics[width=0.8\linewidth]{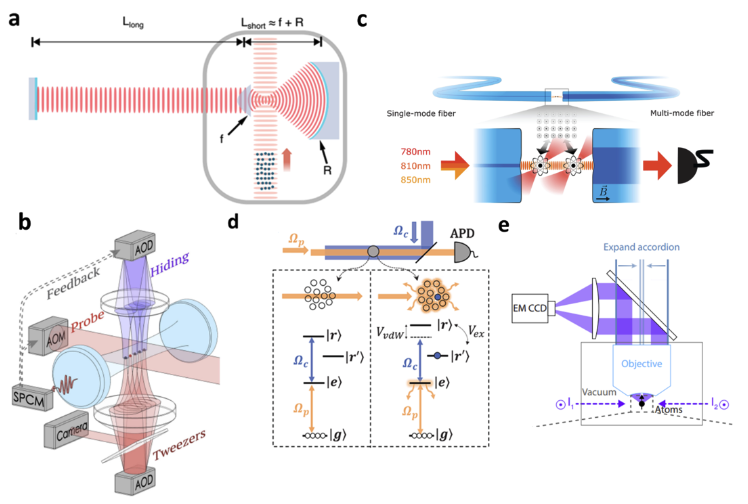}
    \caption{Enhanced atomic qubit readout methods. (a) increased collection efficiency via a low finesse cavity \cite{Shadmany2024},  (Simon group). (b) increased photon emission via a high-finesse, low cooperativity cavity \cite{hu2025site},  (Vuletic group). (c) cavity transmission detection \cite{grinkemeyer2025error},  (Lukin group). (d) Collective, non-destructive readout in free-space via Rydberg EIT \cite{xu2021fast}, enabling $\sim92\%$ readout fidelity and over $\sim90\%$ estimated survival probability in $\sim 6 {\mu s}$ (Vuletic group). (e) alternating resonant fluorescence detection \cite{su2025fast}, enabling 99.6\% fidelity in $\sim 4.8 {\mu s}$, yet with a low survival rate (Greiner group).}
    \label{fig:Vuletic_FastReadout}
\end{figure}

Although there are some ways to significantly increase the readout speed by compromising the survival of atoms \cite{su2025fast}, the two prevalent methods to accelerate the readout process are based on cavities and collective atomic effects. 

\paragraph{Cavity-based readout}

Optical cavities provide a way to realize fast (microsecond-scale) high-fidelity (>99\%) atomic state readout while maintaining high atom survival probabilities (>99\%). In an optical cavity, atomic emission into the photonic mode defined by the cavity is enhanced as compared to emission in free space (Purcell effect). This allows for atomic state detection schemes which require the atom to scatter only a few photons, limiting atom loss out of the trap. Some of the experimental setups are illustrated in Fig.\ref{fig:Vuletic_FastReadout}.  Recent demonstrations include: $\sim$99.5\% state readout fidelity at $\sim$99\% survival probability in $\sim 50 {\mu s}$ (Stamper-Kurn group) \cite{deist2022fast} and 99.4\% readout fidelity with 98.9\% survival probability in $\sim 200 {\mu s}$ (Vuletic group) \cite{hu2025site} in a cm-scale near-concentric cavity. In the Simon group, a recent demonstration realized  99.55\% readout fidelity with 99.89\% survival probability in $\sim 200 {\mu s}$ by using increased collection efficiency via a low finesse cavity \cite{Shadmany2024}. Micro-scale fiber Fabry-Perot cavities (FFPC) with cooperativities of > 100 have been used in the Reichel group to realize 99.92\% \cite{gehr2010cavity} in a read-out time of 100 $\mu s$. More recently, Lukin group used a FFPC to realize 99.95\% readout fidelity with 99.97\% survival probability in $\sim 2.5 {\mu s}$  \cite{grinkemeyer2025error}, which can be narrowed down to sub-microsecond readout at the expense of decreasing the readout fidelity to ~99.0\%, limited by atom collection efficiency. An additional advantage of this readout is that, often, the atom stays in a specific state, e.g. in the $\ket{F=2, m_F=2}$ state of Rb 87 in the Lukin group experiment, so that after readout the atom can be immediately reused by a Raman transfer to the desired qubit state.

A significant downside of the single-cavity approach is that it only allows for a single atom to be read out at a time because the photonic mode couples to all atoms in the cavity. While atoms can be sequentially coherently transported in and out of the cavity mode \cite{deist2022fast}, this approach is relatively slow due to the slow transport times. An alternative approach was recently demonstrated in the Vuletic group \cite{hu2025site}. It involved using a separate laser beam to light-shift select atoms out of cavity resonance. Since switching optical beams on and off is typically much faster than transport, this is a promising approach. 

With the fastest single-atom readout, of e.g. $\sim$1~$\mu$s per atom, a 10,000 atom array can be read out in 10 ms, comparable to global fluorescence-based imaging times. Further increase in speed can be realized by using suitable search algorithms performed on groups of atoms, as recently demonstrated in the Vuletic group \cite{hu2025site}. 

An alternate approach employs cavity arrays for atom-array readout, as demonstrated in the Simon lab~\cite{shaw2026cavity}. Here, up to 600 cavities ~\cite{soper2026stability} have been made degenerate to within their readout-optimized linewidth. Through the use of macroscopic optics and microlens arrays, the spacing between the resonator modes is compatible with the separation of atoms in a typical neutral atom processor, yet all optics remain far (mm to cm) away from the atoms, making this approach quite promising. There is thus the potential to envision a scenario where every atom in a neutral atom computer is read-out, in situ, with its own cavity.

\paragraph{Collective effects}
On the other hand, using collective effects such as electromagnetically-induced transparency (EIT), it is possible to measure the change in the absorption of an atomic ensemble owing to the state of an atom near or inside the ensemble \cite{xu2021fast,vsumarac2026controlling}. The latter method is sensitive to the ensemble shape and the atom-ensemble distance, but $>99\%$ fidelity is expected in a state-of-the-art system, where single atoms will be placed close to large ensembles.

It is also possible to map the state of a single atom onto $N$ auxiliary measurement atoms to achieve a factor of $N$ increase in the photon flux and corresponding reduction in the measurement time  \cite{petrosyan2024fast}. The required repetition code mapping is 
\begin{equation}
\left(c_0|0\rangle+c_1|1\rangle\right)\otimes |00...0\rangle_N\rightarrow  c_0|0\rangle\otimes|00...0\rangle_N +c_1|1\rangle\otimes|11...1\rangle_N. 
\end{equation}
Since the auxiliary atoms are measured in the $z$ basis, phase errors during the mapping, which requires a multi-qubit gate, are inconsequential, which allows for increased effective fidelity of the mapping step. 

In the near future, these two technologies should allow adaptive readout of syndrome qubits utilizing binary searches to find corrupted qubits \cite{hu2025site} and detector multiplexing to further increase SNR and compatibility with existing platforms \cite{zhang2025dualtypedualelementatomarrays}. It is also possible to apply adaptive readout solutions to regular fluorescence detection, by replacing single-photon sensitive cameras with arrays of fast single-photon detectors. Thus, recording detection events with high temporal resolution, as well as recognizing the change in detection statistics, can significantly reduce measurement time. Readout technology as a whole would progress towards state-resolving and loss-detecting methods \cite{scholl2023erasure,ma2023high}, to enable efficient error correction.

\paragraph{Reducing Latency of Classical Controls}
In addition to the time required for qubit state detection via fluorescence imaging readout, another significant contribution to QEC cycle time is the time required to process the obtained data and perform mid-circuit operations by executing the right atom moves, gate pulses, etc. Real-time decoding and feedforward on neutral atom arrays has already been demonstrated within logical qubit algorithms with the help of dedicated Field-Programmable Gate Array (FPGA) based controllers~\cite{bluvstein_logical_2024}, and further developments of this FPGA mid-circuit processing architecture, combined with calculations executed on a dedicated Graphics Processing Unit (GPU) for the computation of e.g. RF waveforms corresponding to atom moves, can reduce the latency of classical controls significantly below the millisecond timescale.

\paragraph{Simultaneous state measurement and cooling}

\begin{figure}[h]
    \centering
    \centering
   \includegraphics[width=0.6\linewidth]{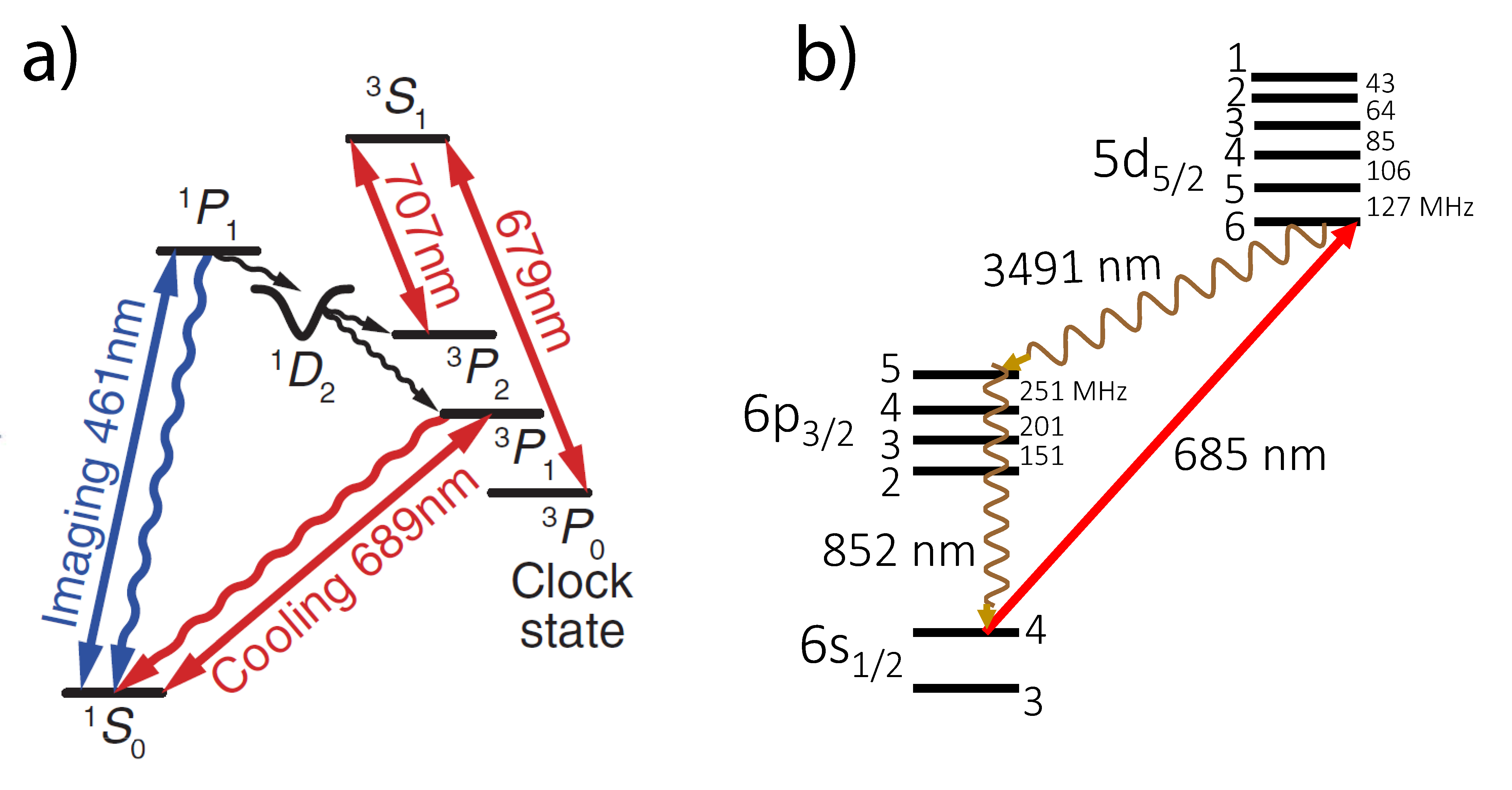}
    \caption{Simultaneous state measurement and cooling using a) a narrow line transition in Sr \cite{covey2019sr}and  b) a  quadrupole transition in Cs \cite{Scott2025}.}
    \label{fig:Cs5d_readout}
\end{figure}

One of the challenges of quantum state measurements is that fast readout is often accompanied by motional heating  which limits the number of readout cycles that can be performed without recooling the atoms.  Early non-destructive  alkali atom demonstrations of qubit state measurements used stretched state cycling on the closed transition between $ns_{1/2}\ket{f=m=I+1/2} \leftrightarrow np_{3/2}\ket{f'=m'=I+1/2+1}$ \cite{Fuhrmanek2011,Martinez-Dorantes2017,Kwon2017}. This requires using $\sigma$ polarized light in a 1D geometry that can only cool along one direction and heats the atoms transversally.  In other atoms there are completely closed transitions which enable cycling for state measurement together with three-dimensional cooling. A prominent example is provided by the $^1S_0 \leftrightarrow {^3}P_1$ transition in alkaline earth like atoms. The $\ket{0}$ state of optical frequency qubits encoded in $\ket{0}=\ket{^1S_0}$ and $\ket{1}=\ket{^3P_0}$ can be detected by cycling on the above transition with essentially no leakage to the $\ket{1}$ state. This was demonstrated with Sr qubits where repeated state measurements with fidelity of 0.981(1) and retention of 0.996(1) were achieved, see Fig. \ref{fig:Cs5d_readout} \cite{covey2019sr}. 

Obtaining similar results with alkali atoms has been challenging due to the lack of fully closed transitions that support 3D cooling. A figure of merit for cycling between ground and excited state hyperfine levels  can be written as $fom\sim (\Delta_{\rm hf}/\Gamma)^2$, where $\Delta_{\rm hf}$ is the excited state hyperfine splitting, and $\Gamma$ is the linewidth. This figure of merit determines how many photons can be cycled without suffering a Raman transition that changes the qubit level. For Rb and Cs the figure of merit when using the $5p_{3/2}$ and $6p_{3/2}$ levels is about 2000, which has proved inadequate for high-fidelity state measurements in free space with 3D cooling. An alternative for Cs is to cycle via the quadrupole transition  from the ground state to  $5d_{5/2}$ \cite{Scott2025}. This is cycling for 
$6s_{1/2}\ket{f=4}\rightarrow 5d_{5/2}\ket{f=6}\rightarrow 6p_{3/2}\ket{f=5}\rightarrow 6s_{1/2}\ket{f=4}$ as shown in Fig. \ref{fig:Cs5d_readout}, and has a much larger figure of merit of about $1.1\times 10^6$ which enabled state identification with fidelity 0.9993(4) and atom retention of 0.9954(5).

\section{Scalable Photonic Control Technologies} \label{sec:lightControl}
To build a scaled neutral atom quantum computer that achieves computational utility, a central problem needs to be solved: how can one manipulate and control atoms individually, at scale, with high precision and fidelity? 

We anticipate that a computer of 10,000 -- 100,000 physical qubits will be required to achieve quantum utility. Conventional bulk-optic devices are reaching limitations in providing a sufficient number of degrees of freedom to control these physical qubits simultaneously. To meet this challenge, a new generation of laser and optical control systems need to be developed, akin to the transition from vacuum tubes to transistors, which enabled the classical computing revolution.

Next generation photonic control needs to support four operations -  quantum gates, dynamic trapping, cooling and imaging. To execute these operations, integrated photonic control must simultaneously satisfy the required light field characteristics, including frequency, intensity, spatial profile and polarization, which may be modulated in time. In addition, the optical signal must meet noise requirements on the frequency and intensity.
Photonic integration offers the potential to bring many of these degrees of freedom to the chip scale, with improved performance, reliability and scale.

In the following sections, the challenges and opportunities for next generation light control as well as proposed research directions are outlined and described.

\subsection{Challenges and Opportunities}

\paragraph{Scaling qubit operations } 
Conventional bulk optic devices provide either slow control speeds ($\sim$ kHz) at very large channel counts ($\sim$ Mega-pixel), e.g. Liquid Crystal on Silicon Spatial Light Modulators (LCoS SLMs) and Digital Micromirror Devices (DMDs), or fast control ($\sim$ MHz) on a limited number ($\sim 100$) of independent control channels, e.g. Acousto-optic modulators (AOMs) combined with Acousto-optic deflectors (AODs). AOMs and AODs are the current standard for realizing frequency and spatial light control for gates and atom motion, but are constrained in reaching large channel counts by size, weight, power, and cost (SWaP-C).

Integrated photonic devices offer the opportunity for precise control of light frequency, phase, intensity, polarization, and spatial profile, enabling independent control of quantum states on a scale compatible with achieving practical utility at $>10^6$ Quops and $\sim$ 10,000 - 100,000 physical qubits. 
Integrated photonic modulators are scalable, due to their small form factor with  modulator sizes <0.1 $\text{mm}^2$. A single wafer may contain hundreds of thousands to millions of modulation channels, delivering SWaP-C advantages. By utilizing photonic chip-based architectures with atomic systems, the local, independent, implementation of single and two qubit gates on large atom arrays is possible. This can enable non-transversal operations that are crucial for most universal error correction codes. Individual addressing also allows for individual calibration of gate pulses. This can overcome fidelity limitations originating from light field inhomogeneities when large arrays of atoms are illuminated globally. 

\paragraph{Atom trapping and motion}
As outlined in Section \ref{sec:atomControl} scaling laser power for atom trapping is a key challenge. Integrated laser systems such as Vertical-Cavity-Surface-Emitting-Laser (VCSEL)~\cite{chen2023deep,warren2018low} arrays are already widely employed in industrial laser systems. NIR wavelength VCSELs operating between 800 nm - 1 um, suitable for trapping, are readily available but need to be validated for use in atomic systems. Another promising candidate for scaling laser power for trapping applications is the Photonic Crystal Surface Emitting Laser (PCSEL), which can support up to 50W of power per device~\cite{yoshida2023high}. Additionally, a variety of gain mediums can be integrated in silicon photonic processes through transfer printing or wafer bonding~\cite{wang2023photonic}.

Another crucial operation of an atom based Quantum Processing Unit (QPU) is dynamic trapping. The shuttling of atoms is needed to establish a dense array and to implement two qubit gates beyond nearest neighbor connectivity. Current protocols for atom rearrangement are constrained by the limited programmability of bulk optics, which cannot independently move tweezers. Furthermore, AODs suffer from aberrations due to finite acoustic velocity which inhibit fast transport~\cite{manetsch2024tweezer}.
Integrated photonic devices such as acousto-optic deflectors on chip~\cite{lin2024optical}, or micro-optical cantilevers \cite{saha2026nanophotonic} as well as actuated vertical grating couplers promise a path towards fully independent two dimensional motion control.

\paragraph{Laser sources for gates}
Reliable, SWaP-C scalable laser sources are a key ingredient enabling atom-based quantum computation at the utility scale. Integrated laser solutions have already  enabled the current generation of commercially available lasers. Utilizing and advancing the state of the art in integrated laser technology will be a key challenge and goal.

Low noise integrated lasers and frequency references can provide the noise performance needed to address atomic transitions and provide precision qubit control at a fraction of the cost and size of bulk ECDLs, Titanium-Sapphire (Ti:Saph) lasers, and Ultra-Low Expansion (ULE) cavities used for laser stabilization.
Chip scale laser designs include extended cavity tunable lasers~\cite{nejadriahi2024a_852nmECL}, self-injection locked lasers~\cite{isichenko2024a_SIL780}, and stimulated Brillouin lasers~\cite{gundavarapu2019a_BrillouinUHQ,chauhan2021a_visibleBrillouin,liu2023a_moleculeSub100mHz}.
\paragraph{Electrical control}
Scalable electronic control of integrated photonic circuits is a central challenge in its own right. At thousands of channels, signal delivery becomes the bottleneck: conventional architectures route $\mathcal{O}(N)$ electrical lines to wire-bond pads along the chip perimeter, whereas the number of modulators grows as $\mathcal{O}(N^2)$ with chip area, so perimeter wiring cannot keep pace with utility-scale channel counts. Signal generation scales just as unfavorably: every channel requires precisely timed, low-noise analog waveforms, and the aggregate data rate and synchronization burden grow with the channel count. Overcoming these bottlenecks will require moving signal generation onto the chip itself, alongside the photonics.

Highly functional photonic integrated circuits have the potential to improve robustness and scalability while reducing footprint and cost. There is also the potential to improve certain performance aspects (e.g. noise, phase stability, control, light quality delivered to the atom).

\subsection{Research Directions}

\subsubsection{VIS-NIR Integrated photonic platforms}
\begin{figure}[h!]
    \centering
    \includegraphics[width=0.95\linewidth]{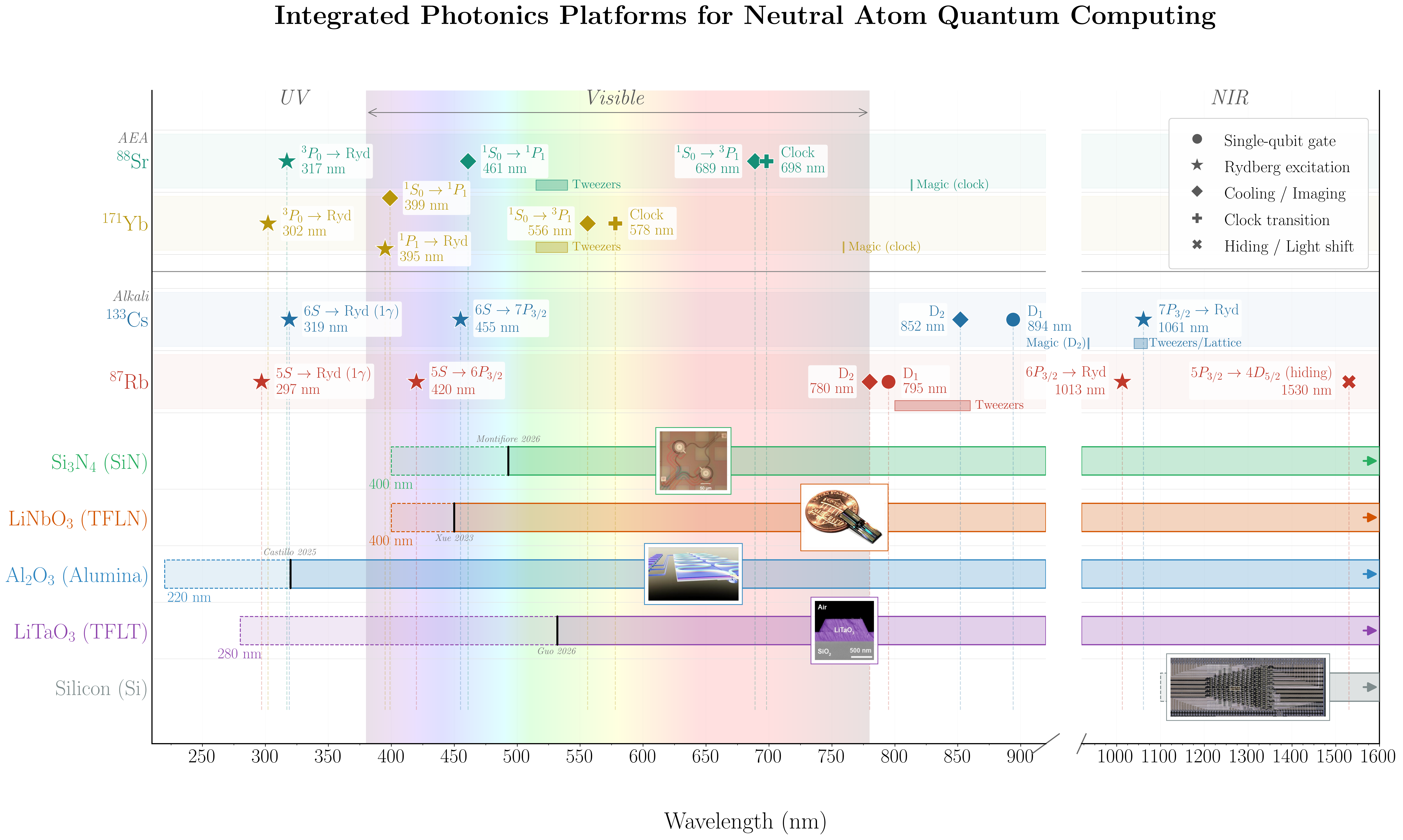}
    \caption{Wavelength ranges for selected operations on prominent species used in neutral atom quantum computation in comparison with operational wavelength ranges for integrated photonic devices. 
    Colored bars for different wave-guided integrated photonics platforms show operational wavelength, bounded by the material (waveguided where available) transparency window (SiN:~\cite{west2019low}, TFLN: \cite{ccabuk1999urbach}, Alumina:~\cite{west2019low}, TFLT~\cite{ccabuk1999urbach}). Vertical bar indicates demonstrated modulation of at least 1 MHz at the stated wavelength. Silicon Nitride: Montifiore~\cite{montifiore2026blue},
    Lithium Niobate: Xue,~\cite{xue2023full}
    Alumina: Castillo,~\cite{castillo_cmos-fabricated_2025}, Lithium Tantalate:~\cite{guo2026robust}. 
    Inset pictures from SiN:~\cite{menssen2023scalable}, TFLN: (unpublished), Alumina:~\cite{castillo_cmos-fabricated_2025}, TFLT~\cite{powell2024dc}. Silicon:~\cite{harris2016large}.
    }
    \label{fig:PIC_platforms}
\end{figure}

The central requirements any photonics platform has to fulfill are:

(1) a transparency window covering wavelengths relevant for atomic species used in neutral atom quantum computing, ideally from ultraviolet (UV) to the near infrared (NIR). Figure \ref{fig:PIC_platforms} shows a comparison of prominent integrated photonic platforms together with relevant wavelengths for optical control of selected alkali and alkali-earth atoms. Most of the relevant wavelengths are in the visible (VIS) and NIR. 

(2) sufficient modulation bandwidth commensurate with the time scales of quantum operations, including gates, motion, readout. The fastest timescale is that of gates and given approximately by the highest Rabi frequency. Some control functions require wide-band frequency response from DC out to 10s or 100s of MHz. In Figure \ref{fig:photonics_modulation} an overview over current architectures and their modulation speed is given.

(3) modulation in a small footprint. The strength of the modulation transfer from electrical to optical is measured by the product of the voltage required to achieve a $\pi$ phase shift of the light in the modulator and the modulator length, $V_\pi \cdot l$. To reach very large channel counts, it will be necessary to co-integrate driving electronics with the optical modulators. Typically, more advanced electronics processes with smaller transistors are designed to operate with smaller voltage swings, so there is a complex trade space between $V_{\pi}$, $l$, and the area of the driving circuits. Critically, the area constrains the circuit complexity: memory size, digital-to-analog converter (DAC) precision, etc..

\paragraph{Silicon Nitride}
Silicon nitride waveguides~\cite{stutius1977silicon,blumenthal2018SiN,sacher2015multilayer} provide a mature and widely established platform for integrated UV-NIR photonics.
A key advantage of SiN is the high power handling and absence of photo-refractive effects. Recently, fast piezoelectric modulators in CMOS-compatible architectures have enabled light modulation at megahertz rates~\cite{stanfield2019cmos,dong2021high,menssen2023scalable}. More recently, acousto-optic modulation of visible light at gigahertz frequencies has also been demonstrated \cite{freedman_gigahertz-frequency_2025}. Typically, gigahertz frequency phase modulation is achieved with bulk electro-optic modulators or with their integrated counterparts that suffer from poor optical power handling. Therefore, this new capability in silicon nitride platforms promises to facilitate scaling of qubit control systems that will require increasing optical power as the number of atoms grows into the thousands and beyond.
\paragraph{Alumina waveguides}

Alumina-based photonics~\cite{mckay2023high,shugayev2024cmos} has recently yielded ultraviolet modulators~\cite{castillo_cmos-fabricated_2025} (see Fig.~\ref{fig:apics}c), transferring the technologies developed in silicon nitride to a platform that operates at wavelengths as low as 320~nm. While recent work performed piezoelectrically controlled strain tuning of alumina ring resonators, future directions can also implement acousto-optic modulators and cantilever/strain modulators to endow the platform with sideband-generation and amplitude-modulation capabilities, as in previous work with silicon nitride~\cite{freedman_gigahertz-frequency_2025,dong2021high,dong_piezo-optomechanical_2022}. Together, these demonstrations of acousto-optic modulation of visible light and fast ring modulation at ultraviolet wavelengths add important capabilities to the growing library of components available in scalable chip-based photonic platforms.

\paragraph{Thin Film Lithium Niobate (TFLN)}
Lithium niobate (LiNbO$_3$) possesses a wide transparency window spanning approximately 400\,nm to 5\,$\mu$m, making it suitable for addressing visible and near-infrared atomic transitions. Its strong electro-optic ($\chi^{(2)}$) effect enables fast, low-voltage modulation. In the thin-film platform, electro-optic modulators with bandwidths exceeding 150\,GHz have been demonstrated~\cite{wang2018integrated}, and recent work has realized large-scale integration of up to 256-channel devices for parallel qubit control~\cite{christen2025integrated}. TFLN is thus a promising candidate for scalable optical addressing in neutral atom arrays, particularly where fast switching and broad spectral coverage are required.

\paragraph{Thin Film Lithium Tantalate (TFLT)}
Lithium tantalate (LiTaO$_3$) has recently emerged as a compelling alternative to lithium niobate for integrated photonics. It possesses a comparable electro-optic coefficient ($r_{33} \sim 30$\,pm/V)~\cite{wang2024litao3nature}. For different crystal compositions light transmission in the DUV, as low as $\sim 280$ nm is reported~\cite{steinberg2009two,yu2025efficient}.
A higher optical damage threshold and reduced photorefractive sensitivity~\cite{powell2025subvolt}  makes TFLT particularly attractive for applications in quantum control at visible light. 
Recent work has demonstrated TFLT modulators with electro-optic bandwidths exceeding 110\,GHz~\cite{wang2024ultrabroadband} and manufacturing using DUV stepper lithography~\cite{wang2024litao3nature}. 

\subsubsection{Integrated Photonics for Gates}
Fast, local, parallel, independent execution of single and two qubit gates at a scale that is capable of addressing 10,000 - 100,000 physical qubits is one of the main ingredients for utility scale quantum computation. Photonic integrated circuits offer MHz - GHz phase and amplitude modulation speeds at component sizes of $\lesssim0.1$ $\text{mm}^2$ per channel.

Taking $^{87}$Rb as a prominent example, 795~nm or 780~nm light drives single-qubit rotations, while 420~nm and 1013~nm light drives two-photon Rydberg blockade gates. These wavelengths fall within the operating windows of several integrated platforms (see Fig.~\ref{fig:PIC_platforms}). Switching bandwidth required for gate operations is a second crucial measure.
The speed of single qubit gates execution is less critical due to the long lived hyperfine ground state.
In Rydberg blockade gates the  timescale is largely dictated by the Rydberg decay rate (T1) and the dephasing rate ($\text{T2}^*$) between the Rydberg state and the hyperfine ground state qubit manifold.
Practically Rydberg Rabi rates are also limited by available laser power. $\sim$ 1-10 MHz modulation bandwidth can yield high fidelity gates > 99.5 $\%$ \cite{Evered_2023}.

\begin{figure}[h]
    \centering
    \includegraphics[width=0.8\linewidth]{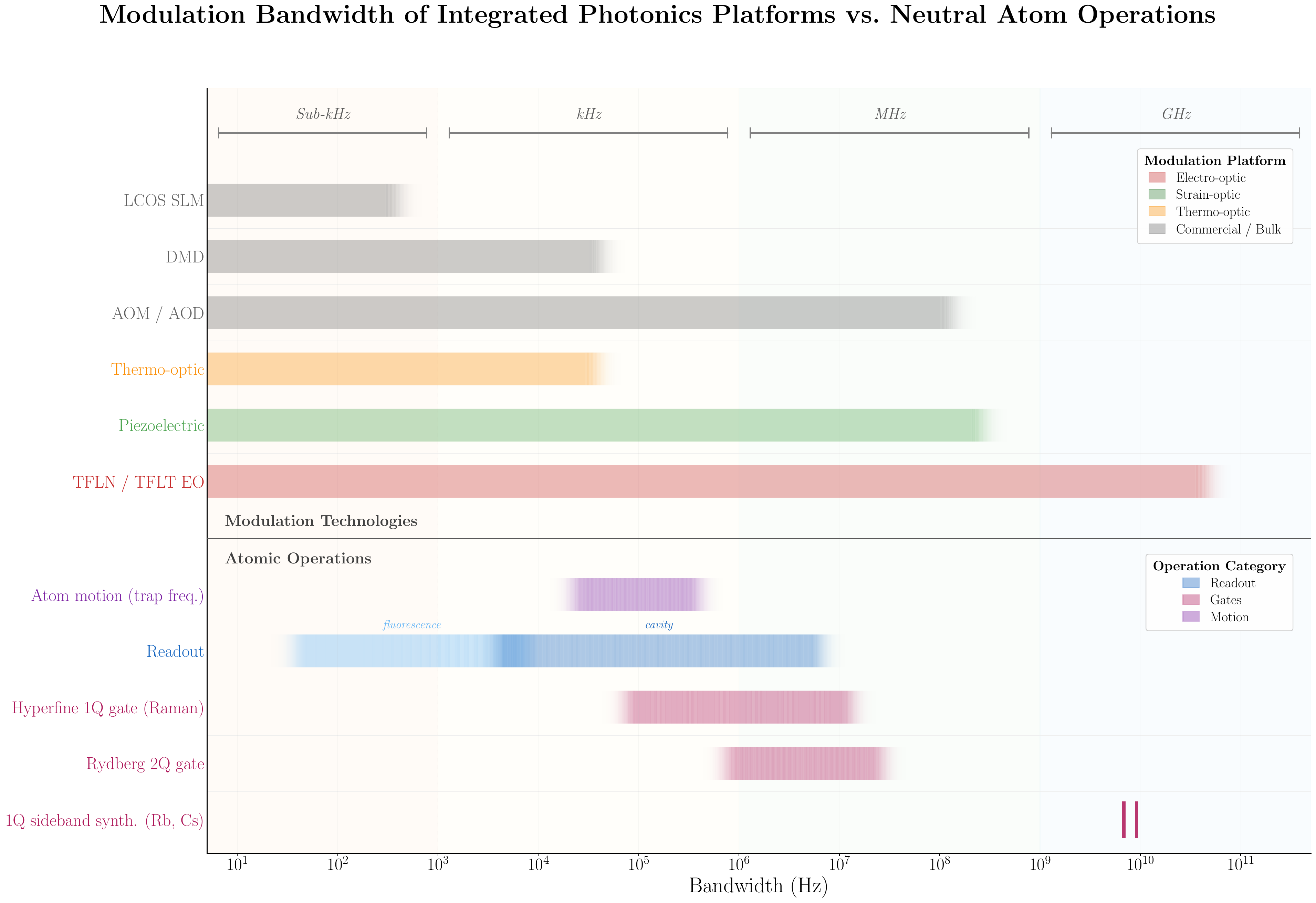}
      \caption{
    Modulation bandwidth of  photonics platforms (top) compared to characteristic bandwidths of neutral atom quantum computing operations (bottom).
    }
    \label{fig:photonics_modulation}
\end{figure}

In Figure \ref{fig:photonics_modulation} an overview of different bandwidth requirements for atomic operations vs. the modulation capabilities of photonics platforms is shown.

Recent research~\cite{menssen2023scalable,palm2023modular_new,christen2025integrated,zhao2025integrated} has demonstrated key capabilities of integrated platforms for quantum control: (C1)
Operation at wavelengths compatible with single and two qubit gates in the near IR and visible wavelengths, (C2) tileable modulator architecture for scaling channel counts, (C3) high modulation extinction > 30 dB, (C4) high repeatability of pulse area, as required for high fidelity gate execution and (C5), modulation speeds > 100 MHz.

\begin{figure}[h]
    \centering
\includegraphics[width=1\linewidth]{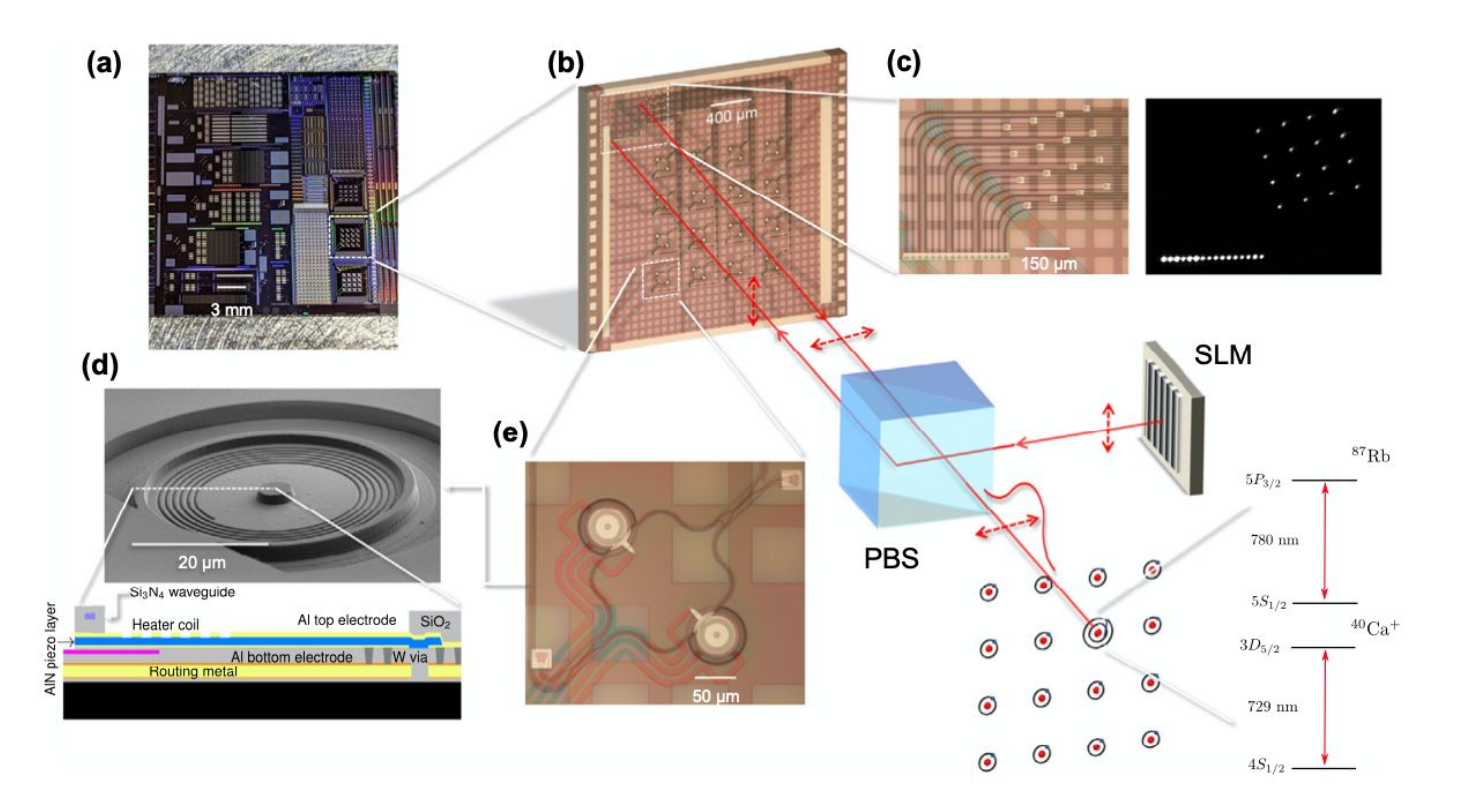}
    \caption{Atom Control Photonic Integrated Circuit (APIC) architecture. (a) Photograph of full reticle. (b) APIC modulator array with a modulator pitch of 420 µm. (c)  out-coupling and  in-coupling area. Chip detail (left) and camera image with light coupled into all ports (right). (d) SEM image of individual ring (top). Schematic cross section of device with piezo-stack and waveguiding layers illustrated. (e) Dual Ring Mach-Zehnder modulator with local in- and out-coupling gratings, which can be used as an alternative to the grating couplers in (c). Bottom right: illustration of setup. An SLM projects light onto the APIC, where the light is modulated and after passing through a PBS imaged onto an array of atoms. The input light path onto the SLM is not shown.}
    \label{fig:mit-apic}
\end{figure}

At MIT these development efforts in photonic integrated devices have been translated to a full quantum control architecture for neutral atoms~\cite{menssen2023scalable,christen2025integrated}. Integration with suitable atomic systems is an ongoing effort. 
At the core of the architecture shown in Fig. \ref{fig:mit-apic} are high speed on-chip modulators that are vertically coupled into free space and then imaged onto atomic qubits. Power distribution to individual channels is achieved through free space holographic fan-in. As opposed to splitting the optical power on-chip, this scheme minimizes power concentrated at a single free-space to chip interface, which could induce damage, and allows for true two-dimensional scaling.
The number of independent optical channels has seen a steady increase, from the 16 channel Silicon Nitride based device in  Fig.~\ref{fig:mit-apic} to the device shown in  Fig.~\ref{fig:apics}a, which shows a 256 channel amplitude modulator in TFLN (unpublished). 

Recently, cantilever based vertical emitting waveguide structures have been demonstrated in piezo-electrically actuated silicon nitride. Fast electrical actuation in two spatial dimensions in combination with fast temporal modulation allows for 2D intensity patterns of light to be generated in a scanning approach~\cite{saha2026nanophotonic}.

In these integrated architectures, waveguides are routed from the modulators to a central vertical coupler array and electrodes are routed to wire bond pads along the perimeter. 
Further scaling of these devices will require co-integration of electronics and unit cell-based approaches as discussed below.

\subsubsection {Scalable Electronic Control}

\begin{figure}[h]
    \centering
    \includegraphics[width=0.8\linewidth]{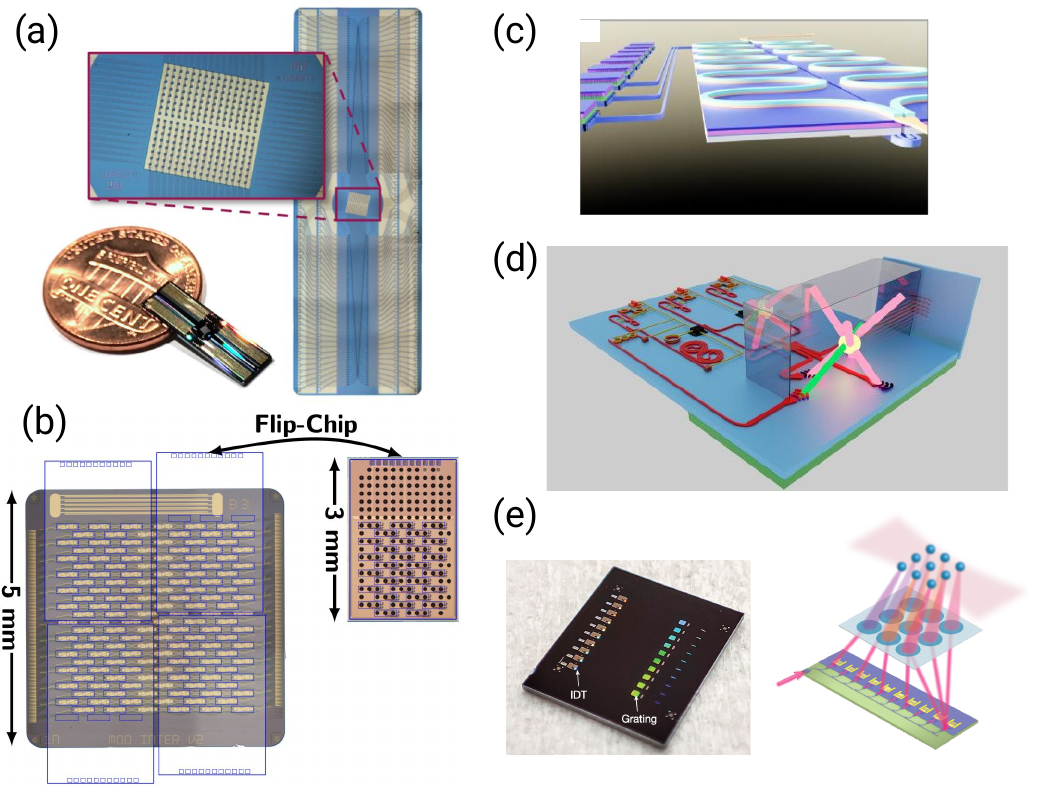}
    \caption{Integrated Modulation Capabilities. (a) LiNbO\textsubscript{3} 256-channel amplitude modulator array for DC-10 GHz control. (b) LiNbO\textsubscript{3} modulator array with bump-bondable contacts for a TSMC ASIC driver to be tiled on top for scalable electronic control. (c) Alumina-aluminum nitride photonic circuit for ultraviolet light control \cite{shugayev2024cmos}.
    (d) Fully integrated MOT and  tweezer array concept~\cite{blumenthal2024enabling}.
    (e) Integrated acousto-optic arrays for dynamic beam steering and tweezer arrangement~\cite{li2023frequency}.
    }
    \label{fig:apics}
\end{figure}

Current state of the art PIC controllers are limited by electrical signal delivery. Conventional architectures rely on wire bonding $\mathcal{O}(N)$ electrical lines routed to the chip's perimeter to control $\mathcal{O}(N^2)$ modulators in the chip's area.  Scaling further by lengthening the perimeter is impractical, and another solution is needed. The proposed solution is to co-integrate the electronics and photonics via flip-chip bump bonding, see Fig. \ref{fig:apics} b. With this architecture, every modulator has its own electronic driver and pulse-sequence memory, resolving two major roadblocks to scaling: 1) rather than routing individual wires to bond pads at the perimeter, voltages for each modulator are generated locally and programmed via a common serial peripheral interface (SPI) and 2) on-chip memory eliminates the data bottleneck for high-speed sequences. 

\subsubsection{Integrated Photonics for Atom Trapping}

Atomic motion is a crucial ingredient in the neutral atom tweezer platform. Fast shuttling is needed for creating defect-free arrays, replacing lost atoms, and entangling operations beyond nearest neighbors. Conventional shuttling techniques utilizing 2D-AODs are limited to patterns that lie in the product space of the frequencies generated for the single-dimensional x and y AODs, so $\mathcal{O}(N)$ parameters are used to control $\mathcal{O}(N^2)$ spots. Using integrated photonic modulator arrays with local control at each pixel to generate complex patterns would asymptotically reduce the number of movement operations to perform a given logical sequence, and allow different zones of the computer to operate independently. Furthermore, typical acousto-optic technologies have rise times of hundreds of nanoseconds. On the other hand, fast integrated modulators with picosecond rise-times have been demonstrated~\cite{yu2022integrated}. The enhanced speed, coupled with the abundance of spatial degrees of freedom, may allow integrated modulators to move atoms at higher velocities than with bulk technologies by careful engineering of the atomic wave packet in motion. For example, techniques such as throwing and catching atoms could be parallelized asynchronously over an array~\cite{hwang2023optical}.

Recent advances in integrated PIC-to-free-space beam emitters have enabled the demonstration of large area $\sim$ mm beam emitters for realizing integrated 3D MOTs~\cite{isichenko2023a_PIBDrubidium}. By incorporating piezo-electric actuators with these emitters, three-dimensional control and dynamic trap depth adjustment becomes possible, enabling fine atom positioning and dynamic loading algorithms for fully populated arrays. A concept of a fully integrated 3D MOT, utilizing integrated photonic beam projection for cooling and tweezer trapping is illustrated in  Fig. \ref{fig:apics} d)~\cite{blumenthal2024enabling}.
\paragraph{Integrated acousto-optic deflectors}

For dynamical beam generation and control, an emerging integrated photonics technology is integrated acousto-optic beam steering (AOBS)~\cite{li2023frequency,lin2024optical} (Fig. \ref{fig:apics} e), which integrates an array of acousto-optic deflectors (AODs) on a PIC platform, typically using a lithium niobate on insulator (LNOI) substrate. Different from conventional AODs using bulk acoustic waves (BAWs) in bulk crystal materials, AOBS uses gigahertz frequency surface acoustic waves (SAWs) generated by planar inter-digital transducers (IDTs) to scatter waveguided optical modes from the surface of the substrate into the free space. The acoustic frequency is typically a few GHz, significantly higher than in conventional AODs, to scatter light out, typically at an angle in the range of 50-70 degrees, from the surface normal. Within the acoustic bandwidth of the IDTs, typically 10-20\% of the central frequency, the scattering angle is continuously tunable by changing the frequency of the microwave signal used to excite the acoustic wave.
Recent results have demonstrated a field-of-view (FOV) of 30 degrees and the generation of 64 beams with diffraction-limited angular resolution of 50 milliarcseconds, while the theoretical limit is more than 200 beams. Each beam can be independently amplitude- or phase-modulated by modulating the corresponding acoustic signal used to generate it. Since each AOBS device's footprint is small, typically 100-200 $\mu$m wide and 1-2 mm long, a large array of them can be integrated on one cm-sized chip. A system of 32 AOBS devices, each emitting 32 beams, can thus generate 1,024 individually steerable and modulated laser beams~\cite{lin2024optical}.
These beams, after shaping and projection, can be used for static or dynamic optical tweezers or gate operations on atomic qubit arrays.

Compared with conventional AODs, which are typically bulky and power-intensive, the integrated AOBS platform achieves over 100-fold miniaturization while remaining compatible with wafer-scale fabrication for scalability. Relative to spatial light modulators (SLMs), AOBS's operation speed of tens of MHz is also three orders of magnitude faster. While earlier versions of AOBS had a relatively low optical efficiency of 2–3\% (the portion of light that is steered from the chip), recent advances in fabrication processes, photonic design, and using resonance enhancement \cite{yu2026resonanceenhancedintegratedacoustoopticbeam}have increased this efficiency to above 25\%.

Since the system is implemented on a lithium niobate platform, additional electro-optic elements—such as frequency shifters and phase modulators—can be integrated monolithically to allow dynamic tuning of optical fields across large qubit arrays. Together, these features establish a high-speed, scalable photonic architecture for quantum control. Furthermore, the technology is expected to be wavelength-agnostic across a broad spectral range from visible with a lower absorption edge around 320 nm~\cite{leidinger2015comparative}, to the near-infrared, relevant to atomic qubits. 

\subsubsection {Integrated Photonics for Stabilized lasers}

\begin{figure}[h!]
    \centering    \includegraphics[width=\linewidth]{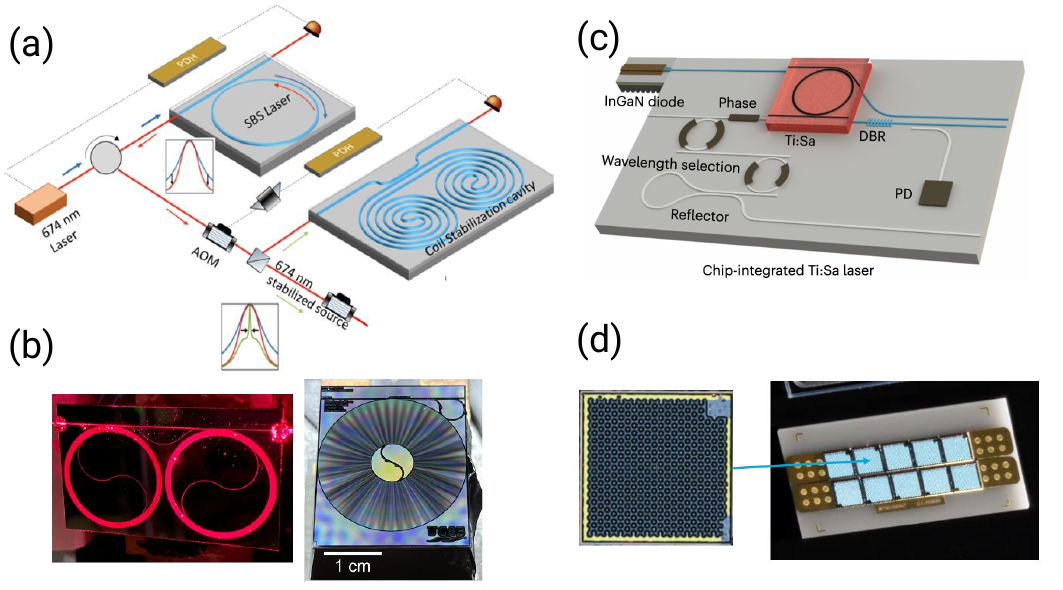}
    \caption{Integrated Laser Capabilities. (a) Two-stage noise reduction with a coil-stabilized stimulated Brillouin scattering (SBS) laser. (b) Coil resonators. 3-meters long operating at 674 nm for the strontium trapped ion clock transition \cite{chauhan2022a_FIOcoil674} and 17-meters long, 250 million Q, operating at 1560 nm \cite{liu2025a_CLEO17mcoil}. (c) Chip-integrated Ti:Saph laser \cite{wang2023photonic}. (d) Large-scale integration of VCSEL arrays \cite{warren2018low}.
    }
    \label{fig:Blumenthal_Fig1}
\end{figure}

Scaling neutral atom quantum computers will require a wide range of laser technology that can address different wavelengths with varying degrees of controllability and performance characteristics. 
\paragraph{Silicon nitride based SBS lasers and reference cavities}
The silicon nitride platform can support visible through NIR wavelengths as well as the stability, phase noise, power and agility requirements, see Fig. \ref{fig:Blumenthal_Fig1}. A primary research thrust is the fully integrated stabilized laser~\cite{heim2026versatile}: a main lasing element, a reference cavity, and integrated modulators for locking the laser to the reference on a single chip, with the stabilized output additionally referenced to an atomic transition (see Fig.~\ref{fig:Blumenthal_Fig1}a). Laser types to be further developed for neutral atom applications include SBS lasers~\cite{gundavarapu2019a_BrillouinUHQ,chauhan2021a_visibleBrillouin,liu2023a_moleculeSub100mHz}, SIL lasers~\cite{isichenko2024a_SIL780}, and ECTL lasers ~\cite{heim2024a_CLEOcoil,nejadriahi2024a_852nmECL}. 
Coil resonators provide a means to narrow the integral laser linewidth and provide carrier stability on timescales (e.g. 100 ms) that implement a holdover cavity for atom transition locking and qubit control. A 17-meter long integrated coil operating at 1550 nm, see Fig. \ref{fig:Blumenthal_Fig1} b) right, that provides a large mode volume to decrease the thermo-refractive noise (TRN) floor enabling phase noise as low as 0.1 Hz$^2$/Hz noise at 10 kHz carrier offset has recently been demonstrated~\cite{liu2025a_CLEO17mcoil}. Such large mode volume cavities have been used to produce among the lowest phase noise microwave signals to date using optical frequency division techniques with spectrally pure lasers~\cite{sun2024a_freqdiv}. The $<$ 50 MHz free-spectral range (FSR) of the coil allows for many laser lock points to align with respect to an atomic transition as closely as possible. 
Another key component will be the study of laser frequency noise engineering, where high frequency noise ($>$ 10 kHz) is first mitigated by an SBS laser, and in a second stage close to carrier noise ($<$ 1 kHz) is mitigated by stabilization of the SBS laser to a coil resonator, both fabricated on the same silicon nitride chip~\cite{song2025a_OFCclocklaser,liu2022_PCLaserStabil}. 
Interfacing these integrated lasers and modulators with atoms and trapped ions has been demonstrated previously~\cite{isichenko2023a_PIBDrubidium,chauhan2024SBSIonQubit}.
For example, a stabilized laser can be locked to a precision rubidium transition via a dual-loop scheme: the laser is locked to a silicon nitride resonator, which is in turn locked by a second feedback loop to the rubidium $D_2$ transition.
By driving the laser stabilization resonator with an arbitrary waveform generator (AWG) it is possible to use programmable frequency sequences to realize higher level functions like sub-Doppler cooling~\cite{isichenko2025a_OFCmodulatorPZT}, which can be further integrated with beam delivery to rubidium 3D-MOTs in the same silicon nitride integration platform~\cite{isichenko2023a_PIBDrubidium}.
\paragraph{Integrated Ti:Saph lasers}
Fully chip-integrated Titanium-Sapphire lasers~\cite{wang2023photonic} 
see Fig.~\ref{fig:Blumenthal_Fig1} c) are another avenue for integrated frequency stabilized lasers. Stabilization is achieved by self-injection locking via an auxiliary ring resonator,
and external distributed Bragg reflector (DBR) feedback. The intrinsic Q-factor of the lasing mode reaches 1.52 M. Integrating Ti:Saph lasers, which are state of the art in conventional bulk optic lasers, on chip, is a substantial advance, helping advance the front of more compact, reliable and cost effective lasers for quantum applications.
\paragraph{VCSEL integrated lasers}
Arrays of integrated vertically emitting lasers (VCSELs) \ref{fig:Blumenthal_Fig1} d) have potential use cases for dynamic trapping and scalable gate operations.
Typical powers for VCSEL can reach several mW per device in the NIR, making them an ideal candidate for trapping of Alkali atom species. Devices can easily be scaled to thousands of channels. While frequency stability requirements for trapping are modest, most gate operations would require narrow linewidth operation. To this end frequency stabilization in VCSELs can be achieved through injection locking \cite{Pfluger:23} to an external laser.
Furthermore, sideband modulation at GHz rates has previously been demonstrated
\cite{Tsygankov:22}, potentially enabling dynamic frequency control. Detailed investigations of noise properties of VCSEL arrays for quantum applications are still outstanding. 

\begin{figure}[h!]
    \centering
    \includegraphics[width=\linewidth]{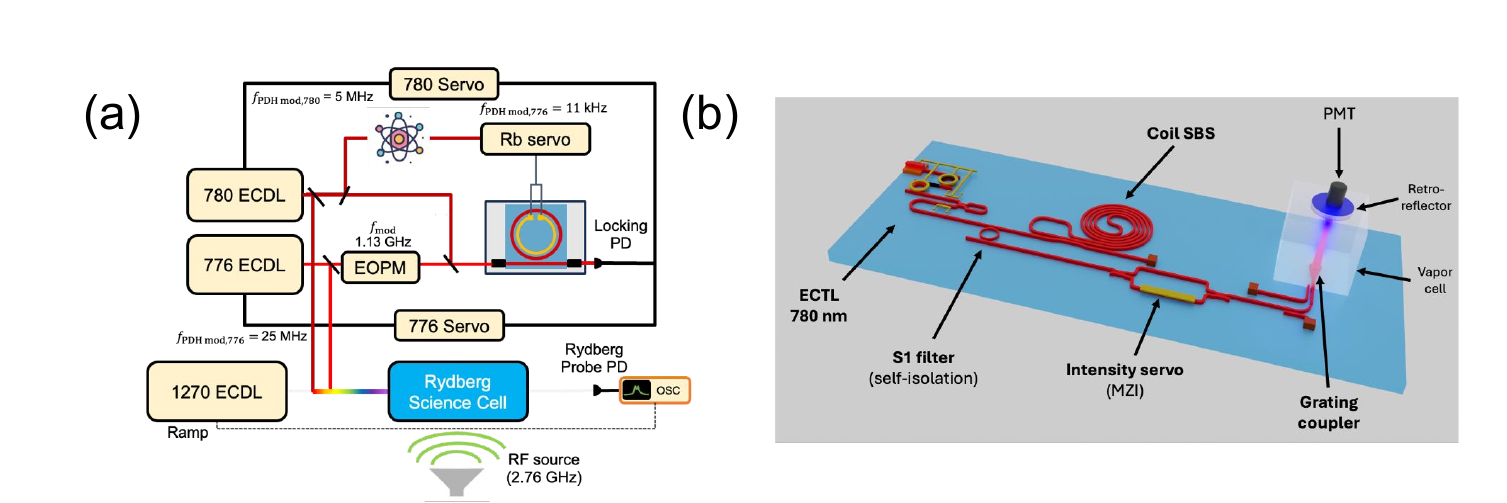}
    \caption{Neutral Atom Technologies based on Silicon Nitride Photonics. (a) Rubidium Rydberg states with frequency agile transfer cavity (unpublished). (b) Integrated rubidium SatSpec with laser stabilization, intensity control servo and beam emitter (unpublished).
    }
    \label{fig:Blumenthal_Fig3}
\end{figure}

\paragraph{Rydberg excitation using integrated lasers}
First studies of Rydberg excitation using integrated lasers and reference cavities are already being pursued as illustrated in \ref{fig:Blumenthal_Fig3} a), where controllable ring resonators are used as a transfer cavity in a 3-photon rubidium Rydberg sensing experiment. The resonator transfer cavity is used to lock the 780 nm probe laser to the  D2 transition, and then the 776 dressing laser is locked to the resonator transfer cavity. The 1260 nm coupling laser is then used to pump excited Rb atoms to the Rydberg state, and electromagnetic sensing is performed by detecting electromagnetic induced transparency (EIT) signals in the probe laser. Work is also under way to implement saturation spectroscopy (SatSpec) on chip, with an ECTL pumping an SBS laser, and an intensity servo used to reduce intensity noise as the beam is delivered to a thermal vapor cell with a surface grating emitter, see Fig. \ref{fig:Blumenthal_Fig3} b). Such atomic reference photonic circuits will be explored to enable large scale Rydberg atom locking and addressing. 

\subsubsection{Optical control combining commercial acousto-optic, electro-optic and spatial light modulation techniques}

Existing commercial devices for optical beam control are limited to either a moderate number of resolvable spots with high temporal bandwidth, e.g. AODs, or many spatial degrees of freedom but low temporal bandwidth, e.g. LCoS SLMs \cite{Bechtold2013}). By combining devices from both categories in a multiscale architecture, it is possible to address a large number of spatial degrees of freedom with high speed.
An example of this type of approach \cite{Graham2023multiscale} uses a 2D AOD scanner to rapidly address subregions of a large SLM device. Each subregion generates a desired spot pattern for controlling a selected group of qubits. An additional 2D AOD scanner then directs the spot pattern onto a desired location inside a larger array. This approach can provide orders of magnitude faster control than the SLM alone. The architecture does not provide the ability to rapidly switch between arbitrary patterns, but is particularly well suited for applying  a sequence of multi-qubit operations across different array regions. Complex gate operations for syndrome extraction on logical qubits provide a prime example of an important use case. In a similar approach, the Thompson group has recently combined acousto-optic modulation with fast DMD to realize fast $\sim$MHz global modulation with slower $\sim$kHz toggles on individual spots \cite{zhang2024scaled}. 

A different approach, allowing sequential individual addressing at MHz speeds by combining EOMs and Virtually Imaged Phase Arrays (VIPA) has recently been demonstrated by the Simon group \cite{wei202610}. This approach relies on frequency multiplexing using fast commercial electro-optic modulation and frequency to spatial mapping using a VIPA.

\section{Quantum Error Correction} 
\label{sec:QEC}
Scalable quantum computation requires correcting errors by encoding logical qubits into physical qubits and performing logical gates \textit{fault-tolerantly}, i.e. without introducing errors which cannot be corrected by the employed code~\cite{gottesman2009introduction}. Neutral-atom quantum computers have now demonstrated most of the subroutines necessary for scalable, fault-tolerant quantum computation~\cite{bluvstein_logical_2024,rodriguez_experimental_2024}. The next frontier for neutral-atom error-correction development is MQuop-scale computation, which requires logical gate fidelities at the $10^{-6}$ level. 

While the most established methods for scalable error correction are based on topological codes, tailored small codes and recently-developed high-rate quantum low-density parity-check (qLDPC) codes offer a pathway to significantly reducing the qubit and gate overheads required. 
 In parallel, decoders must be further optimized to improve both the clock time and the error-correction capabilities of codes. In particular, it is paramount to further adapt decoders to atom loss.

What follows are the identified challenges and opportunities and the possible research directions connected to these questions:
\begin{itemize}
    \item \textbf{Completing the universal gate set:} How can we further reduce qubit and gate overhead of performing a fault-tolerant logical universal gate set?
    \item \textbf{Slowdown due to decoding:} How can we improve the speed and performance of decoding,  in the presence of more complex error correction codes and more complex error channels such as atom loss?
    \item \textbf{Limitations of current codes for near-term fault-tolerant computation:} Can we engineer small or qLDPC codes to perform MQuop logical simulations in the near-term, i.e. with $10^2-10^4$ physical qubits with $99.9\%$ gate fidelity?
\end{itemize}

\subsection{Challenges and Opportunities}

\paragraph{Completing the universal gate set} Transversal gates are logical gates which are applied by performing the physical gate in parallel on the constituent qubits of a code. Due to their structure, they are naturally fault-tolerant, and can be implemented in parallel in neutral atom architectures \cite{bluvstein_logical_2024}. The possibility of easily realizing transversal gates is one of the main advantages of neutral atoms (and trapped ions) compared to superconducting qubit quantum computers, which has to rely on the more costly lattice surgery technique. In particular, these methods can reduce the number of syndrome extraction rounds per logical operation from $O(d)$ to $O(1)$, significantly increasing the logical clock speed~\cite{cain24correlated,zhou2024algorithmic,sahay2025error,wan2024iterative,cain2025fast,serra-peralta2025decoding}. While many quantum codes admit a transversal set of gates, often Clifford operations, 
there are obstructions to a universal set of transversal gates~\cite{eastin_restrictions_2009}. The schemes which have been proposed to completing the universal gate set, e.g. by ``distilling'' a high-fidelity logical T gate from many low-fidelity ones, are significantly more costly than transversal gates, meaning that the physical gate count of fault-tolerant computations are usually dominated by these non-transversal gates. An additional challenge is created by the fact that most schemes only realize a discrete gate set this way: to implement an arbitrary-angle single-qubit gate from the set of Clifford and T (from which arbitrary-angle multi-qubit gates are implemented by conjugation with two-qubit gates), in general one needs a number of T gates which depends on the desired accuracy of the gate; current estimates require around $10$s of T gates to implement a single arbitrary-angle rotation with sufficient precision \cite{nielsen2010quantum,Ross2016}.

Fortunately, there are opportunities here as well. The same transversal Clifford gates mentioned above can be used to reduce the cost of magic state distillation~\cite{zhou2024algorithmic,sales2025experimental}. In addition, an alternative technique -- magic state cultivation~\cite{gidney2024magic}, in which T states are grown in code distance rather than distilled -- has been shown to be as efficient as the implementation of a CNOT gate in lattice surgery (but not as efficient as a transversal CNOT). These methods can be further improved through the use of non-local connectivity in neutral atom systems~\cite{sahay2025fold,vaknin2025magic,chen2025efficient,claes2025cultivating}. Moreover, non-local connectivity can be advantageous to realizing recently developed approaches to preparing $CCZ$-type magic states directly on surface code patches with linear-depth circuits by interfacing with an intermediate \textit{non-Abelian} topological code \cite{davydova2025universal,sajith2026non}. A distinct and very recent approach generates magic states directly on high-rate qLDPC codes. Whereas distillation and cultivation are built on codes encoding a single logical qubit, high-rate codes encode many logical qubits per block and can therefore produce many magic states at once. The idea is to use a code whose non-Clifford gate can be applied in constant depth (transversal in the loose sense of the word): one prepares the logical $|+\rangle$ state and applies this gate to obtain the magic states in a single step. Such codes are highly non-local, making them a natural fit for the reconfigurable connectivity of atom arrays, and concrete examples -- tricycle codes -- have been found with short-depth non-Clifford circuits that achieve favorable space-time efficiencies for implementing non-Clifford operations~\cite{menon2026magic,jacob2025single,li2025transversal}. In addition, a limited set of computations (e.g. quantum simulation) have been shown to yield error-correction advantages even if arbitrary-angle rotations are implemented non-fault-tolerantly \cite{Akahoshi2024,majidy2025,zeng2025error}. These recent developments imply that much is still to be learned about how a universal gate set can be efficiently realized, in particular making use of the unique capabilities of long-range connected hardware such as neutral atoms.
In contrast to utilizing codes with transversal Clifford gate sets such as the $2D$ surface code, an alternative approach to quantum memory involves leveraging \textit{high-rate} quantum low-density paritiy check (qLDPC) codes \cite{panteleev2022asymptotically,panteleev2019degenerate,breuckmann2021quantum,breuckmann2020balanced,leverrier2022quantum}. Such code constructions have the potential to dramatically reduce the qubit overhead of fault-tolerant quantum memory \cite{bravyi2024high} and are naturally suited for implementation on reconfigurable neutral atom-array quantum computers due to their inherently non-local structure. Efficient atom reconfiguration schemes have been developed for the syndrome extraction circuits of certain qLDPC code families of interest including hypergraph product, lifted product, and higher-order balanced product codes \cite{xu2024constant,menon2026magic}. However, performing fault-tolerant quantum computation on such codes introduces additional complexities as essentially all known high-rate qLDPC code families do not host transversal Clifford complete gate-sets. Instead, such codes typically exhibit a small number of symmetries that can be utilized to implement a small subgroup of the Clifford group with transversal, permutation, or fold-transversal automorphism gates \cite{bravyi2024high,breuckmann2024fold,eberhardt2024logical}, while other Clifford operations must be implemented using methods such as generalized lattice surgery \cite{cohen2022low} or homomorphic measurement \cite{huang2023homomorphic}. These latter methods involve introducing additional ancilla qubits to perform fault-tolerant measurement of logical Pauli-product operators. As such, generalized surgery and homomorphic measurement can be considered as methods to perform measurement based computation on qLDPC codes \cite{briegel2009measurement}. A universal gate-set can then be implemented using the ability to measure arbitary Pauli-product logical operators when combined with methods to consume high-fidelity magic-states fault-tolerantly \cite{litinski2019game}. While surgery and homomorphic measurement based methods to perform Clifford logic can introduce substantial space-time overhead compared to approaches based on transversal gates, reducing the overhead of these approaches is a highly active area of research in high-rate quantum error correction, and several recent works have made significant progress towards lowering this overhead, for example, by constructing smaller ancillary gadgets for lattice surgery \cite{williamson2024low,swaroop2024universal,he2025extractors}, measuring several 
logical Pauli-product operators at once \cite{cowtan2025parallel,zheng2025high}, and fast surgery methods which reduce the temporal overhead associated with logical measurement \cite{baspin2025fast,cowtan2025fast}. Crucially, all these improvements construct highly non-local gadgets for performing logical operations on high-rate qLDPC codes, which presents a unique opportunity for quantum computing platforms with reconfigurable or long-range connectivity to enable quantum computing with high-rate codes.

\paragraph{Slowdown due to decoding} A further significant challenge is in the efficient implementation of decoding algorithms, i.e. the task of computing likely error locations from a set of syndrome measurements, which is usually performed on a classical computer.
Optimal decoding for generic stabilizer codes is known to be computationally hard~\cite{Iyer2015}, and even the task of identifying the minimum-weight error in canonical settings, such as the surface code or the color code, is computationally hard~\cite{gu2026colorcodesurfacecode,walters2026minimumweightdecodingcolour}.
The decoder should be able to compute the necessary corrections from the error syndromes on a timescale comparable with gates and measurements on neutral-atom hardware ($100\mu\mathrm{s}$ to $10\mathrm{ms}$). This introduces necessary tradeoffs of finding decoding algorithms which balance speed and accuracy in cases where maximum likelihood estimation becomes impractical~\cite{roffe2020PhysRevResearch}. Furthermore, although the simplest theoretical error models often assume all Pauli errors are equally likely, in practice hardware may bias the noise in a particular direction, and furthermore, different qubits may be subject to different levels of noise. In addition, a large fraction of the errors in neutral atoms are due to atom loss~\cite{Evered_2023}, a non-Pauli error. Decoders must be able to take advantage of this structure in the errors~\cite{baranes2025leveraging,gu2023fault,gu2024optimizing}. In practice errors on different logical qubits may be correlated, a fact which the simplest theories of fault tolerance and error correction do not take into account. Recent work has begun to account for this by correlating the decoding procedures on different logical qubits~\cite{cain24correlated}, but, in general, highly non-local correlated errors are hard to detect and correct~\cite{martinis2020saving}. The addition of these increasingly complicated considerations requires more sophisticated modeling. One opportunity involves 
applying machine learning tools to these problems~\cite{bausch2024learning,bonilla2025neural,gu2026scalableneuraldecoderspractical}, provided that one can gain access to large, high-quality datasets with which to train these models.

The community has identified new opportunities as attention to reducing the practical overhead of decoding has increased. As mentioned above, implementing so-called transversal algorithmic fault tolerance reduces the resources required to implement a logical quantum circuit~\cite{zhou2024algorithmic}, and single-shot quantum error correction reduces the time-complexity required for error correction rounds. In addition, the recent breakthrough results in the discovery of qLDPC codes with favorable scaling in the distance and the number of logical qubits has sparked research in good decoders for these codes as well as practical schemes for logical gates. Finally, it has been shown that decoding with loss-detecting readout when gate errors are dominated by loss can yield higher code thresholds than with Pauli-only errors with the same rate~\cite{baranes2025leveraging}. 

\paragraph{Limitations of current codes for near-term fault-tolerant computation} Beyond the practical hardware challenges involved in efficient decoding and gate implementation, there still remain significant challenges and questions which need resolution in the theory of quantum error correction. As mentioned earlier, non-local correlated errors are hard to correct. Existing code formalisms do not generally include correlations in errors across space and time, nor adaptive frameworks which can take advantage of real-time error characterization. Hardware restrictions impose still further restrictions on which quantum error correcting codes are implementable in practice. The surface code, long a favorite model for theoretical analysis and experimental implementation, is on the one hand convenient because it is implementable in an architecture with 2D nearest-neighbor connectivity, but this structure inherently limits the rate of logical qubit encoding. Codes with better encoding rates, such as new families of high-rate qLDPC codes have better code parameters, but require non-local connectivity. 
This is  hard to engineer in the superconducting setting \cite{bravyi2024}, while neutral atoms have a clear advantage here~\cite{xu2024,cain_shors_2026}.

\subsection{Research Directions}

\subsubsection{Reducing Overhead of Fault-Tolerant Universal Gate Set}

Fault-tolerant quantum computation relies on the ability to perform non-Clifford gates to create a universal gate set. These gates, such as the T gate, are essential for achieving universal quantum computation but are not directly implementable within many error-correcting codes in a fault-tolerant manner. To date, magic state distillation is the dominant approach used for this path (see, e.g., \cite{litinski2019magic} and \cite{gidney2019efficient} for recent overviews). At the same time, novel approaches including code polymorphism  \cite{webster2022universal} and magic state cultivation \cite{gidney2024magic} may provide substantial improvements, and thus are ripe for development. Furthermore, the development of such approaches for high-rate encodings such as qLDPC codes is an important frontier of research, with many recent developments~\cite{menon2026magic,jacob2025single,li2025transversal}. Simultaneously, advances in performing Clifford logic on such codes are particularly important. As discussed in the previous section, improvements to qLDPC code surgery and homomorphic logical measurement protocols in the spirit of Refs. \cite{cowtan2023css,cowtan2024ssip,cowtan2025parallel,he2025extractors,zheng2025high,baspin2025fast,ide2025fault,xu2025fast} will be crucial to realizing architectures based on such high-rate qLDPC codes.

Research into efficient universal gate sets is thus needed.
In all cases, the key metric of interest is reducing the physical overhead while maintaining a target level of implementation fidelity. 

Furthermore, efficient implementation of the universal gate set is only part of the challenge. To understand the practical feasibility of quantum algorithms, it is also crucial to accurately evaluate the resource requirements (number of qubits, gate count, circuit depth, etc.) when implemented fault-tolerantly. This evaluation becomes more sophisticated and potentially more optimistic when considering the integration of decoders and of knowledge of the algorithm into the scheduling, layout, and metrics for the universal gates.

In addition to improved decoders, algorithm design, and universal gate set implementations, there are micro-code improvements that are worthy of investigation. These correspond to making closed-loop iterative improvements of syndrome extraction primitives and magic state distillation and cultivation primitives. A typical error correction circuit involves an extensive number of such micro-code implementations, such as circuit-based, measurement-based, or fusion-based realizations for the surface code, and thus they are worthy of a high degree of optimization for the actual physical system; this in turn informs code and decoder choice.

Finally, a novel direction identified is working directly with fermionic qubits to explore whether there is quantum advantage achievable in the near term \cite{gonzalez-cuadra2023fermionic}. Intrinsically fermionic processors may have natural noise resilience through exploiting the symmetries of the problem and mapping native noise into natural realizations of disorder.  In the long term, it may also be possible to realize error correction and fault-tolerance with logical fermions \cite{schuckert2024,ott2024}, but a number of technical challenges arise in this context that require more investigation by the community.

\subsubsection{Improving Decoder Performance}

Accurate and scalable decoding algorithms are essential for unlocking the full potential of fault-tolerant quantum computing. A key opportunity is to develop decoding strategies that exploit structure in the noise, such as biased dephasing and spatially correlated loss, to significantly raise the effective error thresholds of quantum codes. 
While the details of hardware realizations may differ, there is compelling evidence of the benefits provided by detecting loss (making it a heralded erasure), even when the additional costs of loss detection and imperfect (or delayed) loss detection are taken into account~\cite{gu2023fault,gu2024optimizing,chang2024,baranes2025leveraging}.
Developing a unifying framework of QEC with erasures that is hardware-agnostic (and applicable not only to neutral atoms, but also superconducting circuits) is an important research direction, that will likely result in identifying QEC protocols with higher thresholds.
In addition, mid-circuit syndrome measurements may yield non-binary outcomes, which can provide valuable information about the likelihood of specific errors—information that can be directly utilized by our decoder. For instance, when a discrete-variable code is concatenated with an underlying bosonic code, the multi-valued syndrome outcomes can indicate different likelihood of errors, thereby enhancing the decoding performance of higher-level quantum LDPC codes~\cite{raveendran2021finite} and quantum polar codes~\cite{subramanian2025achievable}.

From a data-centric viewpoint, decoding is a binary classification problem naturally suited to AI-based methods. For memory decoding, AI-based decoding approaches~\cite{Baireuther2018Quantum,Baireuther2019NewJ.Phys.} are increasingly appreciated with their ability to adapt to hardware-specific noise~\cite{lacroix2025Naturec,Varbanov2025Phys.Rev.Research}. A recent demonstration of this potential is Cascade~\cite{gu2026scalableneuraldecoderspractical}, a convolutional neural network decoder that exploits the translational structure shared by many QEC codes. A single architecture decodes both surface codes and high-rate bivariate bicycle codes near-optimally. Crucially, it overcomes a central limitation of belief propagation, the standard qLDPC decoder, whose fixed update rules fail on degenerate error patterns; the learned message-passing rules circumvent these failures. This exposes a ``waterfall'' regime in which logical error suppression improves with code size faster than the naive distance bound suggests: for the $[[144,12,12]]$ Gross code~\cite{bravyi2024} it reaches logical error rates near $10^{-10}$ (per cycle per logical qubit) at a physical error rate of $0.1\%$. Its latency already falls within the real-time decoding budget set by the clock cycle of current neutral-atom arrays, and its local, feed-forward structure leaves room for further speedup on specialized hardware such as FPGAs or ASICs.

While it has recently been established that transversal logical circuits can achieve error thresholds comparable to those of bare quantum memories~\cite{xu2025a}, decoding logical circuits is challenging because of the structure of the underlying decoding hypergraph. This structure is shaped by three factors: (i) the spread of physical errors under transversal entangling gates~\cite{cain2024PhysRevLett}; (ii) correlated noise at the hardware level; and (iii) the intrinsic complexity of quantum LDPC codes~\cite{roffe2020PhysRevResearch}. The recently proposed Multi-Core Circuit Decoder (MCCD)~\cite{zhou2025NatComputSci}  and the work of Ref.~\cite{bonilla2025neural} address this bottleneck through a modular, data-centric architecture in which each logical gate has its own independently trained model. This design enables efficient handling of entangling logical gates and provides a natural path to scalability as experimental gate sets evolve. Looking forward, decoding more extended syndrome sequences—arising from larger code distances and increasingly intricate LDPC constructions—will likely benefit from attention mechanisms, which effectively capture high-order correlations in quantum measurement data~\cite{kim2025scienceadvances}. We also highlight a complementary scalability strategy inspired by ongoing advances in expanding the context windows of large language models.

As one uses more efficient QEC approaches, such as high-rate qLDPC codes~\cite{breuckmann2021quantum} or fast transversal operations~\cite{zhou2024algorithmic}, the resulting decoding problem often becomes more complex. First, unlike the standard surface code where minimum weight perfect matching on edges suffices~\cite{dennis2002topological}, the resulting decoding problem generically involves hyperedges, although this can sometimes be reduced back into regular edges~\cite{cain2025fast,serra-peralta2025decoding,kubica2023efficient,Chamberland2020}. Second, the decoding problem will tend to be more densely connected, increasing the problem size and reducing modularity, which may increase challenges in addressing the backlog problem~\cite{skoric2023parallel,bombin2023modular,terhal2015quantum}.

One possible avenue to mitigate the increasing complexity of the decoding problem is by considering code families capable of single-shot QEC, which guarantees that one can perform reliable QEC even in the presence of measurement errors without the need for repeated measurements~\cite{bombinSingleshotFaulttolerantQuantum2016,campbellTheorySingleshotError2019}.
This property has been demonstrated for topological quantum codes, e.g., the subsystem toric code~\cite{SingleShotSubsystemToricCode}, as well as for quantum Tanner codes~\cite{Gu_2024}.
Single-shot QEC not only helps to tackle the time overhead of QEC; it also provides modularity of the decoding problem in logical circuits, alleviating the need to store the syndrome history.
It is also compatible with any correlated decoding strategies that one may use while implementing entangling logical gates via transversal operations.
Thus, identifying resource-efficient codes with high QEC thresholds that also have the single-shot property is a promising research direction.

\begin{figure}
    \centering
    \includegraphics[width=0.75\linewidth]{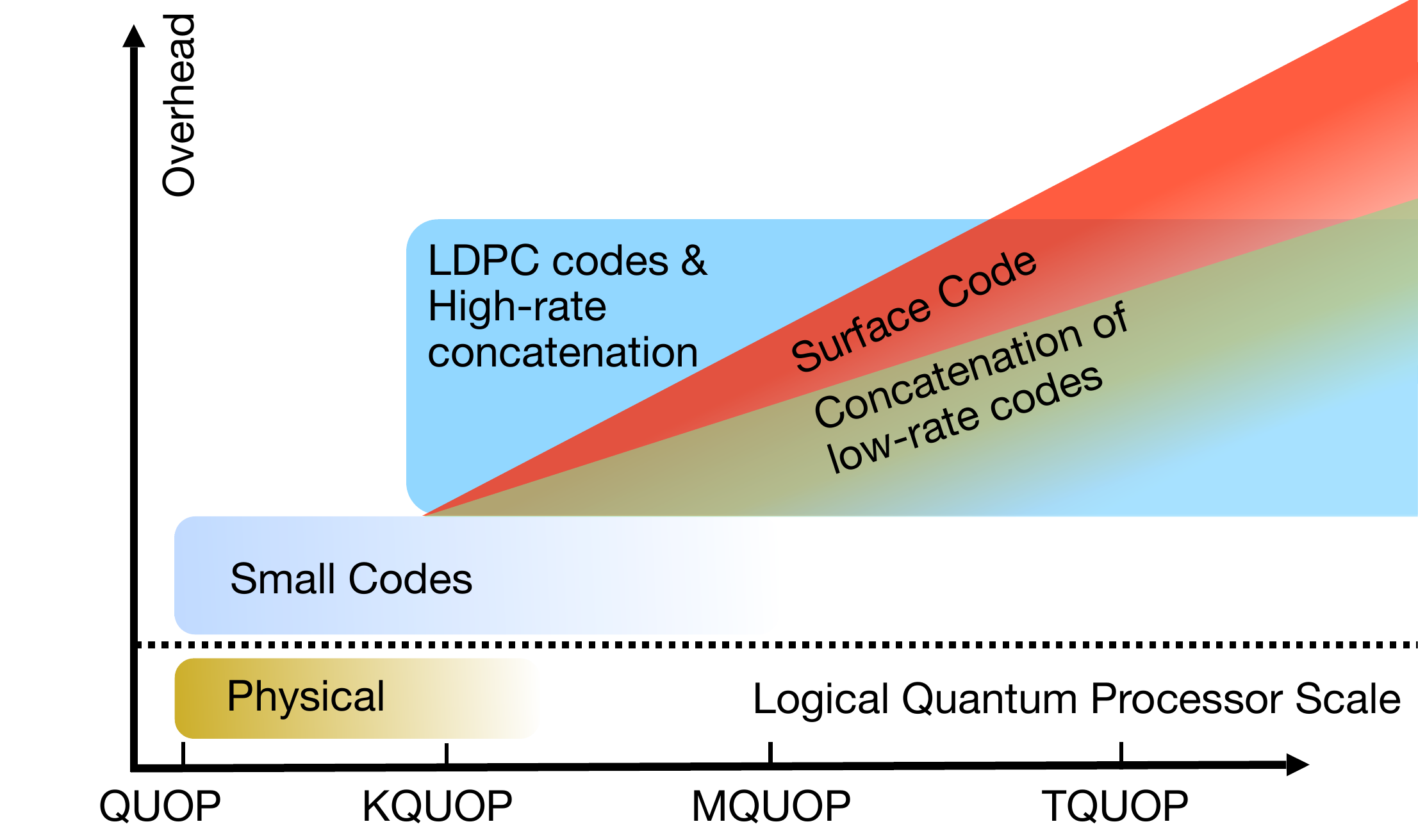}
    \caption{Error-correcting code strategies suitable for a given target logical processor scale. Currently we are in the era of physical qubits or small distance codes for logical computation (e.g., \cite{bluvstein_logical_2024}), but the next generation of quantum processors will require higher-performing codes based on several layers of concatenation (e.g., \cite{yoshida2024concatenate}), the surface code, or LDPC codes \cite{gottesman2013fault}. }
    \label{fig:enter-label}
\end{figure}

\subsubsection{New QEC codes: qLDPC, Fermion-to-Fermion Codes} 

The development of new, low-resource-overhead quantum error correcting codes will also be an important future research direction. To enable MEGAQUOP calculations in the near term, it may not be necessary to use those code families with the highest threshold, but instead use codes which for a given physical qubit budget maximally reduce logical algorithm errors.

Recent work has suggested that high-rate qLDPC families (e.g., hypergraph-product, lifted/quasi-cyclic product, and generalized/bivariate bicycle codes) can reduce physical-qubit counts by close to an order of magnitude relative to the surface code at comparable distance, while still achieving competitive circuit-level thresholds ($>0.5\%$)~\cite{tremblay2022constant,panteleev2019degenerate,scruby2024high,yamasaki2024time,goto2024high,bravyi2024high,tillich2014quantum,leverrier2022quantum,mackay2004sparse}. These results are not only asymptotic: they already appear at block lengths of tens to hundreds, making high-rate codes a promising near-term route to space-efficient logical memories and processors \cite{tremblay2022constant,xu2024constant,bravyi2024}. While early work focused on logical memory, several low-overhead methods for logical gates have matured, including generalized surgery (ancilla-assisted selective Pauli-product measurements) \cite{krishna2021fault,cohen2022low,cross2024improved,williamson2024low,swaroop2024universal,zhang2024time,he2025extractors}, gate teleportation using resource states encoded in the same code \cite{gottesman2013fault,yamasaki2024time,nguyen2025quantum}, homomorphic logical measurements (structure-preserving ancilla blocks) \cite{huang2023homomorphic,xu2025fast}, and symmetry-based gates (fold-transversal and automorphism operations) \cite{kubica2015unfolding,quintavalle2023partitioning,eberhardt2024logical,malcolm2025computing,sayginel2025fault}. Moreover, methods for performing non-Clifford operations have also recently been developed~\cite{xu2025batched,zhang2025constant}. Neutral atom arrays are particularly well-suited to these codes because reconfigurable, non-local connectivity can be realized by atom transport and programmable Rydberg interactions on timescales short compared to coherence times~\cite{bluvstein2022quantum,Evered_2023,xu2024constant,poole2025architecture,pecorari2025high,cain_shors_2026}.

Minimizing logical error rates for a given physical qubit budget also potentially means using different codes for different applications. For example, codes with a non-Clifford transversal gate has been pointed out as being suited for quantum supremacy experiments~\cite{bluvstein_logical_2024,hangleiter_fault-tolerant_2024}. In addition, permutation-based gates can be natively implemented in neutral-atom and ion-trap hardware due to high-fidelity qubit shuttling, as demonstrated in Ref.~\cite{Hong2024}.

Another frontier is finding codes that have logical fermionic and bosonic degrees of freedom, which could reduce the overhead of simulating fermion-boson models which appear in many quantum simulation tasks. For example, logical fermionic degrees of freedom have recently been proposed for material simulation~\cite{schuckert2024}. We can also encode one logical bosonic mode into multiple bosonic modes to correct errors~\cite{noh2020encoding}.

Artificial intelligence methods provide powerful tools for designing improved quantum error-correcting (QEC) codes by efficiently navigating the expansive parameter space of possible code configurations \cite{Su2025}. Machine learning techniques such as reinforcement learning, evolutionary algorithms, and generative neural networks can systematically explore novel code structures optimized for specific hardware noise profiles and operational constraints. For instance, AI and large language models (LLMs) can assist in identifying optimal polynomial selections for complex QEC constructions, such as determining efficient polynomials in the bivariate bicycle code. By training these AI models on both simulated and experimentally derived datasets, it becomes possible to discover previously unexplored QEC codes that minimize overhead, improve error suppression, and leverage hardware-specific features, such as noise bias or unique atomic interactions characteristic of neutral atom quantum computing platforms.

\subsubsection{Fault-tolerant Modular QEC Architectures}
\label{sec:QEC_mod}
Scaling up quantum processors is the central challenge for practical quantum computing. Modular architectures offer a natural path to increasing system size by interconnecting smaller quantum processing units. However, achieving reliable computation in such systems requires fault-tolerant operation not only within each module but also across the links that connect them.

The study of modular quantum error correction predates the emergence of large-scale neutral-atom platforms, with early work focusing on architectures in which logical qubits are distributed across multiple small modules \cite{Oi06,Campbell2007,Liang2007,Met07,Fuj12,Li_2012,Nickerson2013,Nickerson2014,Nigmatullin_2016,Buonacorsi_2019}. In these approaches, inter-module operations are integral to syndrome extraction, and fault tolerance is achieved by coordinating many physical-scale modules. Around 2022, the emergence of large, reconfigurable neutral-atom systems and other platforms capable of hosting tens to hundreds of qubits per module motivated exploration and detailed optimization of a new architectural paradigm in which logical qubits are encoded within individual modules and interconnects serve primarily as a logical interface between them. While this distinction is most clearly reflected in post-2022 work, including \cite{Ramette2024,pattison2024}, a few earlier proposals exhibit elements of this transition. These precursors include \cite{Fow09}, which explored communication between surface-code logical qubits distributed across a network, and hierarchical constructions such as \cite{Li15}, where small code patches within modules act as effective qubits for an outer code spanning multiple modules.

Within the logical-module paradigm, recent proposals fall broadly into two categories for handling interconnect noise: direct interfacing of logical qubits and logical distillation. In direct interfacing approaches, logical qubits are coupled across modules using operations such as lattice surgery or transversal gates \cite{Ramette2024,Andres_Martinez_2024,jacinto2025,stack2025,haug2025,Shalby2025,benito2025,Marton2025,Maeda2025,liu2026,naito2026}. In logical distillation approaches, noisy physical entanglement is injected into encoded blocks and purified at the logical level to enable high-fidelity inter-module operations \cite{pattison2024,marqversen2025,sunami2025}. In parallel, architectures based on distributed logical qubits assembled from smaller modules remain an active area of research, particularly in settings where module size is constrained \cite{Xu22,Sin24,de_Bone_2024,babaie2024,tham2025}.

Direct interfacing approaches aim to implement inter-module logical operations with minimal overhead, using primitives such as lattice surgery, transversal gates, or teleportation between encoded blocks. A central concern in these schemes is the impact of noisy interfaces, where errors are concentrated at the boundary between modules. Extending earlier insights from Refs.~\cite{Fow09,Li15}, Ref.~\cite{Ramette2024} showed that surface codes can tolerate substantial boundary noise without catastrophic degradation: assuming the circuit noise model, teleported logical gates can remain fault-tolerant below non-local thresholds around $10\%$, while the threshold for local operations remains near $\sim 1\%$. This demonstrates that direct logical interfaces can operate in a regime where inter-module operations are significantly noisier than intra-module gates—by roughly an order of magnitude—while still preserving fault tolerance.

Logical distillation approaches instead treat inter-module entanglement as a noisy resource that is purified at the logical level \cite{pattison2024,marqversen2025}. Unlike physical-level distillation, which operates on raw Bell pairs and is fundamentally limited by physical gate errors \cite{Dur2003,Liang2007}, logical distillation acts on encoded states and can suppress errors well below the physical noise. As a result, physical distillation protocols are typically constrained to modest error reduction (e.g., from $\sim 10\%$ to a few times the local gate error rate) and are optimized for resource efficiency under these limitations \cite{Li_2012,Nickerson2013,Nickerson2014,Nigmatullin_2016}. In contrast, logical distillation enables qualitatively different regimes, including constant-rate protocols that can produce extremely high-fidelity logical Bell pairs (e.g., $\sim 10^{-12}$ error) with on the order of tens of physical Bell pairs per output \cite{pattison2024}. This shifts the trade-off from fidelity-limited operation to resource-limited operation, at the cost of increased module size and logical overhead \cite{marqversen2025}.

\begin{figure}[H]
    \centering

    \includegraphics[width=0.75\linewidth]{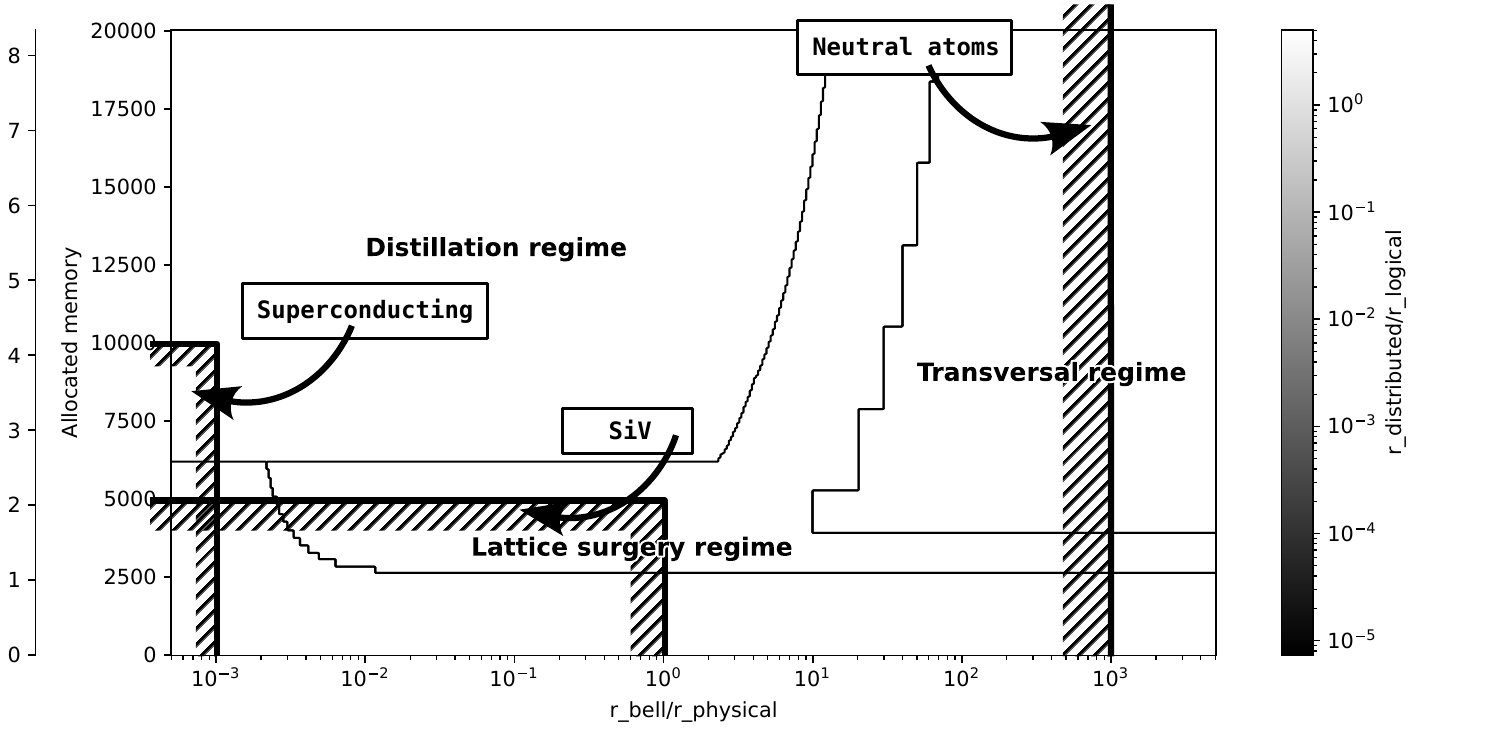}
    \vspace{0.8em}

    \includegraphics[width=0.75\linewidth]{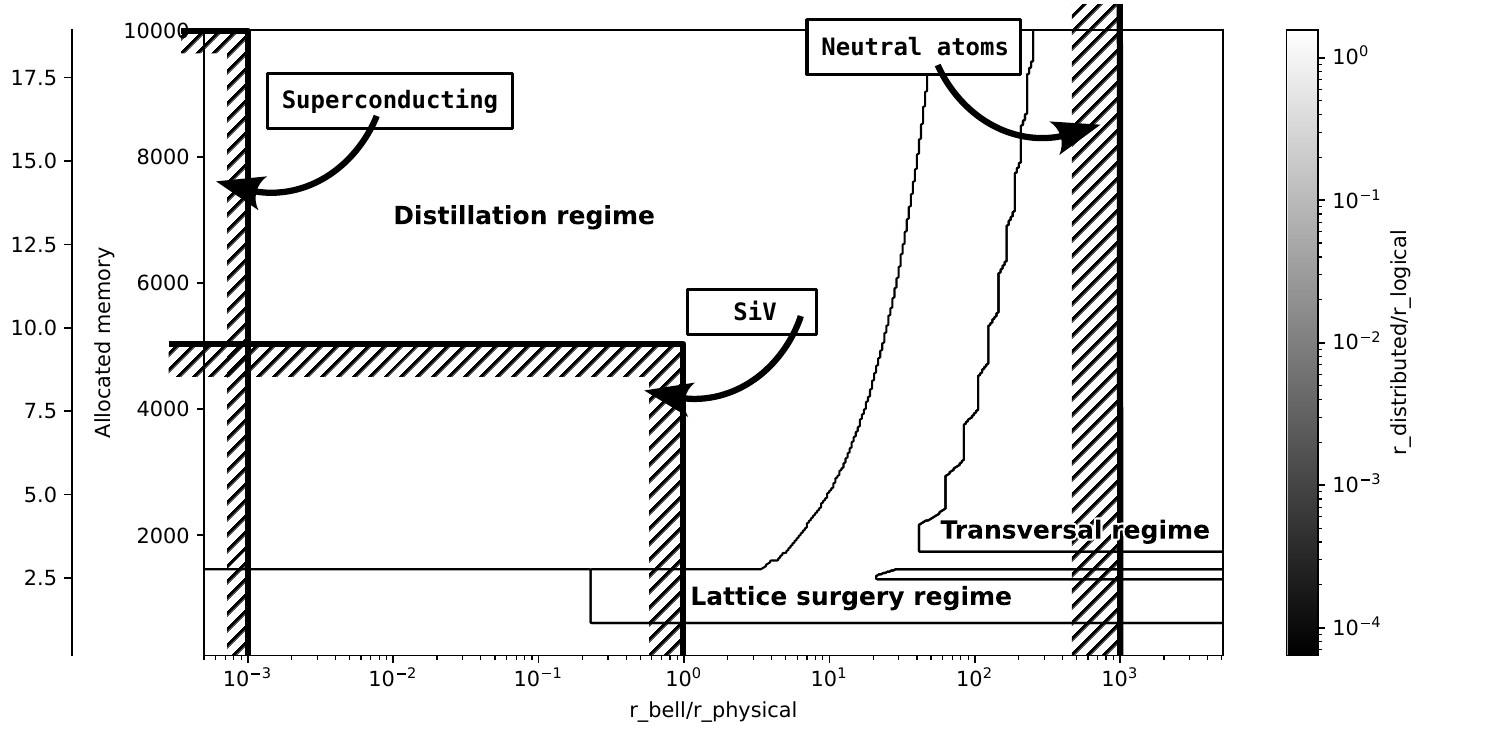}

    \caption{\textbf{Comparing logical Bell pair distribution rate $r_{\mathrm{distributed}}$ as a function of networking memory and physical Bell pair rate $r_{\mathrm{bell}}$.}
    The two plots correspond to (top) $p_{\mathrm{bell}} = 1\%$, $p_{\mathrm{target}} = 10^{-12}$ and (bottom) $p_{\mathrm{bell}} = 5\%$, $p_{\mathrm{target}} = 10^{-6}$. The allocated memory is shown both
in number of logical (right) and physical (left) qubits assuming the surface code.  Rates are expressed in units of the local physical gate rate $r_{\mathrm{physical}}$ and local syndrome extraction rate $r_{\mathrm{logical}}$. Cross-hatch overlays show expected operational regimes for different hardware platforms~\cite{marqversen2025}. Solid lines show boundaries between different optimal QEC architectures. Reproduced from Ref \cite{marqversen2025} with permission. }
    \label{fig:bell_distribution_rate}
\end{figure}

As illustrated in Fig. \ref{fig:bell_distribution_rate}, the performance of these different approaches
depends strongly on hardware-specific parameters such as local gate time and error, qubit coherence time, achievable Bell-pair rate and fidelity, and the number of qubits that can be allocated to communication rather than computation~\cite{pattison2024,marqversen2025}. In particular, allocating more qubit resources for the communication can increase the entangling rate through multiplexing or the fidelity through distillation (see Fig. \ref{fig:bell_distribution_rate}). This creates a three-way trade-off between the number of qubits available for intra-module operations and improving either the speed or fidelity of inter-module operation. The right balance between the two will depend not only on error budgets but also on the physical mechanism used to distribute entanglement, whether through free-space collection or optical cavities. 

Most prior work has focused on linking identical QPUs that implement the same quantum error-correcting code~\cite{ramette2024fault,bonilla2025constant,marqversen2025}. However, large-scale modular systems are unlikely to be homogeneous. If we look to classical distributed systems, we might expect to have architectures where different modules are specialized for different tasks. For example, a memory module may employ a high-rate qLDPC code to maximize storage density~\cite{bravyi2024high}, whereas a magic-state factory may rely on a surface or color code optimized for fast stabilizer extraction and transversal Clifford operations~\cite{litinski2019game}. A computational module may use yet another code chosen for low-depth circuit execution~\cite{xu2025fast}. 

We should, therefore, not expect one optimal code; rather, different tasks will require different optimized codes, further motivating a heterogeneous architecture.

This requires devising how to efficiently connect logical qubits encoded in different codes, with different stabilizer structure, threshold behaviour, and logical clock speed. Logical lattice surgery, transversal teleportation, and distillation-based transfer now represent alternative interface mechanisms whose relative cost depends on the logical error rate, qubit overhead, and general performance across heterogeneous modules. A central research direction is therefore to develop a framework in which these methods are not treated as mutually exclusive, but are selected dynamically according to the role of the modules being linked. 

Going beyond the single link between modules to full modular circuit compilation represents another layer that will need to be addressed. In a heterogeneous fault-tolerant architecture, the cost of a computation is not set solely by the number of entangling gates or by circuit depth, but by the number and timing of logical-level transfers between modules, the degree to which computation can proceed asynchronously, and the availability of memory buffers to e.g.  decouple production of magic states from consumption in the computation. A compiler targeting such a system must therefore account for code transitions, teleportation overhead, and buffering latency, rather than treating all qubits as members of a single monolithic register.

Taken together, these considerations define a broad area of work aimed at understanding how to operate modular systems in which different fault-tolerant codes, different logical roles, and different interconnect methods coexist in a single architecture. The central questions are how to link these modules with minimal overhead, how to allocate qubits between local computation and inter-module connectivity, and how to incorporate these choices into a compiler that can trade qubits, time, and logical error rate in a controlled manner. Addressing these questions is essential for turning modularity from an engineering convenience into a genuine scaling strategy.

\section{Compilation of Quantum Circuits} \label{sec:compilation}
Neutral atom architectures, characterized by reconfigurable qubit layouts and dynamically programmable connectivity, offer potential advantages over fixed-connectivity platforms.
The realization of these hardware-level advantages depends critically on the compiler, which translates high-level program descriptions into optimized, hardware-executable instructions.
Benchmark studies on fixed-connectivity architectures have identified substantial optimality gaps—up to 45× in circuit depth and 330× in gate overhead~\cite{tanOptimalityStudyExisting2021, pingAssessingQuantumLayout2025}—underscoring the considerable opportunity for improvement in quantum compilation more broadly.
Prior compiler work targeting physical-level circuits for neutral atom systems has already demonstrated significant performance gains~\cite{Tan2024compilingquantum}.
In the fault-tolerant quantum computing regime, compiler design is expected to play an even more important role in achieving efficiency and scalability.

\subsection{Challenges and Opportunities}

Current atom array compilation efforts primarily focus on physical circuits~\cite{tan2022qubit, Tan2024compilingquantum, wang_atomique_2024, lin2024reuse, tan_compilation_2025, wang_q-pilot_2024, Schmid24computational, jang25qubit, ludmir24parallax}, but fault tolerance introduces additional complexity.
One major challenge is the lack of a universally accepted, accurate, and efficient cost metric for circuit compilation, especially when different trade-offs related to fault tolerance must be considered.
Additionally, software implementations are fragmented, lacking a unified infrastructure that seamlessly connects different stages of the compilation process.
This fragmentation makes it difficult to optimize large-scale computations and integrate emerging hardware capabilities effectively.  

Quantum error correction (QEC) introduces complications by adding an additional layer of abstraction.
For instance, QEC codes may contain multiple logical qubits per encoding block, imposing additional constraints on compilation strategies.
Additionally, atom movement for generic QEC codes has not been fully explored, limiting the ability to optimize qubit placement and routing. Because atom reconfiguration problems are typically NP-complete even on grids, compilation workflows benefit from modular compilation backends that support multiple algorithmic regimes, from exact solvers to fast heuristics~\cite{Cooper2024a,Cimring2023,ElSabeh2023}.
This gap motivates co-optimizing layout decisions with fast atom-reconfiguration solvers and a batching layer that merges displacement operations for parallel execution under hardware constraints~\cite{Afiouni2025}.
The classical feedforward infrastructure is also underdeveloped, creating a bottleneck for implementing advanced techniques such as correlated decoding~\cite{cain24correlated}.
At the hardware level, control signal generation remains a largely manual and monolithic process, increasing the burden on experimentalists and slowing down progress.  

Emerging hardware features, such as dual-species atom arrays, offer potential advantages by reducing the need for atom movement, but they also introduce new challenges for compilers.
Additionally, compilation support for key capabilities like loss detection and continuous reloading is still lacking, preventing full utilization of neutral atom architectures.
Finally, optimal approaches based on mathematical programming~\cite{Tan2024compilingquantum} offer strong guarantees but do not scale to large neutral atom arrays due to growing complexity.

Addressing these challenges presents significant opportunities.
Standardizing representations at different stages of compilation would improve interoperability and efficiency.
Developing accurate and efficient cost metrics for large-scale fault-tolerant computations would enable better optimization strategies.
Automation of manual tasks, such as pulse generation, could accelerate scientific discovery and reduce the workload on experimentalists.
Providing end-to-end compilation solutions would make fault-tolerant quantum computing more accessible to users with limited QEC expertise.

Integrating artificial intelligence (AI) into the compilation pipeline offers a powerful and largely untapped opportunity to address current limitations.
AI can enhance the compiler's ability to navigate complex optimization landscapes inherent to fault-tolerant quantum computations, with methods like reinforcement learning (RL) and deep learning improving qubit placement, atom routing, and gate scheduling.
Generative models, such as diffusion or transformer-based architectures, can automate pulse design and control signal generation, reducing manual effort and enabling the discovery of optimized, hardware-specific control sequences that boost performance and reliability.

\subsection{Research Directions}
\begin{figure}[tp]
\centering
\includegraphics[width=1\linewidth]{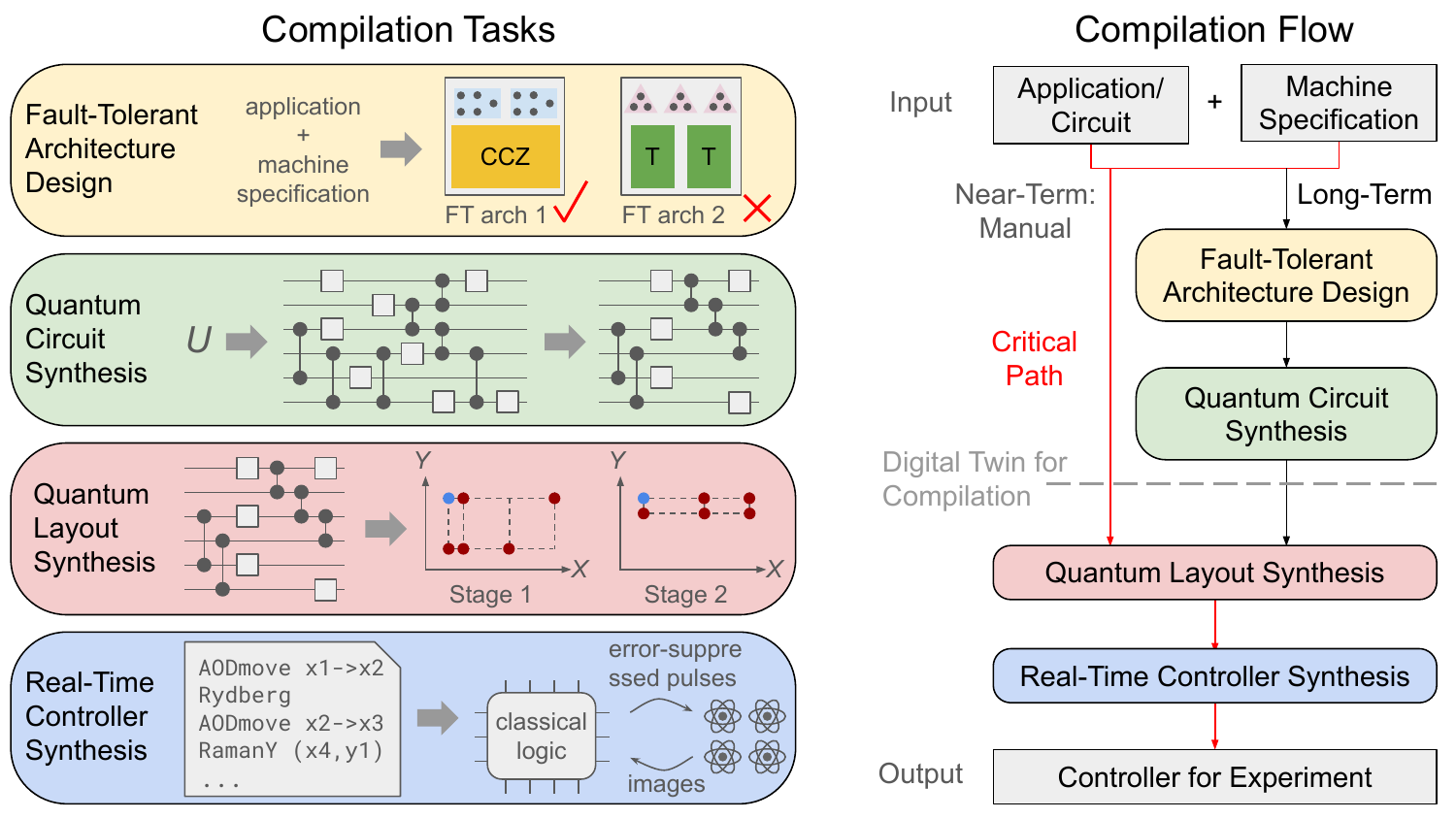}
\caption{Overview of compilation research.
Left: four compilation tasks.
\textit{Fault-tolerant architecture design} selects QEC codes and logical gadgets based on the application and machine specification.
\textit{Quantum circuit synthesis} decomposes and optimizes circuit to the supported gate set.
\textit{Quantum layout synthesis} maps the qubits and gates to spacetime and determines the movements.
\textit{Real-time controller synthesis} transforms machine instructions to error-suppressed pulses and control logic to interact with the quantum system.
Right: the whole compilation flow.
Red arrows represent the near-term critical path.
A digital twin for compilation will be provided to theorists to estimate implementation cost.
}
\label{fig:compilation_overview}
\end{figure}

Based on the above assessment, the research directions in compilation are summarized in Figure~\ref{fig:compilation_overview}.  
The following subsections outline the planned work, roughly ordered by priority.  
For each topic, corresponding machine learning problems can be formulated, and infrastructure can be developed to collect high-quality data to support model training and validation.

\subsubsection{Quantum Layout Synthesis}

Quantum layout synthesis (QLS) maps logical circuit designs to physical execution in spacetime.  
At the logical level, this involves determining how QEC code blocks are placed and moved across the  architectures to implement operations such as logical gates, syndrome measurements, and state preparation.  
To minimize the spacetime volume while maximizing fidelity, QLS should support the use of different QEC codes for different purposes, leverage fault-tolerant operations such as transversal gates, and allow for logical qubit reuse.

The high-level logical operations require physical-level layout synthesis, which entails precise coordination of atom movement, transfers between SLM and AOD traps, and specialized zone (Rydberg, readout, reloading, etc.) operation.  
The objective is to determine spatial and temporal placements of physical qubits and operations that minimize execution time, reduce movement-induced errors, and maximize resource reuse.  
Some initial QLS effort already supports zone-based architectures with transversal gate operations~\cite{lin2024reuse}, but more efforts are needed to cope with the complexities introduced by hardware features such as mid-circuit readout, continuous reloading to mitigate atom loss, and the outlook of future dual-species atom arrays or photonic-enabled individually addressable gates.

To support modularity and interoperability across the compilation pipeline, open intermediate representations will be developed to encode layout decisions and associated metadata at both the logical and physical levels.
The Zoned Architecture Intermediate Representation (ZAIR) provides a strong foundation for such efforts~\cite{lin2024reuse},
defining abstractions for qubit location, movement, and gate operations.
In addition to this, Bloqade-Shuttle and  Bloqade-Geometric from QuEra offer a practical IR for describing neutral-atom-specific operations and architectures, including site locations, atom rearrangement, and Rydberg excitation. These intermediate representations should expose atom-movement plans and relevant hardware constraints so that reconfiguration solvers and batching routines can be invoked as modular back-end components within the compilation workflow~\cite{Afiouni2025}.
Such standardized intermediate representations allow for clear interfaces between different stages of the compiler, from logical circuit synthesis and scheduling to physical layout.
They also facilitate analysis, debugging, and optimization, enabling compiler passes to operate independently yet coherently within a shared semantic layer.
Furthermore, open intermediate representations create opportunities for external integration.
For example, digital twins or machine learning models can interact with the layout synthesis process in a transparent and robust manner.
Ultimately, these intermediate representations are essential for building a scalable and extensible compilation ecosystem, where advances in algorithmic optimization or hardware can be rapidly incorporated without disrupting the overall compiler architecture.

\subsubsection{Real-Time Controller Synthesis}

Quantum systems are fundamentally controlled by peripheral classical electronics, which generate waveforms for opto-electronic control instruments and perform classical computations to determine feedforward operations required for atom displacement, logical gates and quantum error correction.  
While layout synthesis abstracts time to capture operation dependencies, real-time controller synthesis determines the precise timing of both quantum and classical processes.

A key goal is to automate pulse generation, incorporating error suppression techniques such as dynamical decoupling sequences and composite pulses to mitigate decoherence and crosstalk, thereby improving overall fidelity.  
Replacing manual efforts, the compiler will generate control sequences tailored to qubit species, gate types, and hardware capabilities in neutral atom systems.

Prior to running quantum circuits, hardware calibration is required---a process that presents opportunities for automation.
AI algorithms can be developed to infer optimal control parameters, correct for hardware drift, and adapt to device-specific variations in real time, such as through automatic tuning of control pulses demonstrated in Ref.~\cite{liang2022variational}.
These models are expected to reduce manual tuning while improving consistency and performance across different neutral atom configurations.

Another objective is the automation of real-time classical computing hardware configuration to support adaptive control tasks, including QEC decoding and feedforward control.
FPGA/GPU-based systems capable of low-latency classical computation and low-overhead data transfer between acquisition and control peripherals can be integrated in collaboration with experimentalists~\cite{Dadpour2025}. Key challenges include achieving deterministic timing with minimal jitter, implementing efficient shared-memory access and memory-transfer pathways, sustaining high-bandwidth device-to-device transfers to prevent buffer overflows, maintaining compatibility across evolving peripherals, algorithms, and applications, and scaling to multi-GPU deployments. An attractive option is developing an end-to-end, integrated real-time acquisition-and-control system spanning the full device stack, from data acquisition to low-latency processing and actuation.
These systems will be co-developed with customized, accelerated decoding algorithms in coordination with QEC experts and will be closely interfaced with the quantum control stack to enable rapid responses to measurement outcomes and dynamic adjustments during execution.

Together, these efforts will yield robust real-time controllers and decoding capable of driving quantum system operation across a wide range of fault-tolerant quantum applications.

\subsubsection{Quantum Circuit Synthesis}

Quantum circuit synthesis (QCS) maps high-level representations of computation to the architecture’s native gate set under a cost model that captures non-Clifford overhead, circuit latency, qubit footprint, and error rates.
The cost model may be a QEC-independent one that prioritizes non-Clifford metrics such as T-count or T-depth in a Clifford+T representation, or QEC-dependent, which evaluates implementations under a specific code by accounting for logical operators, error-correction scheme, and space-time volume~\cite{WangDAC2026ExactT,vandaele2025lower}.

Beyond optimizing circuits in the logical gate set, the synthesizer must determine when and where to insert syndrome extraction, balancing adequate error correction with circuit duration.
Additional optimization can be achieved by leveraging application-specific structures, such as don't-care conditions in quantum circuits~\cite{wang2024quantumstatepreparationcircuit}.
The synthesis process should also consider addressability constraints in high-rate QEC codes, where transversal gates sometimes have to act on all logical qubits in a QEC block symmetrically and simultaneously, instead of acting on selected subset of logical qubits~\cite{nguyen2024quantum}.

\subsubsection{Fault-Tolerant Architecture Co-Design}

Demonstrated neutral atom array architectures feature multiple functional zones for storage, entanglement, and readout~\cite{bluvstein_logical_2024}.
Future extensions may include zones for continuous reloading and quantum networking mentioned in other sections.
In fault-tolerant quantum computing, these physical zones may further split into logical architectural areas, such as magic state factories or arithmetic modules within the entanglement zone.

A robust compilation flow enables co-design of the physical and logical architectures with the target application.
This scenario is analogous to FPGAs, which consist of interconnected programmable classical logic units.
Just as an FPGA fabric can be optimized based on quantitative evaluation to determine the size of lookup tables (LUTs), clustering structure, etc.~\cite{li03architecture}, the quantum physical architectures can be tailored by adjusting the size and configuration of functional zones.
Designing quantum logical architectures parallels assigning FPGA regions to implement specific IP cores and route the computation between them.
Initial study has been conducted on superconducting qubit based platforms with such a compiler-driven methodology for some NISQ applications~\cite{lin22domain}.
Extending this paradigm to neutral atoms, Ref.~\cite{lin2024reuse} demonstrates compiler support for multi-entanglement zone and multi-AOD architectures and shows how compilation results with metrics such as movement duration, hardware utilization, and achievable parallelism, can be used as quantitative indicators of architecture quality to guide device-level design.

In the long term, compilation research aims to provide an end-to-end service for users without QEC expertise.
The compiler should automatically design fault-tolerant architectures by selecting suitable QEC codes, syndrome extraction strategies, and non-Clifford protocols based on the application and the hardware specification.
Given the diversity of QEC codes, trade-offs in encoding rate, noise threshold, and gate compatibility must be evaluated to select the best fit for the task and hardware.
For example,~\cite{bluvstein2025architecturalmechanismsuniversalfaulttolerant} demonstrates a heterogeneous code architecture where different codes are employed for different computational roles.
In their scheme, Clifford gates are implemented with a 2D color code,
while non-Clifford gates are realized by a 3D color code with gate teleportation.
The other example for a heterogeneous code architecture is to have a code with high encoding rate, e.g., Bivariate bicycle (BB) code, for long-term qubit storage and surface code for active computation~\cite{stein2025hetec}.
The study demonstrates that physical qubit reduction through using BB code comes with the cost of longer circuit duration due to the extra overhead for conversion between two codes.
These examples highlight the opportunities for compilers to optimize circuit performance by architecture designs.

\section{Networking Quantum Processors} \label{sec:network}

Scaling quantum processors is a common challenge across platforms. For neutral-atom arrays, scaling to 10,000 physical qubits is feasible in the near term. With specialized lenses and ultra-high-power lasers, further scaling, potentially up to 100,000 controllable atoms per device, may be possible. Continued scaling will eventually require the integration of multiple devices into a modular architecture. This approach, which is also being explored in other platforms \cite{Monroe2014,ang2022,Jameson2024}, has practical appeal because it simplifies the challenge of scaling to designing a quantum processor that can distribute entanglement to other modules. Modular quantum computing requires fast and high-fidelity quantum network channels between multiple quantum processor units (QPUs) facilitated by long-range atom transport or scalable atom-photon interfaces. Although these technologies remain significantly slower and lower fidelity than local entangling operations (Rydberg gates), looming size limits make this a crucial challenge for continued scaling of atom-array quantum processors. 

\subsection{Challenges and Opportunities}

Endowing neutral atom quantum processors with a quantum communication interface has the potential to unlock a new exponential trajectory for scaling, while also promising to improve the fundamental operating performance of individual modules by boosting logical clock rates and qubit readout speed. 
In addition, the powerful combination of computing and networking capabilities enables new applications such as blind quantum computing~\cite{Fitzsimons2017} and quantum computational sensing~\cite{khan2025}, through quantum network nodes capable of distributing multipartite entanglement, performing fault-tolerant remote entanglement distribution, enabling in-node entanglement swapping and fusion, and generating non-classical photonic states.

Developing fault-tolerant optical interconnects within fault-tolerant modular architectures is a highly active field of research \cite{ang2022,Ramette2024,pattison2024,jacinto2025,stack2025,Sinclair2025,marqversen2025,Shalby2025,sunami2025,liu2026} and it is too early to state exactly what the minimum hardware requirements will be to achieve fault-tolerance. Different QEC architectures support different trade-offs between interconnect speed, interconnect fidelity, and the resources within each module dedicated to inter-module communication. For instance, since non-local atom-atom Bell pairs lack the desired fidelity for many modular QEC protocols, one approach to increase their fidelity is to perform local operations to distill multiple low-fidelity non-local Bell pairs into fewer high-fidelity non-local Bell pairs \cite{Bennett1996}.  The number of non-local Bell pairs required to perform a non-local error-corrected logical gate can vary by multiple orders of magnitude, depending on the QEC scheme \cite{pattison2024}. 

To scale networked QPUs, interconnects must operate with sufficient rate and fidelity to support fault-tolerant operations across modules, exceeding QEC thresholds that are substantially more stringent than those required for quantum communication applications. While coherent transport between separated QPUs within the same vacuum chamber may enable further scaling beyond 100,000 atoms \cite{Sinclair2025,xue2026}, most proposals expect that distant entanglement generation will be mediated by atom-photon entanglement. In such approaches, photon loss is the dominant limitation: heralding prevents loss from introducing errors, but renders entanglement generation intrinsically probabilistic and therefore rate-limited. As a result, practical operation relies on repeated or parallel entanglement attempts to overcome low single-shot success probabilities. 

A challenge closely linked to photon loss is recoil heating: atom-photon entanglement generation inherently involves scattering photons. Each scattering event excites motion along a given mode with probability set by the Lamb-Dicke parameter squared of that mode. Two features make recoil heating a key consideration in photon-mediated entanglement of neutral atoms. First, current two-qubit gate fidelities are sensitive to the atoms being in the motional ground state~\cite{robicheaux2021photon,pagano2022error}. Second, it is challenging to cool an atom's motion without decohering its internal-state qubit. With Lamb-Dicke parameters typically 0.1--0.5 in neutral atom traps, heating per scattering event is substantial. This heating rate limits the number of attempts per re-cooling cycle, motivating the development of more robust gate schemes, efficient atom-photon interfaces, parallelization to amortize re-cooling, and motion-tolerant two-qubit gates~\cite{Li2024,Petrosyan2017}.

Current experimental progress is concentrated along three directions: improving atom-photon entanglement fidelity through more robust atom-photon entanglement protocols,  increasing entanglement rates through more efficient atom-photon interfaces, and exploring multiplexing across many atoms or cavity modes. These directions align closely with intrinsic strengths of the neutral-atom platform. Optical transitions in the near-infrared and telecom bands enable high-finesse cavities and low-loss photonic links; large-scale arrays enable massive spatial multiplexing; and the diversity of atomic species and level structures supports tailored atom–photon entanglement protocols.

\subsection{Research Directions}

\begin{table}[h!]
\centering
\begin{tabular}{||c c c c c c||} 
 \hline
 Year & Group & Description & Qubit & Rate $(s^{-1})$ & Fidelity  \\ [0.5ex] 
 \hline\hline
 2004 & \cite{Matsukevich2006} Kuzmich & free-space, entanglement &  $^{87}$Rb & - & - \\ 
 \hline
 2007 & \cite{Moehring2007} Monroe & free-space, entanglement &  $^{171}$Yb$^+$ & 0.0021 & 0.63 \\ 
 \hline
 2009 & \cite{Olmschenk2009} Monroe & free-space, teleportation &  $^{171}$Yb$^+$ & 1.4e-3 & 0.90 \\ 
 \hline
 2012 & \cite{Ritter2012} Rempe & cavity, teleportation &  $^{87}$Rb & 30 & 0.85 \\ 
 \hline
 2013 & \cite{Nolleke2013} Rempe & cavity, teleportation &  $^{87}$Rb & 0.1 & 0.88 \\ 
 \hline
 2013 & \cite{Slodifmmode2013} Blatt & free-space &  $^{133}$Ba$^+$ & 0.25 & 0.64 \\ 
 \hline
 2013 & \cite{Hucul2015} Monroe & free-space &  $^{133}$Ba$^+$ & 4.5 & 0.78 \\ 
 \hline
 2017 & \cite{Rosenfeld2017} Weinfurter & loophole free&  $^{87}$Rb & 0.011 & ? \\ 
 \hline
 2020 & \cite{Stephenson2020} Lucas & free-space &  $^{87}$Sr$^+$ & 182 & 0.94 \\ 
 \hline
 2022 & \cite{vanLeent2022} Weinfurter & 33 km entanglement &  $^{87}$Rb & 1/85 & 0.622 \\ 
 \hline
 2022 & \cite{Krutyanskiy2023} Northup & cavity, 230 m &  $^{40}$Ca$^+$ & 0.49 & 0.88 \\ 
 \hline
  2022 & \cite{Nadlinger2022} Lucas & free-space, DIQKD &  $^{88}$Sr$^+$ & 63 & 0.96 \\ 
 \hline
  2024 & \cite{Jameson2024} Monroe & free-space, pol. &  $^{133}$Ba$^+$ & 250 & 0.94 \\ 
 \hline
 2024 & \cite{Saha2025} Monroe & free-space, time-bin &  $^{133}$Ba$^+$ & 0.35 & 0.97 \\ 
 \hline
\end{tabular}
\caption{Table of experimental demonstrations of \textbf{remote atom-atom and ion-ion entanglement}. Reported fidelitys are measured and do not correct for separately characterized error channels. }
\label{table:1}
\end{table}

\begin{table}[h!]
\centering
\begin{tabular}{||c c c c c c c||} 
 \hline
 Year & Group & Description & Atom Qubit & Distance & Rate $(s^{-1})$ & Fidelity  \\ [0.5ex] 
 \hline\hline
 2004 & \cite{Blinov2004} Monroe & NA=0.23 & $^{111}$Cd$^+$ & $<1$ km &  0.3 & 0.87 \\ 
 \hline
 2006 & \cite{Volz2006} Weinfurter & NA=0.38 & $^{87}$Rb & $<1$ km & 0.2 & 0.87 \\
 \hline
 2007 & \cite{Wilk2007} Rempe & C=1.3 & $^{87}$Rb & $<1$ km & ? & 0.86 \\
 \hline
 2009 & \cite{Weber2009} Rempe & C=1.3 & $^{87}$Rb & $<1$ km & 12 & 0.93 \\
 \hline
 2012 & \cite{Stute2012} Blatt & C=1.75 & $^{40}$Ca$^+$ & $<1$ km & 40.5 & 0.974 \\
 \hline
 2018 & \cite{Bock2018} Eschner & NA=0.4 & $^{40}$Ca$^+$ & $<1$ km  & 27.6 & 0.955 \\
 \hline
 2019 & \cite{Krutyanskiy2019} Lanyon & C=1.47 & $^{40}$Ca$^+$ & $50$ km  & 1 & 0.90 \\
 \hline
 2019 & \cite{Crocker19} Monroe & NA=0.6 & $^{138}$Ba$^+$ & $<1$ km  & ? & 0.93 \\
 \hline
 2020 & \cite{vanLeent2020} Weinfurter & NA=0.5 & $^{87}$Rb & 20 km  & 0.6 & 0.785 \\
 \hline
 2021 & \cite{Kobel2021} Kohl & C=0.056 & $^{171}$Yb$^+$ & $<1$ km & 63 & 0.901 \\
 \hline
 2021 & \cite{Schupp2021} Lanyon & C=1.47 & $^{40}$Ca$^+$ & $<1$ km & 62.5  & 0.966 \\
 \hline
 2021 & \cite{Connell2021} Erik & NA = 0.68 & $^{171}$Yb$^+$ & $<1$ km & freq. conv.  & 0.924 \\
 \hline
 2022 & \cite{Hannegan2022} Quraishi & NA = 0.6 & $^{138}$Ba$^+$ & $<1$ km & 143  & 0.93 \\
 \hline
 2023 & \cite{Drmota2023} Lucas & NA = 0.6 & $^{88}$Sr$^+ \rightarrow ^{43}$Ca$^+$ & $<1$ km & -  & 0.977 \\
 \hline
 2024 & \cite{Krutyanskiy2024} Lanyon & C=1.47 & $^{40}$Ca$^+$ & $101$ km  & 3 & 0.96 \\
 \hline
 2024 & \cite{Zhou-memory-2024} Weinfurter & NA=0.5 & $^{87}$Rb & $101$ km & 0.0038 & 0.7 \\
 \hline
 2024 & \cite{Hartung2024} Rempe & C=3.6 & $^{87}$Rb & $<1$ km & - & 0.86 \\
 \hline
 2025 & \cite{Li2025} Covey & NA=0.6 & $^{171}$Yb & $<1$ km & - & 0.9 \\
 \hline
 2026 & \cite{Safari2026} Saffman & NA=0.61 & $^{87}$Rb & $<1$ km & 12 & 0.93 \\
 \hline
 2026 & \cite{hwang2026} Saffman & NA=0.55 & $^{133}$Cs & $<1$ km & 0.3 & 0.94 \\
 \hline
\end{tabular}
\caption{Table of experimental demonstrations of \textbf{atom-photon and ion-photon entanglement}. Reported fidelity's are measured and do not correct for separately characterized error channels. Cavity cooperativities are calculated using $C = g^2/\kappa \gamma$, where $\kappa$ and $\gamma$ are amplitude decays.}
\label{table:2}
\end{table}

\subsubsection{High Fidelity Atom-Photon Entanglement}

Currently, the fidelity of remote entanglement lags significantly behind the fidelity of entanglement generated within QPUs using Rydberg gates. It is possible this gap may not need to close, as multiple QEC architectures exist \cite{Ramette2024,Sinclair2025,pattison2024} with Bell pair infidelity thresholds an order of magnitude higher than the local gate thresholds. However, per Table \ref{table:1}, even a relaxed infidelity threshold of 10\% for remotely entangled neutral atoms has not yet been achieved. Furthermore, error rates should be well below threshold (typically $10\times$ below) to reach a regime where errors are efficiently suppressed leading to a target atom–atom Bell pair infidelity of $\sim 1\%$. Reaching this regime is a prerequisite for scalable architectures in which additional QPUs can be connected while efficiently suppressing errors.

To improve the fidelity of remote entanglement between neutral atoms, recent work has focused on improving atom–photon entanglement fidelity (see Table \ref{table:2}). Sources of infidelity include photon distinguishability arising from atom motion or Doppler shifts, unwanted transitions, polarization errors, false heralding, and decoherence of the matter qubit after heralding. These limitations are being addressed through novel atom–photon entanglement protocols, advanced cooling techniques to suppress motional dephasing, and the use of in-vacuum polarization optics to reduce polarization errors. Two recent works, by the Covey and Saffman groups, have achieved inferred atom–photon Bell state fidelities at or above 95\% \cite{li2025parallelized,Safari2026}, representing significant progress towards the target of $\sim 99\%$ fidelity required for fault-tolerant operation.

\begin{figure}
    \centering
    \begin{subfigure}{0.9\textwidth}
        \centering
        \includegraphics[width=0.7\textwidth]{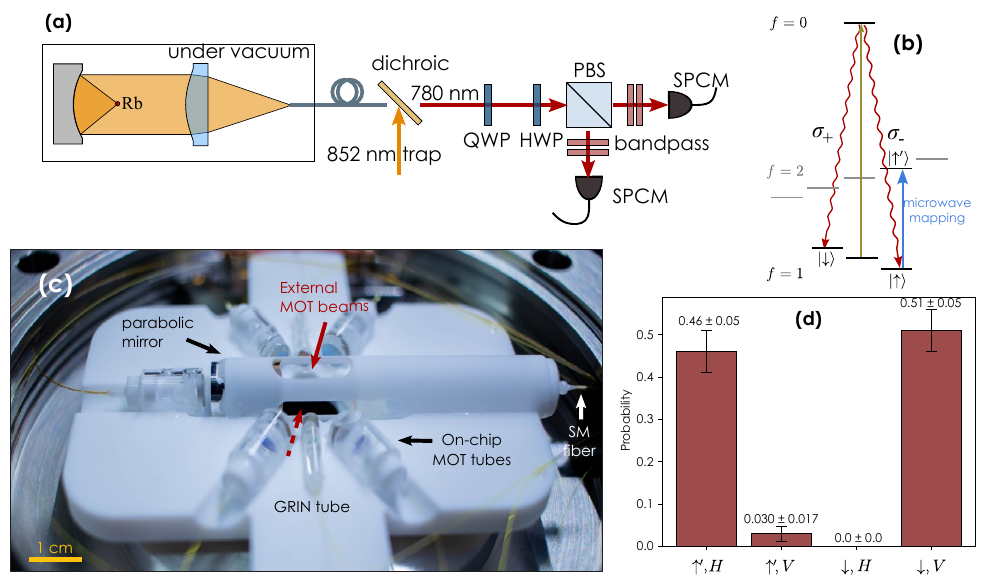}
        \caption{}
        \label{fig:FigureParabolicOnChip_combined}
    \end{subfigure}
    
    \vspace{1em}
    
    \begin{subfigure}{\textwidth}
        \centering
        \includegraphics[width=0.7\textwidth]{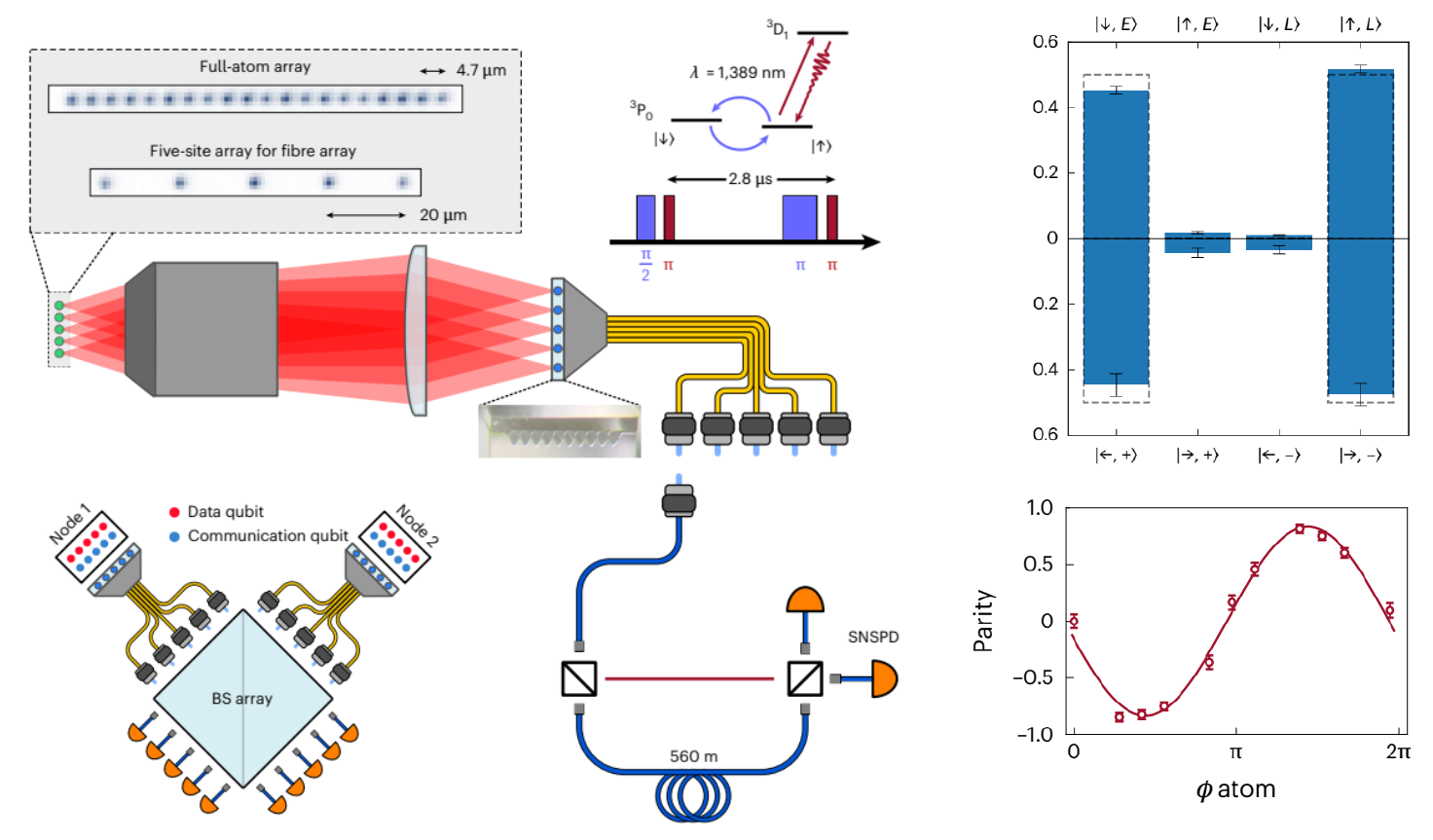}
        \caption{}
        \label{fig:FigureYbCovey_combined}
    \end{subfigure}
    
    \caption{\textbf{Advancing hardware enables higher atom-photon entanglement fidelity:} 
    (i) (a) A schematic of the Saffman group's optical setup for trapping a single Rb atom and characterizing atom-photon entanglement, (b) the relevant level structure and the mapping transition to the qubit basis $(\downarrow , \uparrow')$ with long coherence time, (c) an in-vacuum setup built from pre-aligned mm-scale optics on a Macor platform yielding a photon generation and collection efficiency of 9\% after single-mode fiber coupling, (d) and measured joint probabilities in the \(x\)-basis yielding a raw atom-photon entanglement fidelity of 0.93. 
    (ii) Parallelized telecom networking with Yb, showing a networking protocol and architecture based on a microscope objective, together with the measured atom-photon Bell-state fidelity and parity oscillations.}
    \label{fig:NetworkingNodeComparison}
\end{figure}

In polarization-based entanglement protocols, the heralded atomic qubit is often initially encoded in magnetically sensitive Zeeman states. After photon detection, this matter qubit can therefore dephase during the time required for feed-forward, state mapping, or subsequent operations. In alkali atoms, the corresponding coherence time is on the order of $100~\mu$s, directly limiting both atom-photon and heralded atom-atom entanglement fidelities~\cite{Zhou-memory-2024}. To mitigate this effect, the Saffman group maps the initially generated atomic qubit to a ``magic'' qubit (Fig.~\ref{fig:FigureParabolicOnChip_combined}) that is insensitive to magnetic-field fluctuations at a bias field of $3.23$~G for $^{87}$Rb~\cite{Harber2002}. This extends the measured coherence time to $T_2^*=3.2$~ms in $^{87}$Rb~\cite{Safari2026}, and to $T_2^*=14$~ms in $^{133}$Cs~\cite{hwang2026}. As a result, residual atomic-qubit dephasing contributes only about $0.5\%$ to the entanglement infidelity; at lower temperatures, magnetically insensitive states can reach coherence times of a few seconds~\cite{Treutlein2004}.

Another research direction is exploring new atom–photon entanglement schemes, particularly those employing time-bin qubits rather than the traditional polarization-based encoding.  Time-bin photonic encoding can reduce sensitivity to polarization errors and optical path fluctuations, making it particularly suitable for long-distance entanglement distribution or for large modular quantum computing systems with many interconnects. Nevertheless, time-bin photonic encoding introduces its own set of challenges, including reduced fidelity due to recoil-induced distinguishability and timing errors. Recent experiments using time-bin encoded photonic qubits have been realized in the Monroe group with trapped ions \cite{Saha2025} and in the Covey group with Yb atoms \cite{li2025parallelized} and theoretically explored \cite{Menon_2020,Apolin2026recoilderror,Yu2026recoiltiming,kikura2025tamingrecoil}.

In the Covey experiment, a raw atom–photon entanglement fidelity of 0.9 and a corrected fidelity of 0.95 between a time-bin encoded photonic qubit and highly coherent nuclear spin qubits in ytterbium-171 were achieved. These results are also notable as the first experimental demonstration of atom–photon entanglement directly generated in the telecom band, enabling compatibility with low-loss fiber-based quantum networks. The experimental setup, depicted in Fig.~\ref{fig:FigureYbCovey_combined}, utilized an NA$=0.63$ objective; however, further improvements in fidelity are anticipated with increased photon collection efficiency, for example through the use of an optical cavity, which would reduce photon detection errors that currently dominate the residual infidelity.

Taken together, these results highlight how advances in atom–photon interfaces, encoding schemes, matter-qubit coherence, and optical control are converging to suppress both coherent and incoherent error mechanisms, pushing entanglement fidelities toward the thresholds required for fault-tolerant distributed computing.

\subsubsection{High Rate Remote Entanglement Generation}

Photons can mediate remote atom-atom entanglement following different protocols \cite{Beukers2024}. The most commonly-considered one for modular quantum computing is "meet-in-the-middle", in which two photons entangled with one atom at each node must be detected. In this protocol, the success probability scales with the square of the efficiency in each link, i.e. with the square of the single-photon collection and detection efficiency: $P_{\rm succ} = \frac{1}{2}\eta^2_{\rm coll}\eta^2_{\rm det}$. The overall rate of entanglement generation is then $R=P_{\rm succ}R_{\rm att}$, where $R_{\rm att}$ is the rate at which entanglement is attempted.

The current record on remote entanglement generation rates for neutral atoms is 30 $s^{-1}$ \cite{Ritter2012} (see Table~\ref{table:1}). In contrast, directly interfacing surface code logical qubits through remote lattice surgery requires approximately $\sqrt{N}$ Bell pairs per QEC cycle, where $N$ is the number of physical qubits~\cite{Sinclair2025}. At 1 millisecond code cycles, this requires 10 to 100 kHz Bell pair rates to reach the MQuop to TQuop regime. These rate requirements may increase if physical distillation is required \cite{Nigmatullin_2016}, or they may be significantly decreased if physical qubits can be injected into logical Bell pairs for more efficient logical distillation strategies\footnote{These modular QEC architectures are described in more detail in Section \ref{sec:QEC_mod}.} \cite{pattison2024,marqversen2025,sunami2025}. Regardless of architectural choice, there remains a large gap between current capabilities and the rates required for scalable fault-tolerant operation.

The primary obstacles to high-rate remote entanglement generation are low photon collection efficiency and limited attempt rates. 
Using a lens to collect photons typically limits this efficiency to percent-level, resulting in $\sim 10^{-4}$ probability of successful pair detection. The leading approach to increasing collection efficiency is the use of optical cavities \cite{Covey2019,Sinclair2025,Young2022a,Covey2023,Li2024,sunami2025scalable}, which can achieve efficiencies exceeding 50\%, thereby increasing entanglement rates quadratically with interface improvement.

In a network node with a single neutral atom, multiple factors limit the attempt rates, including the overall experiment cycle time (which can be on the order of seconds), the number of times a single atom can be reused, the reset time after a failed attempt, or the duration of the atom-photon entanglement protocol. While a burst rate of a few 10 s of kHz is possible~\cite{vanLeent2022}, the average attempt rate is typically closer to Hz, limited mainly by atom reloading time. This highlights the importance of multiplexing to improve the entanglement rate. Several multiplexing schemes have been proposed, including temporal, frequency, and spatial multiplexing, which require parallel control of many qubits and, for spatial multiplexing, scalable
atom–photon interfaces.  
Temporal multiplexing proposals predict rates as high as 100 kHz using optimized cavity designs and large atomic reservoirs \cite{Young2022a,sunami2025scalable,Li2024,Covey2023}, extending into the tens of MHz range when cavity arrays are used \cite{Sinclair2025}. 

\textbf{Efficient Atom-Photon Interfaces}

With photon loss as a key limitation to both fidelity and especially speed, increasing the efficiency of atom-photon interfaces directly increases the probability of successful entanglement generation per attempt. Even with a high-NA objective, only a limited fraction of the scattered photons is collected into the desired spatial mode. Imperfect mode matching and alignment-sensitive single-mode fiber coupling introduce additional losses, so the usable fiber-coupled photon collection efficiency is often reduced to the percent level.

To address this challenge, the Saffman group has developed a parabolic-mirror-based monolithic architecture in which the same optics are used both to form the dipole trap and to collect the scattered photons (Fig.~\ref{fig:FigureParabolicOnChip_combined}). A key advantage of this design is that the optical path of the collected fluorescence is intrinsically matched to that of the trapping beam, making the system largely insensitive to optical misalignment and thereby improving mode matching to the single-mode fiber. With this system, they achieve an overall single-photon collection and fiber-coupling efficiency of $9\%$ while maintaining high single-photon purity, characterized by $g^{(2)}(0)=0.006 \pm 0.006$ \cite{Safari2026}. These results show that a cavity-free architecture can reach a regime of efficient photon generation and high entanglement fidelity while retaining a simple light-matter interface.

\begin{figure}
    \centering
    \includegraphics[width=1\linewidth]{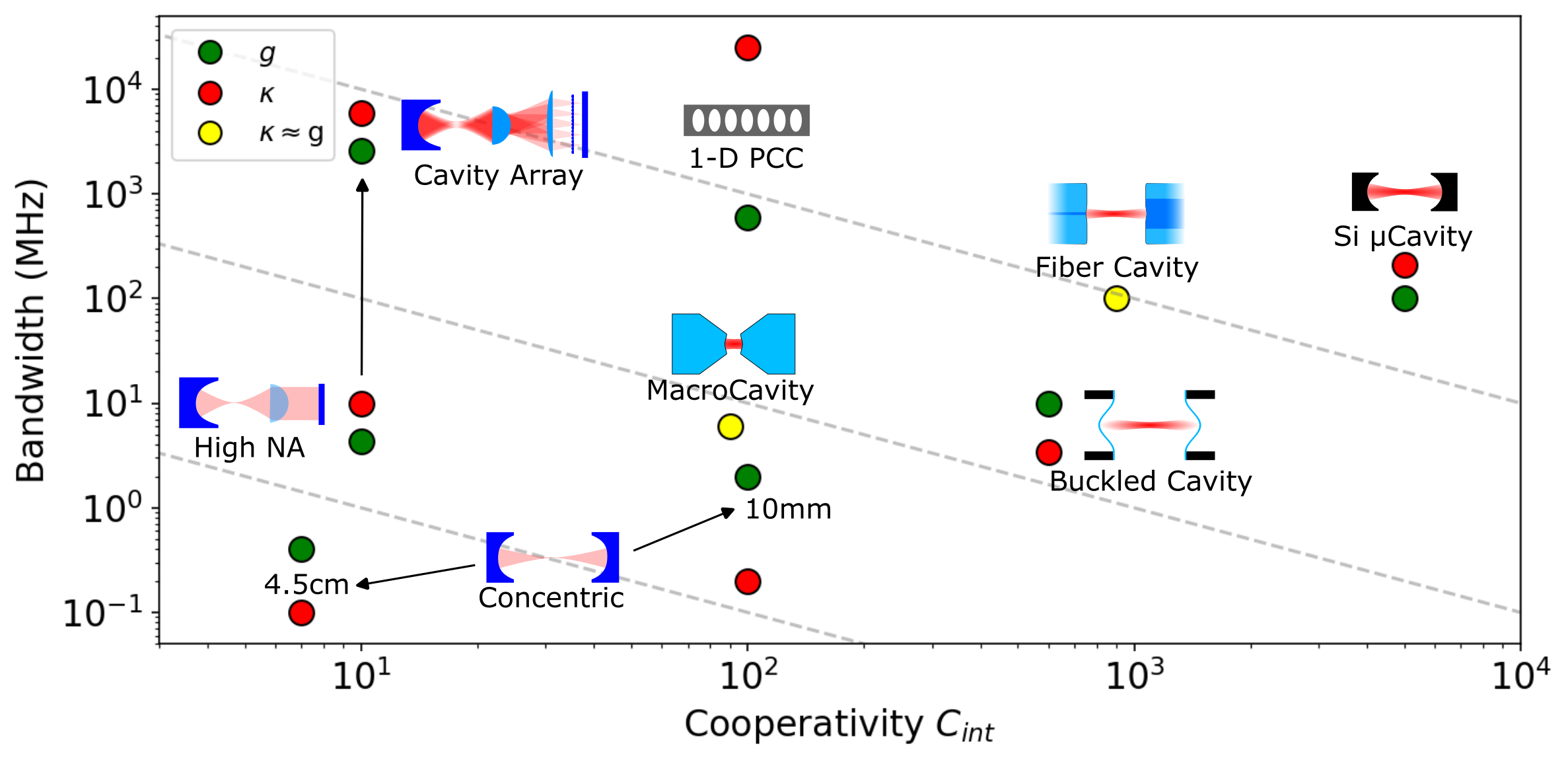}
    \caption{\textbf{Comparison of cQED platforms for quantum networking.} We define a metric, intrinsic single-atom cooperativity $C_\text{int} = 4g^2/\kappa_{sc}\gamma$. This number indicates the maximum possible cooperativity of a cQED system, limited only by cavity scattering. This sets a fundamental limit on the collection efficiency for a given cQED system. Green and red markers denote the atom-cavity coupling rate $g$ and cavity field decay rate $\kappa$, respectively, while yellow markers indicate the strong-coupling regime where $\kappa \approx g$. Dashed lines indicate contours of constant cooperativity. Cavity types shown include near-concentric Fabry-P\'{e}rot cavities (assuming a scattering limited Finesse of $10^6$) at 10~mm~\cite{hu2025site} and 4.5~cm~\cite{Deist2022} lengths, a high numerical aperture (NA) cavity~\cite{shadmany2025cavity}, a cavity array with bandwidth equal to $600 \times$ the high-numerical aperture cavity \cite{soper2026stability}, a macroscopic Fabry-P\'{e}rot cavity~\cite{welte2018gate}, a one-dimensional photonic crystal cavity (1-D PCC)~\cite{samutpraphoot2020strong}, a buckled microcavity~\cite{ding2026high}, a fiber Fabry-P\'{e}rot cavity~\cite{grinkemeyer2025error}, and a silicon microcavity (Si $\mu$Cavity)~\cite{wachter2019silicon}. Together, these platforms span several orders of magnitude in both bandwidth and cooperativity, illustrating the trade-offs between cavity size, photon confinement, and coupling efficiency relevant to atom-photon entanglement generation.}
    \label{fig:cavityOverview}
\end{figure}

However, further improvements in collection efficiency are possible using resonantly enhanced atom–photon interfaces. Optical cavities are therefore widely regarded as the leading technological candidate for boosting the entanglement rate.
Optical cavities span centimeter, micro-, and nano-scale regimes, each offering distinct advantages and challenges see Fig.~\ref{fig:cavityOverview}. Centimeter-scale Fabry–Pérot and bow-tie cavities are straightforward to implement with commercial components, providing moderate cooperativity (C$\sim$1–20) and good photon collection efficiency. These systems enhance emission into a well-defined spatial mode, improving both collection efficiency and fiber coupling. Large 1D and 2D atom arrays can be coupled to the cavity mode enabling parallel atom–photon interfaces while preserving Rydberg interactions without strong decoherence from nearby surfaces. To maintain strong coupling in these larger cavities, several groups, such as the Vuletić, Stamper-Kurn, Zeiher, and Covey groups, use a near-concentric geometry, in which a small mode waist is obtained by operating near the edge of stability \cite{Deist2022,hu2025site,DeSantis2026}. However, this design requires larger mirrors and longer cavity lengths, which can reduce optical access along directions typically used for Rydberg excitation and limit repetition rates through reduced cavity bandwidth, partially offsetting gains in collection efficiency.

Micro-scale cavities, such as fiber-Fabry Perot cavities (FPFC) \cite{hunger2010fiber} or micro-mirror cavities \cite{Hartung2024,ding2026high}, achieve stronger coupling (with C > 100 for FPFC) and fast, high-fidelity single-atom readout. Similarly to large-scale cavities, they offer free space atom transport and addressing, but typically support only 1D atomic arrays coupled to the cavity mode. Micro-scale cavities are currently being explored by the Lukin and Reichel groups.

Nano-scale platforms, including photonic-crystal cavities~\cite{dhordjevic2021entanglement,Menon2024}, micro-ring resonators~\cite{Chang2020effiecntcoupling,Kim2019,Xinchao2024}, whispering-gallery-mode resonators~\cite{Bechler2018,Ohana2024,Will2021}, and nanofiber cavities~\cite{Kato2015,Nayak2019,sunami2025scalable} provide extreme light confinement and very large cooperativities ($C\sim 70$ \cite{dhordjevic2021entanglement}). In addition, the large photon outcoupling rates demonstrated in nanophotonic cavities ($\kappa \sim  2\pi$ 3.6 GHz  \cite{dhordjevic2021entanglement})  are particularly suitable for quantum networking. 
Among these platforms, nanofiber cavities uniquely leverage the ultralow propagation loss of heat-and-pull fibers, enabling millimeter-scale nanofiber regions that accommodate hundreds of atoms within a single device \cite{takahata2026} while retaining a high intrinsic single-atom cooperativity, $C_\mathrm{int}\sim100$~\cite{Ruddell2020,horikawa2025}. Such atom capacities are projected to enable near-MHz-rate operation through time multiplexing,~\cite{sunami2025scalable}, as discussed in the later section.

Compared to large-scale cavities, micro- and nano-scale cavities offer distinct advantages and challenges. First, due to the high light confinement, these cavities allow for high-fidelity atom-photon entanglement \cite{Covey2019,Menon_2020,Hartung2024,sunami2025scalable} and also for cavity-mediated atom-atom entanglement with all-to-all connectivity \cite{welte2018gate,Ramette2022,grinkemeyer2025error,dhordjevic2021entanglement}. 
Nevertheless, the errors of deterministic cavity-mediated atom-atom entanglement typically scale as 1/$\sqrt{C}$ (or better with error detection \cite{grinkemeyer2025error,Borregaard2015,Nagib2024,Ramette2025},) limiting the maximum achievable fidelity and requiring either the more commonly-considered heralded protocols, or new fabrication techniques to increase the cooperativity.
Second, the small form factor allows for the implementation of cavity arrays, enabling multiplexed optical interfaces, as discussed in the next section. Nonetheless, several challenges remain. First, trapping and imaging atoms near surfaces remain technically demanding due to scattering, electric and magnetic field noise, and polarization control challenges. Second, for small-scale cavities, a major challenge is the integration of coherent Rydberg-mediated operations. Since optical cavities are typically constructed from dielectrics that can accumulate electrical charges, the resulting charging can induce dephasing in the Rydberg state. 
Although coherent Rydberg operations have been demonstrated as close as 100 ~$\mu$m from a SiN nanophotonic cavity ~\cite{ocola2024control}, drifting patch charges motivate multiple efforts to mitigate patch charges across a broader range of cavity platforms. These include developing highly-transparent, low-loss conductive coatings ~\cite{wang2025can} and using long-range coherent transport to separate Rydberg gate zones from cavity surfaces ~\cite{dhordjevic2021entanglement,ocola2024control,sunami2025scalable}.

The exploration of new optical \textbf{cavity fabrication methods} can address several challenges in integrating atom arrays with optical cavities. For example, more advanced fabrication techniques can improve the finesse of the cavity and consequently increase atom–cavity coupling strengths \cite{wachter2019silicon,jin2022micro}, thus enhancing the emission in a well-defined mode and increasing the efficiency of photon collection per entanglement attempt. These methods can also ensure pristine surface quality, thereby reducing surface charge fluctuations. Furthermore, protective coatings applied during fabrication can mitigate charge fluctuations arising from cavity surfaces, enabling more coherent Rydberg operations near optical interfaces.

Traditional cavity-fabrication methods are generally limited to producing mirrors with relatively large radii of curvature (>5mm), which in turn leads to longer cavity lengths. Unfortunately, longer cavities naturally reduce the achievable bandwidth due to the larger mode size and the lower cavity decay rate, resulting in slower photonic interfaces. Microcavities, by contrast, can offer significantly higher bandwidth with smaller mode volume and larger cavity decay rate, but their small radii of curvature cannot be produced using conventional shaping and polishing techniques. For this reason, it has been of great interest to develop methods for creating a smaller radius of curvatures while maintaining high-quality mirror shape and surface roughness.

Many new fabrication approaches have been developed to create small-ROC mirrors while still maintaining high finesse. One of the earliest such methods is CO$_2$-laser ablation \cite{hunger2010fiber}, often performed on the tip of an optical fiber, which can achieve low surface roughness and small radii of curvature on the order of $10$s to $100$s of microns. Over the past decade, this technique has matured substantially, enabling reliable production of a wide range of curvatures as large as several millimeters and even allowing nearly arbitrary surface shaping \cite{ott2016millimeter}, supporting the realization of high-bandwidth, high-efficiency atom–photon interfaces.

While CO$_2$ ablation remains powerful, other methods have emerged that offer greater scalability by leveraging established microfabrication processes. Isotropic etching in silicon, for example, enables the creation of highly spherical micromirrors with excellent uniformity over large wafers. Surface roughness can then be improved beyond what CO$_2$ ablation can typically achieve using well-developed silicon polishing techniques. More recently, high-quality micromirrors have also been demonstrated in ULE glass using resist reflow, which provides similarly low surface roughness together with access to a wide range of radii of curvature ranging from 100s of microns to several mms \cite{jin2022micro}, supporting both high cooperativity and scalable integration.

The field continues to advance with innovative techniques based on controlled buckling of thin films, which can produce precise curvature in a highly scalable fashion \cite{ding2026high}. These approaches also benefit from the use of super-polished flat substrates, which are significantly easier to fabricate than polished curved surfaces. As a result, they have enabled the demonstration of exceptionally high finesse values, exceeding 900,000, even at the shorter wavelengths relevant for quantum applications. Collectively, these novel fabrication methods are opening new opportunities to integrate microcavities into a wide variety of quantum platforms, not only neutral atoms but also trapped ions, solid-state emitters, and other emerging quantum systems, with direct implications for increasing entanglement generation rates in networked architectures.

\textbf{Multiplexing Atom-Photon Entanglement Generation}

Multiplexing provides a route to increasing entanglement generation rates by parallelizing attempts across multiple temporal, spatial, or spectral modes, effectively increasing the total attempt rate without requiring faster single-atom cycling. 

\textbf{Temporal multiplexing} utilizes the strengths of the neutral atom platform, namely the large number of individually-addressable atoms in tweezers, while overcoming some of the challenges, namely the very long cooling and state preparation times. Here, idle atoms are prepared and cooled while at the same time other atoms are engaged in atom-photon entanglement attempts. 

The rate of successive attempts on an individual atom is limited by the details of the atom-photon entanglement protocol, including the spontaneous emission timescale or the bandwidth of the optical cavity. The number of times an atom can be reused is determined by the heating rate, often requiring re-cooling after a number of attempts or transitioning to the next atom \cite{Li2024}. This process can be optimized by placing many atoms in the same optical cavity and using individually addressed Stark shifts to shift all but the desired one out of resonance with the cavity \cite{hu2025site}. Therefore, often the predicted attempt rates are limited by the switching, reset, or recooling times. However, certain optical cavity designs with long optical path lengths and high finesse may necessitate even slower rates \cite{Schupp2021}. As a result, the overall entanglement generation rate is constrained by both the per-attempt success probability and the repetition rate. 

Ultimately, the long cycle times and pulsed operation of atom array experiments will constrain the overall achievable entanglement rates. The recently demonstrated continuous-reloading technique of neutral atom arrays \cite{li25continuous,Chiu_2025} could be extended to atom arrays interfaced with optical cavities, providing a continuously refilled reservoir \cite{Li2024} or a continuous flow of atoms into an optical cavity \cite{Sinclair2025}, enabling higher effective duty cycles and continuous entanglement generation.

Recently, the Rempe group demonstrated temporally multiplexed atom–photon entanglement generation by coupling an array of atoms to an optical cavity \cite{Hartung2024}. In this system, multiple atoms share a common photonic interface, enabling sequential emission attempts that approach the cavity-limited repetition rate. In addition, various designs and techniques have been proposed utilizing temporal multiplexing and focused on increasing the number of atoms and minimizing down times between atom-photon entanglement attempts. \cite{huie2021multiplexed,Sinclair2025,Li2024,sunami2025scalable}.

\textbf{Spatial multiplexing} can improve upon temporal multiplexing by matching the  attempt rate across many channels in parallel. The resulting remote entanglement rate in principle increases linearly with the number of parallel channels. This approach requires multiple copies of the remote-entanglement generation hardware, for example arrays of optical cavities, fibers, and photonic Bell state analyzers.

In free space, spatial multiplexing is naturally achieved by collecting light simultaneously from all atoms. While the collection efficiency for a single atom is quite small in practice ($\sim 1\%$) this approach can outperform optical cavities for large arrays >40,000 atoms \cite{Sinclair2025}. The Covey group successfully demonstrated small-scale parallel atom–photon entanglement from an array of Yb atoms when collecting photons with a microscope objective and coupling them into a 5-site fiber array \cite{li2025parallelized} [see Fig.~\ref{fig:FigureYbCovey_combined}].

For cavities, spatial multiplexing could also be achieved by utilizing \textbf{cavity arrays}. Micro- and nano-scale cavities are particularly suited to this approach. In addition to providing spatial multiplexing, these systems are amenable to fully integrated approaches in which dense multi-cavity arrays can be incorporated on a single chip and seamlessly coupled to bus waveguides and optical-fiber networks \cite{Kato2019}.

\begin{figure}[H]
    \centering

    \begin{subfigure}[t]{0.48\textwidth}
        \centering
        \includegraphics[width=\textwidth]{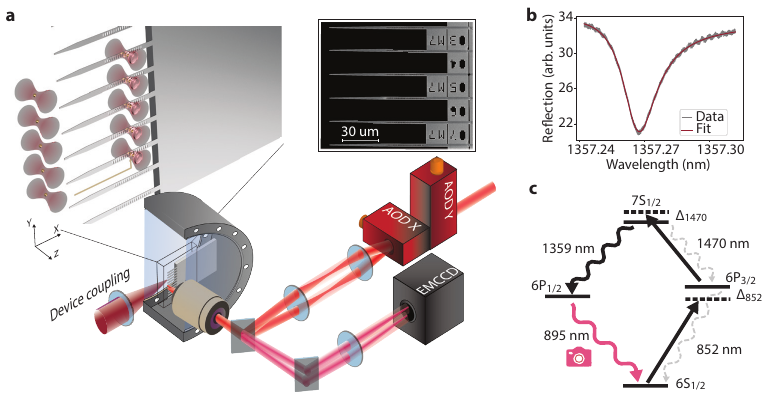}
        \caption{}
        \label{fig:BernienNanophotonic}
    \end{subfigure}
    \hfill
    \begin{subfigure}[t]{0.48\textwidth}
        \centering
        \includegraphics[width=\textwidth]{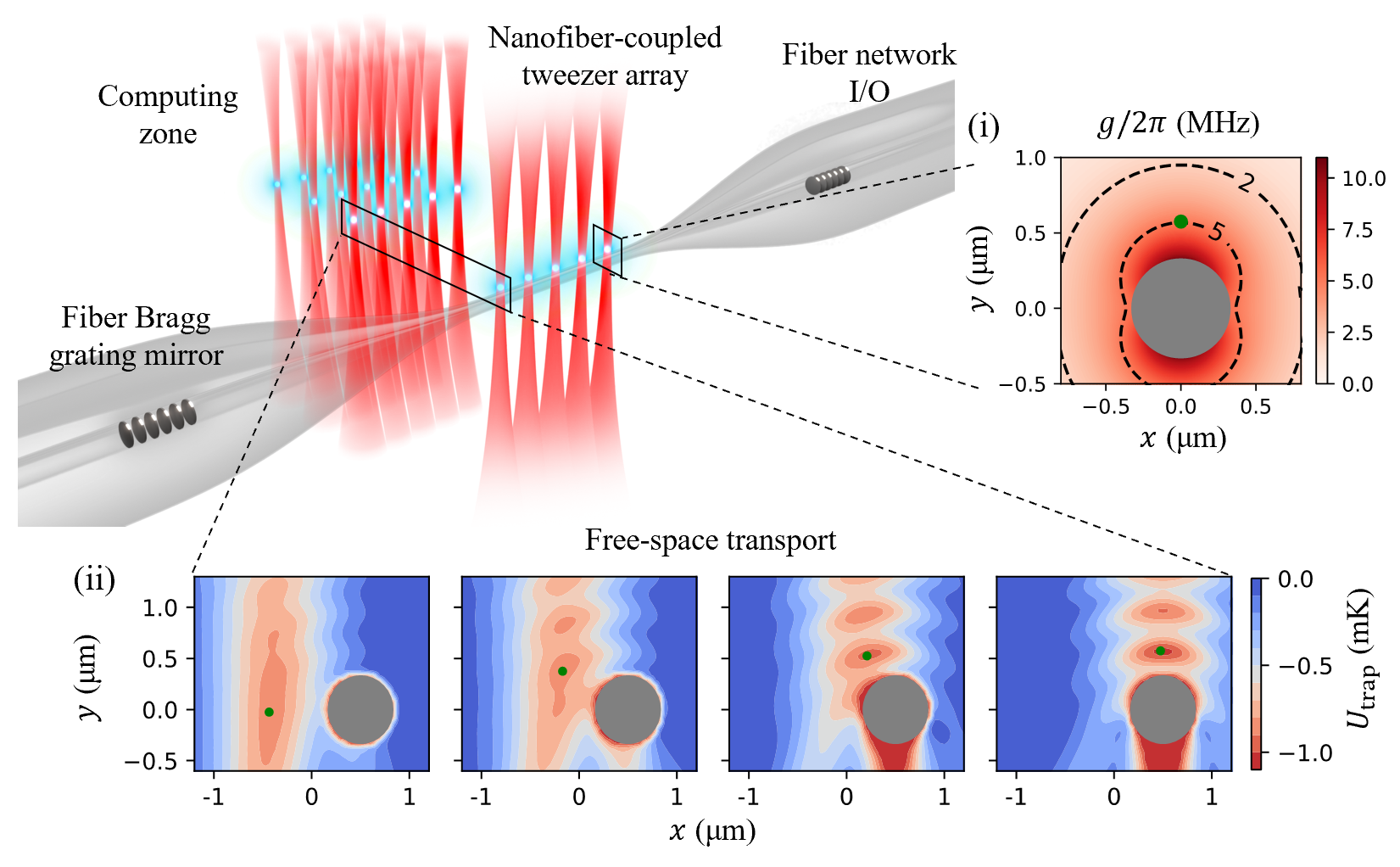}
        \caption{}
        \label{fig:NanoQT}
    \end{subfigure}

    \vspace{1em}

    \begin{subfigure}[t]{0.48\textwidth}
        \centering
        \includegraphics[width=\textwidth]{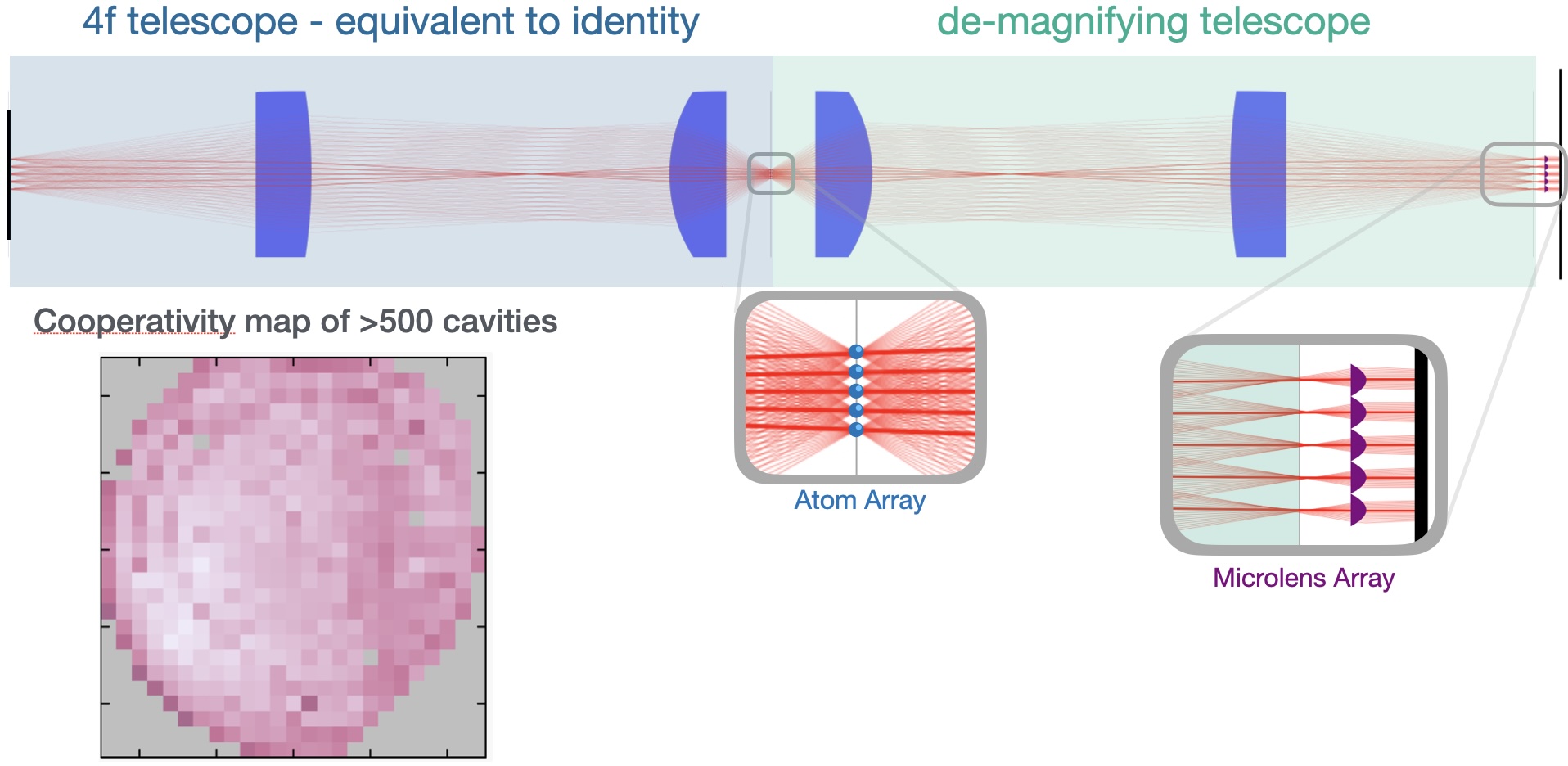}
        \caption{}
        \label{fig:SimonCavityArray}
    \end{subfigure}
    \hfill
    \begin{subfigure}[t]{0.48\textwidth}
        \centering
        \includegraphics[width=\textwidth]{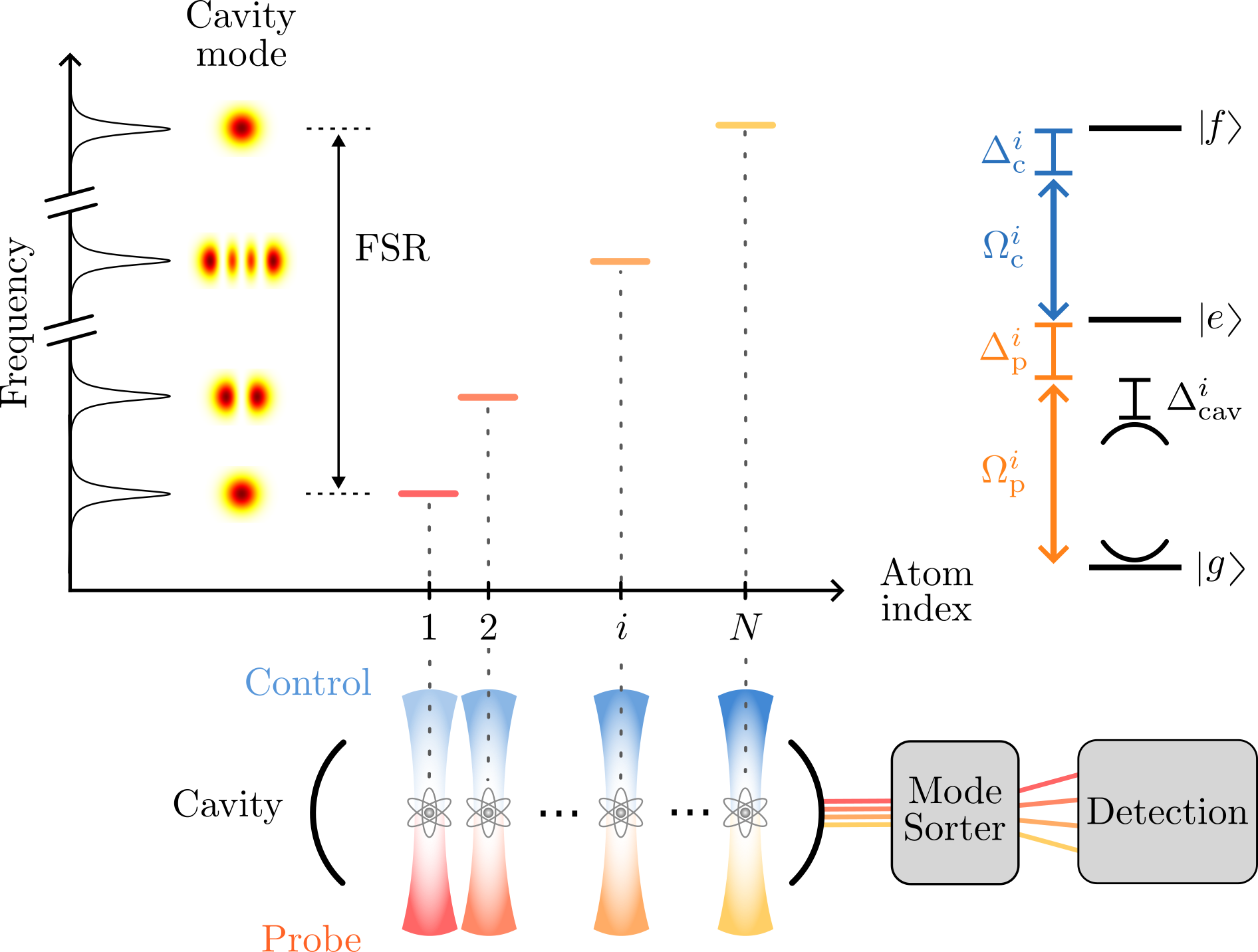}
        \caption{}
        \label{fig:VuleticFrequencyMultiplex}
    \end{subfigure}

    \caption{\textbf{Cavity-based quantum networking platforms with multiplexing capabilities.}
    (i) \textbf{Atom array—nanophotonic chip platform.} Schematic of an optical tweezer array manipulating single cesium atoms near a chip hosting nanophotonic devices, integrated into a UHV chamber with optical control and imaging components. Also shown are cavity reflection spectra and a level structure for background-free imaging.
    (ii) \textbf{Nanofiber cavity-based atom-photon interface.} A nanofiber cavity formed by fiber Bragg gratings supports strong evanescent coupling between atoms and guided modes. Optical tweezers load atoms into a standing-wave trap near the fiber surface, enabling high cooperativity and time-multiplexed networking.
    (iii) \textbf{Cavity array architecture.} A microscope-based cavity array with dual high-NA imaging systems, retro-reflecting mirrors, and an in-cavity microlens array enables large-scale cavity-resolved atom-photon interfaces.
    (iv) \textbf{Cavity-mode multiplexing.} Site-selective light shifts tune different atoms into resonance with distinct cavity modes, allowing frequency-multiplexed photon emission and downstream mode sorting.}
    \label{fig:multiplexing_combined}
\end{figure}

A nanophotonic cavity array, depicted in Fig. \ref{fig:BernienNanophotonic}, was recently realized in the Bernien group, showcasing integration with tweezer arrays and background-free imaging near the nanophotonic structure \cite{Menon2024}. Similarly, nanofiber cavity arrays have been proposed for integration with a reconfigurable optical tweezer array providing efficient transport of atoms between the network zone and the computing zone, supporting a continuous supply of entangled atoms for high-rate quantum networking, as shown in Fig.~\ref{fig:NanoQT}~\cite{sunami2025scalable,kikura2025tamingrecoil,kikura2025passivequantuminterconnects}. Finally, silicon micromirrors can be fabricated with sizes of a few tens of microns offering a scalable and flexible geometry for designing cavity arrays for integration with a nearby Rydberg atom quantum processor \cite{ding2026high}. 

\textbf{Lens arrays} integrated in a large optical cavity is an alternative approach to multiplexed optical cavities that has been demonstrated in the Simon group. They implement a novel optical cavity design to generate an array of small-waist independent cavity modes \cite{shadmany2025cavity,soper2026stability}. This approach leverages the high-NA optics already in use for  atom array state-manipulation, imaging, and transport, combined with retro-reflecting end mirrors, to generate an array of cavities whose modes are stabilized against aberrations by a microlens array (see Fig.~\ref{fig:SimonCavityArray}). This ``one cavity per atom'' approach enables highly parallelized, \emph{in situ} networking (and cavity-enhanced readout) within the main atom array, absent the need for additional transport to cavities. Furthermore, it ensures that all optical elements remain mm$\sim$cm from the atom array, fully mitigating the effects of the dielectric surfaces. The relatively low  cooperativities  $C\approx 10$, combined with low cavity linewidths due to the long lengths of these resonators, limits the per-cavity readout speed compared to their nano-fabricated counterparts. The low finesse requirements of this approach mean that in-cavity frequency conversion~\cite{taneja2025light} to/from telecom is potentially possible.

\textbf{Cavity-mode multiplexing}~\cite{aqua2026mode} is another recently introduced approach to scalable photonic interfaces, in which multiple modes of a single optical cavity are used as independent interaction channels. Individual atoms can be selectively coupled to different cavity modes using site-resolved control fields that tune the atomic transitions into resonance with the desired modes (see Fig.~\ref{fig:VuleticFrequencyMultiplex}). At the cavity output, the different modes can be efficiently separated either spectrally, using high-resolution dispersive elements~\cite{limbach2019fully,wei202610}, or spatially via multi-plane light conversion~\cite{fontaine2019laguerregaussian,fontaine2021hermitegaussian}, allowing independent detection and routing of photons emitted into each mode. In this way, a single cavity can support many atom–cavity interactions simultaneously across different modes.

By lifting the single-mode constraint and enabling parallel interactions, cavity-mode multiplexing relaxes the conventional tradeoff between cavity length and interaction rate. It therefore enables the use of larger cavities that can host many atoms, reduce the need for extensive atom transport, and remain compatible with the geometry of state-of-the-art atom arrays. These features make it a promising route toward high-rate entanglement distribution in networks of atomic quantum processors.

\textbf{Spectral multiplexing} provides another route to increasing remote-entanglement rates by distributing photons from many communication qubits over distinct optical frequency channels. In a recently proposed architecture~\cite{Nejabati2025}, reconfigurable quantum interfaces convert atom-generated photons into selected dense-wavelength-division-multiplexing (DWDM) channels while also allowing temporal-mode matching, after which wavelength-selective switches route the photons between QPUs. This approach combines parallel Bell-pair generation with compatibility with telecom networking components, and is therefore another promising direction for scaling neutral-atom interconnects beyond single-channel entanglement distribution.

\subsubsection{Integration with External Devices and Disparate Quantum Systems}
\begin{figure}[H]
    \centering
    \includegraphics[width=1\linewidth]{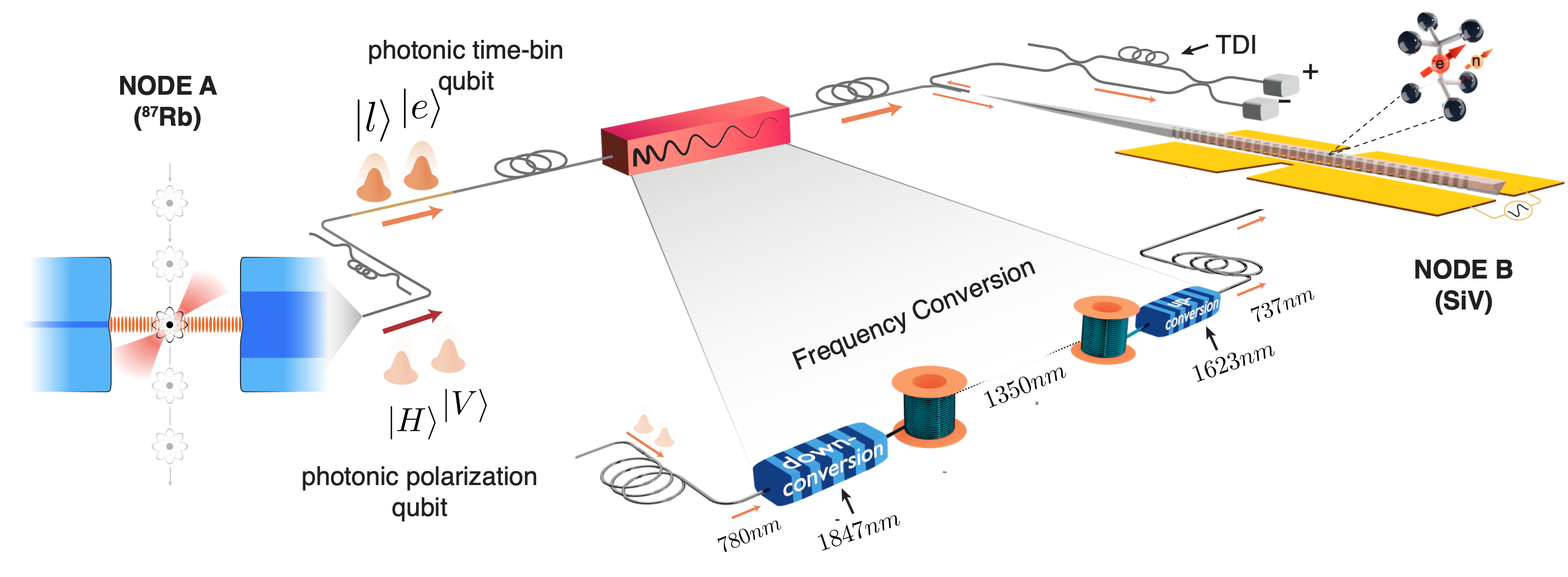}
    \caption{\textbf{Proposed experiment demonstrating entanglement between disparate quantum systems:} the polarization-based photonic qubit is emitted from an atom in a fiber cavity while maintaining entanglement with the atom. The photonic qubit undergoes a transformation using a time-delay interferometer with a PBS from polarization-based to time-bin encoding to allow interfacing with the SiV in a nanocavity after frequency conversion from 780 nm to 737 nm via 1350 nm. The photonic time-bin qubit then undergoes reflection-based spin-photon entanglement with the electron spin of the SiV generating a GHZ state between a photon, Rb atom and electron spin of the SiV. Finally, the photonic qubit is measured in the X basis using a time-delay interferometer (TDI), leaving two matter qubits entangled across two labs.
}
    \label{fig:networking}
\end{figure}

In addition to networking identical modules, interfacing disparate quantum system involving other quantum information processing platforms is an emerging research field. Such heterogeneous integration could, in principle, utilize the strengths of each platform and open up new research directions. 
Networking disparate quantum devices necessarily requires a shared mode of communication.  Photonic transport of quantum states is the only feasible candidate for long-distance links. Furthermore, the low attenuation in telecom fiber and existing deployed infrastructure strongly motivate using optical wavelengths in the telecom band for long-distance communication. This has recently been demonstrated through the entanglement of solid-state defects over kilometers of deployed telecom fiber \cite{knaut2024entanglement}. In this and other work, frequency conversion is used to translate photons into the telecom band. Importantly, this same frequency conversion can bridge the gap between disparate quantum systems, which will almost certainly operate at differing wavelengths. For one example of a proposed experiment, see Fig. \ref{fig:networking}. Frequency conversion, as well as methods for converting between different types of photonic qubits (e.g.\ polarization, time-bin), are well established, making photons an efficient and well-motivated carrier for generating entanglement between disparate quantum systems. We also note that many atomic species commonly used for quantum computing (Rb, Cs, Yb) have native telecom-band transitions, and several research groups are pursuing atom-photon entanglement using telecom-band photons \cite{li2025parallelized}.

While in a modular quantum computing architecture, the modules are most likely to be in the same physical location, related networking applications motivate long-distance quantum communication, requiring quantum repeaters. Optical defects in diamond, especially the silicon-vacancy center defect, are exciting candidates for hardware-agnostic quantum repeaters. The SiV defects are good quantum memories and can easily be made to interface with telecom wavelength photons \cite{knaut2024entanglement}. Utilizing these systems as a memory in between networked quantum systems can extend the range of communication as a quantum repeater. Using the SiV as a small memory node can also enhance networking rates by storing a photon from one client in the memory while waiting for another. Upon successful storage of photons from both clients, the memory performs a Bell-state measurement realizing distributed entanglement. The existence of a small memory register thus removes the constraint that photons from both clients arrive simultaneously at the Bell-state analyzer, and as a result can enhance networking rates by over an order of magnitude. In addition to high-fidelity entanglement in the telecom band, these systems offer a clear path towards device-agnostic distributed entanglement, enabling the integration of external and disparate quantum devices.

\subsubsection{Modular architectures based on transportable atom arrays}
\begin{figure}[H]
    \centering
    \includegraphics[width=0.7\linewidth]{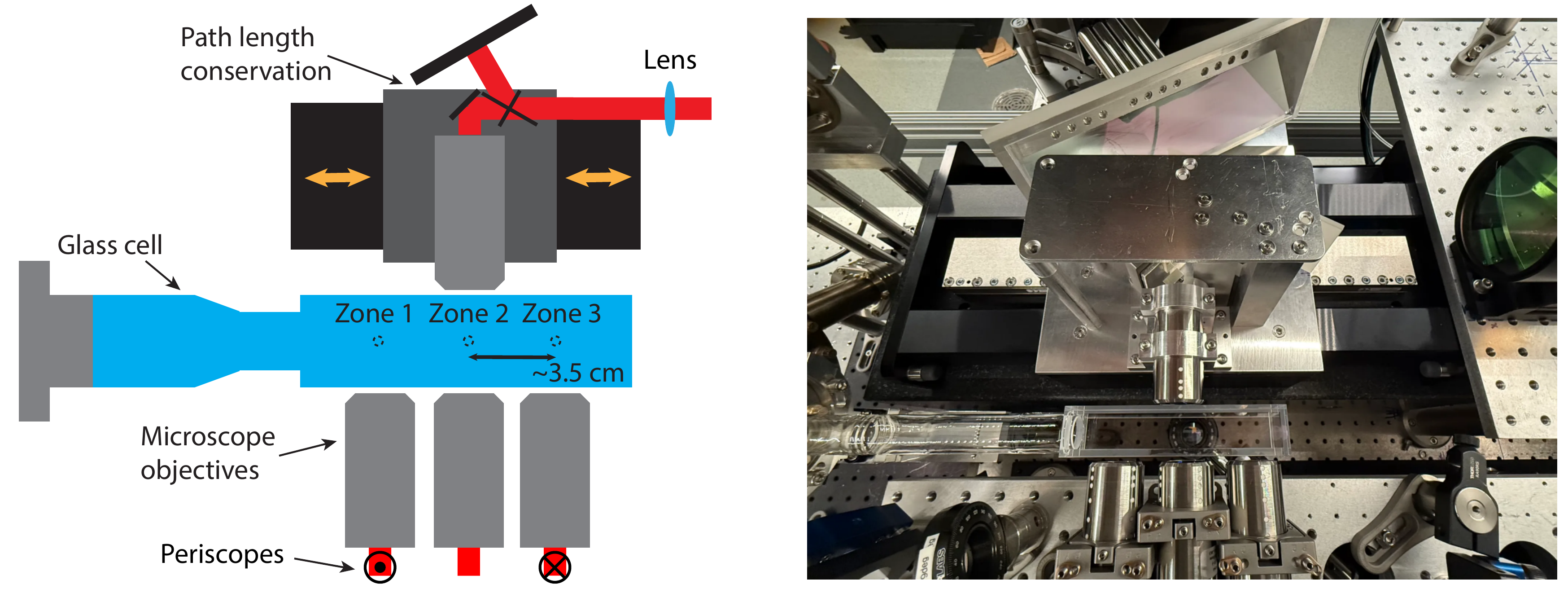}
    \caption{\textbf{A modular architecture based on transportable atom arrays:} An ``array of arrays" is co-located within a single vacuum chamber (e.g. glass cell) via an array of microscope objectives. A transportable atom array, projected from opposite the static modules but lying on the same plane, carries large atom arrays between the static modules such that parallelized Rydberg-mediated gates can be performed between them.}
    \label{fig:FigureCsCovey}
\end{figure}

Optical interconnects are required when the network nodes span distances of meters or kilometers. However, for the purpose of building a modular quantum processor, it may be desirable to co-locate all the modules as close together as possible within a single vacuum chamber. One example might be to have an array of microscope objectives that are spaced by $\lesssim4$ cm, thereby generating an ``array of arrays". Each module could contain 10's of thousands of qubits, and thus a 10-site array of modules could potentially reach a qubit count of $\gg100,000$. This vision is illustrated in the left side of Fig.~\ref{fig:FigureCsCovey}.

The Covey group is pursuing this alternate vision through an experimental effort to develop transport between atom array modules by transporting the microscope objective itself across $20$ cm using air-bearing linear translation stages (see right-side of Fig. \ref{fig:FigureCsCovey}). Their approach is enabled by preserving the optical path length between the objective and the preceding lens. Optical characterization has demonstrated that an array of approximately $8000$ tweezers can be translated over $15$ cm with no measurable change in performance. They have further demonstrated $100$ nm of bidirectional repeatability in tweezer position during maximum acceleration/deceleration and maximum velocity, comparable to the atomic wavefunction at typical operating temperatures and tweezer depths. 

Entanglement between modules will be performed by `fly-by' Rydberg-mediated gates requiring only 100's of ns, during which relative motion is negligible even at speeds of $\sim$m/s. At top speed, the dynamic objective will cross adjacent modules in tens of milliseconds. Parallelizing over arrays of thousands of atoms, this approach could yield entanglement distribution rates of $\gtrsim10^5$ per second between adjacent modules \cite{xue2026}. The slowdown due to turning around at each end of the travel range can be used for atom reloading, and could be obviated by potentially using a circular motion profile for the transportable array(s).

\section{Conclusions and Outlook} \label{sec:conclusions}
This is a very special time in the field of quantum computation, with many platforms having achieved landmark breakthroughs over the last decade. With the neutral atom logical quantum processor architecture in particular, a path is now emerging for reaching the 10,000 to 100,000 physical qubit level. Scalable optical control methods enabling generation and manipulation of hundreds of long-lived logical qubits in the deep circuit regime, and with quantum networking taking us even further. Combined with efforts in quantum algorithm design, new approaches to hardware-efficient quantum error correction, and optimized quantum circuit compilation, early examples of practical quantum advantage may soon be within reach. Nevertheless, there are significant technical challenges to be overcome along the way, requiring broad participation from the entire quantum science and engineering community to make the future breakthroughs that will continue to drive progress forward.

\section*{Acknowledgments}
We thank all the participants of the NSF NQVL Town Hall on Advancing 
Quantum Computing with Neutral Atoms, held at MIT Endicott House in 
January 2025, for stimulating discussions on many of the ideas contained 
in this strategic plan. This work was supported by the U.S. National 
Science Foundation under Award Nos.\ 2410716 and 2533041 through the 
National Quantum Virtual Laboratory (NQVL) program.

\section*{Disclaimer}
This publication was prepared by a multi-institutional group of researchers affiliated with universities, national laboratories, and commercial entities active in neutral atom quantum computing. The authors include individuals employed by or affiliated with Pasqal, QuEra Computing Inc., Infleqtion, planqc GmbH, and Nanofiber Quantum Technologies, Inc. (NanoQT), as well as researchers at academic institutions. The views expressed herein represent the scientific assessments of the individual authors and do not necessarily reflect the official positions, policies, or endorsements of any affiliated company, university, funding agency, or government body. The mention of specific commercial products, platforms, software tools, or company names is for scientific clarity and does not constitute an endorsement or recommendation by the authors, their institutions, the National Science Foundation, or any other funding agency. No proprietary or commercially sensitive information has been disclosed. The authors have endeavored to present technical assessments impartially, but readers should be aware that some authors have financial interests in companies developing neutral atom quantum computing technologies. Funding sources and potential conflicts of interest for individual authors are available upon request.
\printbibliography

\end{document}